\documentclass[journal]{IEEEtran}

\usepackage{array, calc, color, multicol}
\usepackage{algorithm}
\usepackage[noend]{algorithmic} %, algorithm2e}
\usepackage{graphics, epstopdf, graphicx, epsfig, psfrag, epsf, subfigure}
\usepackage{amssymb, amsfonts, amsmath, times}
\usepackage{multicol,balance, verbatim}
\usepackage{textcomp, mathrsfs, stmaryrd, setspace}

\newcommand{\ie}{{\em i.e., }}
\newcommand{\eg}{{\em e.g., }}

\newcommand{\state}{I$^2$NC-state~}
\newcommand{\stateless}{I$^2$NC-stateless~}
\newcommand{\cope}{COPE~}

\newtheorem{example}{Example}

\DeclareGraphicsExtensions{.eps, .pdf, .png, .jpg}   % optional
\newcommand{\Jset}{\mathcal{J}}
\newcommand{\Sset}{\mathcal{S}}
\newcommand{\Kset}{\mathcal{K}}
\newcommand{\Aset}{\mathcal{A}}
\newcommand{\Pset}{\mathcal{P}}
\newcommand{\Cset}{\mathcal{C}}
\newcommand{\Nset}{\mathcal{N}}
\newcommand{\Hset}{\mathcal{H}}
\newcommand{\Iset}{\mathcal{I}}

\newcommand{\Qset}{\mathcal{Q}}

\begin{document}

\title{Intra- and Inter-Session Network Coding \\ in Wireless  Networks
\thanks{H.~Seferoglu is with the Laboratory for Information and Decision Systems (LIDS), Massachusetts Institute of Technology. Email: {\tt hseferog@mit.edu}. Mail: 77 Massachusetts Avenue, Room 32-D671, Cambridge, MA 02139.}
\thanks{A.~Markopoulou is with the Electrical Engineering and Computer Science Department, University of California, Irvine. Email: {\tt athina@uci.edu}. Mail: CalIT2 Bldg, Suite 4100, Irvine, CA 92697.}
\thanks{K.~K.~Ramakrishnan is with AT\&T Labs Research. Email: {\tt kkrama@research.att.com}. Mail: 180 Park Avenue, Building 103
Florham Park, NJ 07932.}
}

\author{Hulya Seferoglu, {\em Member}, {\em IEEE}, Athina Markopoulou, {\em Member}, {\em IEEE}, K.~K.~Ramakrishnan, {\em Fellow}, {\em IEEE}}

\maketitle

\pagestyle{empty}
\thispagestyle{empty}

\begin{abstract}

In this paper, we are interested in improving the performance of constructive network coding schemes in lossy wireless environments. We propose I$^2$NC - a cross-layer approach that combines inter-session and intra-session network coding and has two strengths. First, the error-correcting capabilities of intra-session network coding make our scheme resilient to loss. Second, redundancy allows intermediate nodes to operate without knowledge of the decoding buffers of their neighbors. Based only on the knowledge of the loss rates on the direct and overhearing links, intermediate nodes can make decisions for both intra-session (\ie how much redundancy to add in each flow) and inter-session (\ie what percentage of flows to code together) coding. Our approach is grounded on a network utility maximization (NUM) formulation of the problem. We propose two practical schemes, I$^2$NC-state and I$^2$NC-stateless, which  mimic the structure of the NUM optimal solution. We also address the interaction of our approach with the transport layer. We demonstrate the benefits of our schemes through simulations.
\end{abstract}


\section{\label{sec:intro}Introduction}
Wireless environments lend themselves naturally to network coding (NC), thanks to their inherent broadcast and overhearing capabilities. In this paper, we are interested in wireless mesh networks used for carrying traffic from unicast sessions, which is the dominant traffic today. Network coding has been used as a way to improve throughput over such wireless environments. Given that optimal inter-session NC  for unicast is still an open problem, constructive approaches are used in practice \cite{chou-unicast, cope, tiling, poison_antidote, BFLY}. One of the first practical wireless NC systems is COPE \cite{cope} - a coding shim between the IP and MAC layers that performs one-hop, opportunistic NC. COPE codes packets from different unicast sessions, and relies on receivers being able to decode these using overheard packets. This way, COPE combines multiple packets by using information on overheard packets which are exchanged through transmission reports and effectively forwards multiple packets in a single transmission to improve throughput. In order for COPE to work in a multihop network, nodes must cooperate to (i) exchange information about what packets they have overheard and also (ii) code so that all one-hop downstream nodes can decode. This must be done at every hop across the path of a flow and cross-layer optimization approaches can be used \cite{hs_netcod_2010} to further boost the performance.

One important problem that remains open, and is the focus of this paper, is COPE's performance in the presence of non-negligible loss rates. The reason is that intermediate nodes in COPE require the knowledge of what their neighbors have overheard, in order to perform one-hop inter-session NC. However, in the presence of medium-high loss rate, although each node fully cooperates to report what it has overheard, this information is limited, possibly corrupted, and/or delayed over lossy wireless channels. COPE turns off NC if loss rate exceeds a threshold with default value 20\% \cite{cope}. However, this does not take full advantage of all the available NC opportunities. To better illustrate this key point, let us discuss the following example.

\begin{figure}[t!]
\vspace{-10pt}
\centering
\includegraphics[scale=0.32, angle=-90]{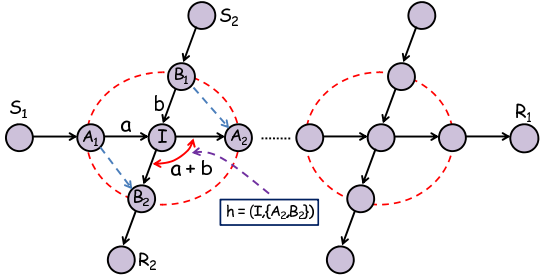}
\vspace{-15pt}
\caption{\label{fig:one-hop} Example of a unicast flow (from $S_1$ to $R_1$) traversing multiple wireless hops. Each node performs (intra- and inter-session) NC. The neighborhood of $I$ is shown here in detail. (Two unicast flows, $S_1-R_1$ and $S_2-R_2$, meeting at intermediate node $I$. $I$ receives  packets $a,b$ from  nodes $A_1, B_1$, respectively. It can choose to broadcast $a$, $b$ or $a+b$ in a single transmission to both receivers. The next hops $A_2, B_2$ can decode $a+b$ because they overhear packets $b$ and $a$ transmitted from $B_1, A_1$, respectively.)}
\vspace{-15pt}
\end{figure}

\begin{example}\label{ex1}
Let us consider Fig.~\ref{fig:one-hop}, and focus on the neighborhood of node $I$, \ie only the packets transmitted via $I$, from $A_1$ to $A_2$ and from $B_1$ to $B_2$. This forms an ``X'' topology which is a well-known, canonical example of one-hop opportunistic NC \cite{chou-unicast, cope}. In the absence of loss, throughput is improved by $33.3\%$, because $I$ delivers two packets in three transmissions (with NC), instead of four (without NC). Let us re-visit this example when there is packet loss. Assume that there is loss only on the overhearing link $A_1-B_2$, with probability $\rho_{\{A_1,B_2\}}=0.3$, and all other links have no loss. In this case, $70\%$ of the packets can still be coded together, and throughput can be improved by $26\%$,  which is still a significant improvement. Even at higher loss rate, \eg $\rho_{\{A_1,B_2\}}=0.5$, inter-session coding still improves throughput up to $20\%$ %$16.6\%$.
This is under the assumption that $I$ knows the exact state of $A_2,B_2$, \ie what packets were overheard, and thus $I$ is able to decide what packets to code together so as to guarantee decodability at the receivers. However, at high loss rates, cooperation among nodes becomes difficult. This is why COPE turns off the coding functionality when loss rate is higher than a threshold with default value 20\%, thus not taking full advantage of all coding opportunities.
\hfill $\blacksquare$
\end{example}

We propose a solution to this problem with a design which combines intra- and inter-session NC over wireless mesh networks. We use intra-session NC to combine packets within the same flow and introduce parity packets to protect against loss. Then, we use inter-session NC to combine packets from different (already intra-session coded) flows, and thus increase throughput. Our approach for combining intra-session with inter-session NC, which we refer to as I$^2$NC, has two key benefits. First, it can correct packet loss and still perform inter-session NC, even in the presence of medium-high loss rates, thus improving throughput. Second, the use of intra-session NC makes all packets in the session equally beneficial. Thus, I$^2$NC eliminates the need to know the exact packets that have been overheard by the neighbors of intermediate node $I$. It is sufficient to know the loss probabilities of overheard and transmitted packets. In our scheme, this information is reported by each node to the nodes in its neighborhood which makes NC possible even at higher loss rates.

Adding redundancy in this setting is non-trivial, since a flow is affected not only by loss on its direct links, but also by loss on overhearing links. This affects the decodability of coded packets. Therefore, the amount of redundancy needed to be determined carefully.

{\em Example 1 - continued:} Consider again the neighborhood of $I$ in Fig.~\ref{fig:one-hop}. Flow 2 (originated from $S_2$) is affected not only by loss on its own path $B_1-I-B_2$, but also by loss on the overhearing link $A_1-B_2$, which affects the decodability of coded packet $a+b$ at $B_2$. In order to protect flow 2 from high loss rate on the overhearing link $A_1-B_2$, $I$ may decide either to add redundancy on flow 2, or to not perform coding, or a combination of the two. On the other hand, $I$ may also decide to add redundancy on flow 1 (originated from $S_1$), to correct loss on the overhearing link $A_1-B_2$, thus helping $B_2$ to receive $a$ and decode $a+b$.
\hfill $\blacksquare$

Therefore, a number of questions need to be addressed in the design of a system that combines both intra- and inter-session NC. In particular:
\begin{itemize}
\item[Q1:] {\em How to gracefully combine intra- and inter-session NC}? We propose a generation-based design, and specify the order we perform the two types of coding.
\item[Q2:] {\em How much redundancy to add in each flow?} We show how to adjust the amount of redundancy after taking into account the loss on the direct and overhearing links. We implement the intra-session NC functionality as a thin layer between IP and transport layer.
\item[Q3:] {\em What percentage of flows should be coded together} and what parts should remain uncoded? We design algorithms that make this decision taking into account the loss characteristics on the direct and overhearing links. We implement this and other functionality (\eg queue management) performed with or after inter-session NC as a layer between MAC and IP.
\item[Q4:] {\em What information should be reported to make these decisions}? We propose two schemes: {I$^2$NC-state}, which needs to know the state (\ie overheard packets) of the neighbors; and {I$^2$NC-stateless}, which only needs to know the loss rate of links in the neighborhood.
\end{itemize}

Our approach is grounded on a network utility maximization (NUM) framework \cite{tutorial_doyle}. We formulate two variants of the problem, depending on available information (as in Q4 above). The solution  of each problem decomposes into several parts with an intuitive interpretation, such as rate control, NC rate, redundancy rate, queue management, and scheduling. The structure of the optimal solution provides insight into the design of our two schemes, I$^2$NC-state and I$^2$NC-stateless.

We evaluate our schemes in a multi-hop setting, and we consider their interaction with the transport layer, including TCP and UDP. We propose a thin adaptation layer at the interface between TCP and the underlying coding, to best match the interaction of the two. We perform simulations in {\tt GloMoSim} \cite{glomosim}, and we show that our schemes significantly improve throughput compared to COPE.

The structure of the rest of the paper is as follows. Section~\ref{sec:related} presents related work. Section~\ref{sec:system} gives an overview of the system model. Section~\ref{sec:opt2} presents the NUM formulation and solution. Section~\ref{sec:algs} presents the design of the {I$^2$NC} schemes in detail. Section~\ref{sec:performance} presents simulation results. Section~\ref{sec:conclusion} concludes the paper.

\section{\label{sec:related}Related Work}
{\em COPE and follow-up work.} This paper builds on COPE, a practical scheme for one-hop NC across unicast sessions in wireless mesh networks \cite{cope}, which has generated a lot of research interest. Some researchers tried to model and analyze COPE \cite{proutiere_scheduling}, \cite{how_many_packets_infocom_2008}, \cite{sudipta_sengupta_infocom_2007}. Some others proposed new coded wireless systems, based on the idea of COPE \cite{practical_nc_dong}, \cite{BFLY}. In \cite{medard_ita}, the performance of COPE is improved by looking at its interaction with MAC fairness. Our recent work in \cite{hs_netcod_2010} improves TCP's performance over COPE with a NC-aware queue management scheme. This paper also improves COPE by adding intra-session redundancy with a cross-layer design and reducing the amount of information that is needed to be exchanged among nodes cooperatively, \ie nodes no longer need to know the exact packets overheard by their neighbors and can operate
only with knowledge of the link loss rates.

{\em NUM in coded systems.} The NUM framework can be applied in networks, to understand how different layers and/or modules (such as flow control, congestion control, routing, etc.) should be restructured when NC is used. Although the approach is general, the parts and interpretation of the distributed solution is highly problem-specific. For NUM to be successful, the optimization model must be formulated so as to capture and exploit the NC properties.
This is highly non-trivial and problem-specific. A body of work has looked at the joint optimization of NC of unicast flows, formulated in a NUM framework.

Optimal scheduling and routing for COPE are considered in \cite{proutiere_scheduling} and \cite{sudipta_sengupta_infocom_2007}, respectively. 
A linear optimization framework for packing butterflies is proposed in \cite{poison_antidote}. A re-transmission scheme for one-hop NC is proposed in \cite{rayanchu_nc_fec}. Forward error correction over wireless for pairwise NC is proposed in \cite{nc_fec_pairwise}, \cite{ronasi_nc_fec}, which are also the most closely related formulations to ours. Our main differences are that we consider: (i) multiple flows coded together instead of pairwise, (ii) local instead of end-to-end redundancy, and (iii) the effect of losses over direct and overhearing links, to generate the right amount of redundancy.

{\em Dealing with wireless loss.} Recent studies of IEEE 802.11b based wireless mesh networks \cite{MPLOT_ref1}, \cite{MPLOT_ref2}, have reported packet loss rates as high as 50\%. 
Dealing such level of loss in wireless networks is a hard enough problem on its own, which is further amplified by NC. There is a wide spectrum of well-studied options for dealing with loss, \eg using redundancy and/or re-transmissions, locally (MAC) and/or end-to-end (transport layer). Local re-transmissions increase end-to-end delay and jitter, which, if excessive, may cause TCP timeouts or hurt real-time multimedia. Furthermore, the best re-transmission scheme for network coded packets varies with the loss probability\footnote{ \scriptsize We have observed through simulations that if a network coded packet is lost for one receiver but received correctly for other receiver(s), it is better to re-transmit the same network coded packet for low loss rates. However, it is better to combine the packet which is lost in the previous transmission with new packets for high loss rates.
} and it is hard to switch among re-transmission policies when the loss rate varies over time. Re-transmission also requires state synchronization to perform inter-session NC, which is not reliable at all loss rates. We follow an alternative approach of local redundancy because we are interested in keeping delay low and we want to eliminate the need for knowing the state of neighbors.

There is extended work on TCP over wireless. One key problem is the need to distinguish between wireless and congestion loss and have TCP react only to congestion; this is possible \eg through Explicit Congestion Notification (ECN). When re-transmissions exceed the delay budget, end-to-end redundancy may also be used to combat  loss on the path \cite{LT_TCP}. The error-correcting capabilities of intra-session NC have recently been used in conjunction with the TCP sliding window in \cite{NC_meets_TCP}. In contrast, we focus on one-hop inter-session coding rather than end-to-end intra-session coding.

\section{\label{sec:system}System Overview}
We consider multi-hop wireless networks, where intermediate nodes perform intra- and inter-session NC ({I$^2$NC}). Next, we provide an overview of the system and highlight some of its key characteristics.

\subsection{Notation and Setup}

\subsubsection{Sources and Flows} Let $\Sset$ be the set of unicast flows between source-destination pairs in the network. Each flow  $s \in \Sset$ is associated with a rate $x_{s}$ and a utility function $U_{s}(x_{s})$, which we assume to be a strictly concave function of $x_{s}$.

\subsubsection{Wireless Transmission}
Packets from a source (\eg $S_1$ in Fig.~\ref{fig:one-hop}) traverse potentially multiple wireless hops before being received by the receiver (\eg $R_1$). We consider a model for interference described in \cite{gupta_interference_model}: each node can either transmit or receive at the same time, and all transmissions in the range of the receiver are considered as interfering.

We use the following terminology for wireless. A hyperarc $(i,\Jset)$ is a collection of links from node $i \in \Nset$ to a non-empty set of next-hop nodes $\Jset \subseteq \Nset$. A hypergraph $\Hset=(\Nset,\Aset)$ represents a wireless mesh network, where $\Nset$ is the set of nodes and $\Aset$ is the set of hyperarcs. For simplicity, $h=(i,\Jset)$ denotes a hyperarc, $h(i)$ denotes node $i$ and $h(\Jset)$ denotes the set of nodes in $\Jset$, {\em i.e.}, $h(i)=i$ and $h(\Jset) = \Jset$. We use these notations interchangeably in the rest of the paper. Each hyperarc $h$ is associated with a channel capacity $R_h$. Since $h$ is a set of links, $R_h$ is the minimum  capacity of all the links in the hyperarc, \ie $R_h = \min_{j \in h(J)}\{R_{i,j}\}$ s.t. $i \in \Nset$. In the example of node $I$ in Fig.~\ref{fig:one-hop}, $h = (I, \{B_2,A_2\})$ is one of the hyperarcs, and its capacity is $\min \{R_{\{I,B_2\}}, R_{\{I,A_2\}}\}$.

Note that with both intra- and inter-session NC, it is possible to construct more than one code over a hyperarc $h$. Let $\Kset_{h}$ be the set of inter-session network codes over a hyperarc $h$. $\Sset_{k} \subseteq \Sset$ be the set of flows coded together using code $k \in \Kset_{h}$ and broadcast over $h$.\footnote{Note that we consider constructive inter-session NC, {\em i.e.}, network codes $k \in \Kset_{i,\Jset}$ as well as $h=(i,\Jset)$ is determined at each node with periodic control packet exchanges or estimated through routing table.}

Given $\Hset$, we can construct the conflict graph $\Cset = (\Aset, \Iset)$, whose vertices are the hyperarcs of $\Hset$ and edges indicate interference between hyperarcs. A clique $\Cset_{q} \subseteq \Aset$ consists of several hyperarcs, at most one of which can transmit without interference, \ie a transmission over a hyperarc interferes with transmissions over other hyperarcs in the same clique.

\subsubsection{Loss Model} %Consider one-hop transmission.
A flow $s$ may experience loss in two forms: loss $\rho_{h}^{s}$ over the direct transmission links; or loss $\rho_{h,k}^{s,s'}$ of antidotes\footnote{Following the poison-antidote terminology of \cite{poison_antidote},  we call ``antidotes'' the packets of flows $s'$ that are coded together with $s$, and thus are needed for the next hop of $s$ to be able to decode.  {\em E.g.}, in Fig.\ref{fig:one-hop}, $a$ is the ``antidote'' that $B_2$ needs to overhear over link $A_1-B_2$, to decode $a+b$ and obtain $b$.} on overhearing links. These two types of loss have different impact on network coded flows.

First, let us discuss loss on the direct links. A flow $s$ transmitted over hyperarc $h$ experiences loss with probability $\rho_{h}^{s}$. This probability is different per flow $s$, even if several flows are coded and transmitted over the same hyperarc $h$, because different flows are transmitted to different next hops, thus see different channels. For example, in Fig.~\ref{fig:one-hop}, $\rho_{(I,\{B_2,A_2\})}^{S_1}$ is equal to the loss probability over link $I-A_2$ and $\rho_{(I,\{B_2,A_2\})}^{S_2}$ is equal to the loss probability over link $I-B_2$.

Second, let us discuss the effect of lost antidotes on the overhearing link. Consider that flow $s$  is combined with flow $s'$ s.t. $s \neq s'$, and that some packets of  flow $s'$ are lost on the overhearing link to the next hop of $s$.  Then, coded packets cannot be decoded at the next hop and flow $s$ loses packets, with probability $\rho_{h,k}^{s,s'}$. For example, in Fig.~\ref{fig:one-hop}, packets from flow $S_1$ cannot be decoded (hence are lost) at node $A_2$ due to loss of antidotes from flow $S_2$ on the overhearing link $B_1-A_2$.

In our formulation and analysis, we assume that $\rho_{h}^{s}$ and $\rho_{h,k}^{s,s'}$ are i.i.d. according to a uniform distribution. However, in our simulations, we consider a Rayleigh fading channel model. The loss probabilities are calculated at each intermediate node as explained later in this section.

\subsubsection{Routing}
Each flow $s \in \Sset$ follows a single path $\Pset_{s} \subseteq \Nset$ from the source to the destination, which is pre-determined by a routing protocol, {\em e.g.}, OLSR or AODV, and given as input to our problem. Note that the nature of wireless networks is time varying, \ie nodes join and leave the system dynamically. In such cases, the routing protocol actively determines new paths which are used as input to our problem. It is not critical that the paths remain fixed, neither from a theoretical nor from a practical point of view, as explained in the following sections. Also, note that several different hyperarcs may connect two consecutive nodes along the path. We define $H_{h,k}^{s}=1$ if $s$ is transmitted through hyperarc $h$ using network code $k \in \Kset_{h}$; and $H_{h,k}^{s}\! =\! 0$, otherwise.

\subsection{Intra- and Inter-session Network Coding}

Next, we give an overview of how an intermediate node performs intra- and inter-session NC. The implementation details are provided in Section \ref{sec:algs}.

\subsubsection{Intra-session Network Coding (for Error Correction)}
Consider the commonly used generation-based NC \cite{practical_NC}: packets from flow $s \in \Sset$ are divided into generations (note that we use ``generation'' and ``block'' terms interchangeably), with size $G^{s}$. At the source $s$, packets within the same generation are linearly combined (assuming large enough field size) to generate $G^{s}$ network coded packets. Each intermediate node along the path of flow $s$ adds $P^{s}$ parity packets, depending on the loss rates of the links involved in this hop. At the next hop, it is sufficient to receive $G^{s}$ out of $G^{s}+P^{s}$ packets. The same process is repeated at every intermediate node until the receiver receives $G^{s}$ error-free packets, which can then be decoded and be passed on to the application.

There are many ways to generate parities ($P^s$) in practice. We use generation based intra-session NC \cite{practical_NC} for this purpose. Although one could use various coding techniques, such as Reed-Solomon or Fountain codes, using intra-session NC has several advantages. First, it has lower computational complexity. Second, in systems like COPE that already implement inter-session NC, it is natural to incrementally add intra-session NC functionality. Moreover, in this setting, hop-by-hop intra-session coding (in which redundant packets are generated at each hop) is clearly a better choice than end-to-end coding for dealing with loss. In terms of performance, hop-by-hop coding achieves higher end-to-end throughput (thanks to introducing less redundancy than end-to-end coding), without adding high complexity (and thus delay) to the intermediate nodes. Furthermore, in terms of system implementation, our hop-by-hop scheme requires minimal modifications on top of the inter-session NC, which is already implemented.

\subsubsection{Inter-session Network Coding (for Throughput)}
\begin{figure}
\centering \vspace{-5pt}
\includegraphics[width=7cm, angle=-90]{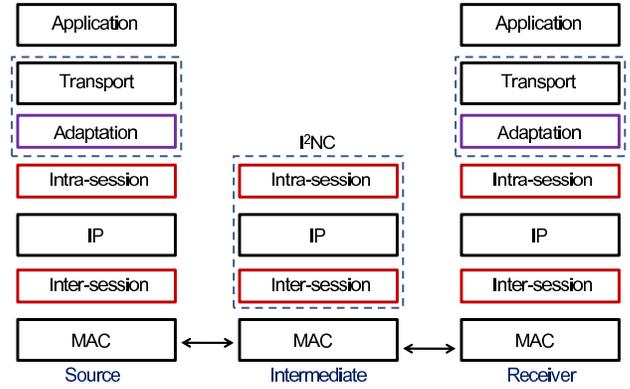}
\vspace{-20pt}
\caption{Operations taking place at end-points and intermediate nodes.}
\label{fig:protocol_stack}
\vspace{-10pt}
\end{figure}

After an intermediate node has added redundancy ($P^s$) to flow $s$, it treats all ($G^s+P^s$) packets as indistinguishable parts of the same flow. Inter-session NC is applied on top of the already intra-coded flows, as a thin layer between MAC and IP (similar to COPE), shown in Fig.~\ref{fig:protocol_stack}. We design two schemes, {I$^2$NC-state} and {I$^2$NC-stateless}, depending on the type of information that is needed to make network coding decisions. We define as state of a node the information about which exact packets have been overheard at that node. %; overheard packets always belong to flows that are of no interest to the node.

{\bf I$^2$NC-state:} First, we assume that intermediate nodes use COPE \cite{cope} for inter-session coding. Each node $i$ listens all transmissions in its neighborhood, stores the overheard packets in its decoding buffer, and periodically advertises the content of this buffer to its neighbors. When a node $i$ wants to transmit a packet, it checks or estimates the contents of the decoding buffer of its neighbors. If there is a coding opportunity, the node combines the relevant packets using simple coding operations (XOR) and broadcasts the combination to $\Jset$. The content of the decoding buffers needs to be exchanged, in order to make NC decisions, \ie state synchronization is required.

{\bf I$^2$NC-stateless:} Second, we design an improved version of COPE, which no longer requires state synchronization. The key idea is to exploit the fact that the redundancy already introduced by intra-session coding makes all $G^s+P^s$ packets in a generation equally important.\footnote{It no longer matters which exact packets a node has. As long as a node has any $G^s$ out of $G^s+P^s$, it can decode with high probability. As long as it knows the percentage of received packets it can make coding decisions.} In this improved scheme, each node $i$ still listens to all transmissions in its neighborhood and stores the overheard packets.\footnote{Note that when inter-session network coded packets are overheard, they are not stored in the ``decoding buffer'', but discarded.}
The node periodically advertises the loss rate for each received and overheard flow, which is then provided as input to the intra-session NC module to determine the amount of redundancy needed. In particular, the loss rates are calculated at each intermediate node as one minus the ratio of correctly received packets over all the packets in a generation. Also, the loss rate over overhearing links is calculated as effective loss rate. {\em E.g.,} in Fig.~\ref{fig:one-hop}, the loss rate at node $A_2$ is calculated as follows. If  $G^{S_1}+P^{S_1}$ packets are sent by $B_1$ and at least $G^{S_1}$ packets are received at $A_2$, then the loss rate is set to 0. If $G^{S_1}-u$ packets are received by $A_2$ such that $u \leq G^{S_1}$, then the loss rate is set to $u/G^{S_1}$. The loss rates calculated for each generation are advertised to other nodes in the neighborhood. Then, each node calculates its loss probabilities ($\rho_{h}^{s}$ and $\rho_{h,k}^{s,s'}$) as weighted average of the loss rates it has received.

In summary, there is a synergy between intra- and inter-session NC. Intra-session makes the process sequence agnostic, which allows inter-session coding to operate using only information about the loss rates, not about the identity of the packets. The loss rates can be used as input for tuning the amount of redundancy in intra-session NC. In terms of implementation, the two modules are separable: an intermediate node first performs intra-session, then inter-session NC.

\section{\label{sec:opt2}Network Utility Maximization Formulation}
\subsection{I$^2$NC-state Scheme}
\subsubsection{Formulation}
Our objective is to maximize the total utility function by optimally choosing the flow rates $x_s$ at sources $s \in \Sset$, as well as the following variables at the intermediate nodes: the fraction $\alpha_{h,k}^{s}$ (or ``traffic splitting'' parameters, following the terminology of \cite{opt_multicast_nc}) of flows inter-session coded using code $k \in \Kset_{h}$ over hyperarc $h$; and the percentage of time $\tau_{h,k}$ each hyperarc is used. 
\begin{align} \label{opt:eq1}
\max_{\boldsymbol x, \boldsymbol \alpha, \boldsymbol \tau} \mbox{ } & \sum_{s \in \Sset} U_{s}(x_{s})  \nonumber \\
\mbox{s.t.} \mbox{ }  & \frac{H_{h,k}^{s} \alpha_{h,k}^{s} x_{s}}{1-\rho_{h}^{s}} + \sum_{s' \in \Sset_{k}-\{s\}} H_{h,k}^{s'}\alpha_{h,k}^{s'}x_{s'}\rho_{h,k}^{s,s'} \leq R_{h}\tau_{h,k},  \nonumber \\
& \mbox{ } \forall  h  \in \Aset, k \in \Kset_{h}, s \in \Sset_{k} \nonumber \\
& \sum_{h(\Jset) | h \in {\Aset}} \sum_{k \in \Kset_{h} \mid s \in \Sset_{k}} \alpha_{h,k}^{s}  = 1, \mbox{ }  \forall s \in \Sset, i \in \Pset_s \nonumber \\
& \sum_{h \in \Cset_q} \sum_{k \in \Kset_{h}} \tau _{h,k} \leq \gamma,  \mbox{ }   \forall \Cset_q \subseteq \Aset
\end{align}
The first constraint is the capacity constraint for each flow $s \in \Sset_{k}$. It is well-known, \cite{minimum_cost_multicast}, that NC allows flows that are coded together in code $k \in \Kset_{h}$, to coexist, {\em i.e.}, each have rate up to the rate allocated to that code $k$. The right hand side, $R_{h}\tau_{h,k}$, is the capacity of hyperarc $h$; $\tau_{h,k}$ is the percentage of time hyperarc $h$ can be used for transmitting the $k$-th network code. $\tau_{h,k}$ is determined by
scheduling in the third constraint, taking into account interference: all hyperarcs in a clique interfere and should time-share the medium. Therefore, the sum of the time allocated to all hyperarcs in a clique should be less than an over-provisioning factor, $\gamma \le 1$. The second constraint is the flow conservation: at every node $i$ on the path $\Pset_s$ of source $s$, the sum of $\alpha_{h,k}^{s}$ over all network codes and hyperarcs should be equal to 1.  Indeed, when a flow enters a particular node $i$, it can be transmitted to its next hop $j$ as part of different network coded and uncoded flows.

The first constraint is key to our work because it determines how to deal with loss on the direct ($\rho_{h}^{s}$) and overhearing ($\rho_{h,k}^{s,s'}$) links and how large a fraction ($\alpha_{h,k}^{s}$) of flow rate ($x_s$) to code in the $k$-th code over hyperarc $h$. Let us discuss the left hand side in more detail.\footnote{Note that our formulation has two novel aspects, compared to prior work, which allow us to better handle loss and parities. First, we allow for flows coded together to have different rates (in the first constraint in Eq.~(\ref{opt:eq1})). Second, we allow for loss rates of each link to be specified separately, even for links in the same hyperarc.}

The first term refers to the direct link of flow $s$. $H_{h,k}^{s} \alpha_{h,k}^{s} x_{s}$ is the fraction of flow rate $x_s$ allocated to code $k$ and hyperarc $h$. It is scaled by $1-\rho_{h}^{s}$ to indicate that we use redundancy to protect against loss that flow $s$ experiences with probability $\rho_{h}^{s}$. $(H_{h,k}^{s} \alpha_{h,k}^{s} x_{s})/(1-\rho_{h}^{s})$ is the total rate of flow $s$, including data and redundancy.

The second term refers to loss on the overhearing links. $\sum_{s' \in \Sset_{k}-\{s\}} H_{h,k}^{s'}\alpha_{h,k}^{s'}x_{s'}\rho_{h,k}^{s,s'}$ is the amount of redundancy (via intra-session coding) added by the intermediate node on flow(s) $s'$ to protect flow $s$ against loss of antidote packets. These antidotes come from other flows ($s' \in \Kset_{h}$) that are coded together with flow $s$, reach the next hop for flow $s$ through the overhearing links, and are needed to decode  inter-session coded packets.

{\em Example 1- continued.} In Fig.~\ref{fig:one-hop}, let us consider flow 2 from $B_1$ to $B_2$, as the flow of interest. The intermediate node $I$ adds redundancy to $S_2$ to protect against loss rate $\rho_{(I,\{B_2,A_2\})}^{S_2}$ on the direct link  $I-B_2$. It also adds redundancy to flow 1 to protect against loss rate $\rho_{(I,\{B_2,A_2\})}^{S_2,S_1}$ of antidotes coming to $B_2$ from flow 1 over the overhearing link $A_1-B_2$.

\subsubsection{Optimal Solution}
To solve Eq.~(\ref{opt:eq1}) we follow a similar approach proposed in \cite{cc_multicast_nc}. First, we relax the capacity constraint in Eq.~(\ref{opt:eq1}), and we have the Lagrangian function:
\begin{align} \label{opt:eq1_Lagrange1}
&  \!\!L(\boldsymbol x, \boldsymbol \alpha, \boldsymbol \tau, \boldsymbol q)= \sum_{s \in \Sset} U_{s}(x_{s}) \!- \!\!\!\sum _{h \in \Aset} \sum_{k \in \Kset_{h}} \sum_{s \in \Sset_{k}} q_{h,k}^{s} \Bigl( \frac{ H_{h,k}^{s} \alpha_{h,k}^{s} x_{s}}{1-\rho_{h}^{s}} \nonumber \\
&  +  \sum_{s' \in \Sset_{k}-\{s\}} H_{h,k}^{s'} \alpha_{h,k}^{s'}x_{s'}\rho_{h,k}^{s,s'} - R_{h} \tau_{h,k} \Bigr)\!,
\end{align}
where $q_{h,k}^{s}$ is the Lagrange multiplier, which can be interpreted as the queue size for $k$-th network code at hyperarc $h$ for flow $s$. We define $\rho_{h,k}^{s,s'} = 0$ if $ s = s', \forall s, s' \in \Sset $ and we rewrite $\sum_{k \in \Kset_{h}}$ $\sum_{s \in S_{k}}$ as $\sum_{s \in \Sset}$ $\sum_{k \in K_{h} \mid s \in \Sset_{k}}$. The Lagrange function is  $ L(\boldsymbol x, \boldsymbol \alpha, \boldsymbol \tau, \boldsymbol q) = \sum _{s \in \Sset} ( U_{s}(x_{s}) - x_{s} \sum _{h \in \Aset} \sum_{k \in \Kset_{h} \mid s \in \Sset _{k}} H_{h,k}^{s} \alpha_{h,k}^{s} ((q_{h,k}^{s})$ $/(1 - \rho_{h}^{s})  + \sum _{s' \in \Sset_{k}} q_{h,k}^{s'} \rho_{h,k}^{s',s}) ) + \sum _{h \in \Aset} \sum _{k \in \Kset_{h}} \sum_{s \in \Sset_{k}} q_{h,k}^{s}R_{h}\tau_{h,k}$. It can be decomposed into several intuitive parts (rate control, traffic splitting, scheduling, and queue update), each of which solves the optimization problem for one variable.

\textbf{Rate Control.} First, we solve the Lagrangian w.r.t $x_s$:
\begin{equation} \label{opt:eq1_rateControl1}
\textstyle x_s = ({U'_{s}})^{-1} \left( \sum _{i \in \Pset_{s}} Q_{i}^{s} \right),
\end{equation} where $({U'_{s}})^{-1}$ is the inverse function of the derivative of $U_{s}$, and $Q_{i}^{s}$ is the occupancy of flow $s$ at node $i$ and expressed as
\begin{equation} \label{opt:eq1_Q_i_s}
\textstyle Q_{i}^{s} = \sum _{h(J) \mid h \in \Aset} \sum _{k \in \Kset_{h} \mid s \in \Sset_{k}} H_{h,k}^{s} \alpha_{h,k}^{s} Q_{h,k}^{s},
\end{equation} where $Q_{h,k}^{s}$ is the queue size of flow $s$ associated with hyperarc and network code pair $\{h,k\}$:
\begin{equation} \label{opt:eq1_Q_h_k_s}
\textstyle Q_{h,k}^{s} = \frac{q_{h,k}^{s}}{1-\rho_{h}^{s}} + \sum _{s' \in \Sset_{k}-\{s\}} q_{h,k}^{s'}\rho_{h,k}^{s',s}
\end{equation}

\textbf{Traffic Splitting.} Second, we solve the Lagrangian for $\alpha_{h,k}^{s}$. At each node $i$ along the path ({\em i.e.}, $i \in \Pset_{s}$), the traffic splitting problem can be expressed as follows:
\begin{align} \label{opt:eq1_trafficSplit}
\min_{\boldsymbol \alpha} & \textstyle \sum_{h(J)|h \in \Aset} \sum_{k \in K_{h} | s \in \Sset_{k}} \alpha_{h,k}^{s} H_{h,k}^{s} Q_{h,k}^{s} \nonumber \\
\mbox{s.t. } & \textstyle \sum_{h(J)|h \in \Aset} \sum_{k \in K_{h} | s \in \Sset_{k}} \alpha_{h,k}^{s} = 1. %, \mbox{   } \forall i \in \Pset_{s}.
\end{align} Let us assume that
%$\Qset_{h,k}^{s} = \frac{q_{h,k}^{s}}{1-\rho_{h}^{s}} + \sum _{s' \in \Sset_{k}-\{s\}} q_{h,k}^{s'}\rho_{h,k}^{s',s}$ and
$E_{i}[Q(t)]$ is the maximal $\overset{-}{Q}_{i}(t)$ at time $t$ such that $\overset{-}{Q}_{i}(t) = \frac{1}{|\Aset_{i}^{'}(t)|} \sum_{\varphi \in \Aset_{i}^{'}(t)} H_{h,k}^{s}Q_{h,k}^{s}(t)$
%Q_{i}(t)$
with $\Aset_{i}^{'}(t) := \{\varphi = \{h(\Jset),k\} | \alpha_{h,k}^{s} > 0 \mbox{  or  } H_{h,k}^{s}Q_{h,k}^{s} \leq \overset{-}{Q}_{i}(t), h(\Jset) \in \Nset \mbox{ s.t. } h \in \Aset, k \in \Kset_{h}  \}$. At each node $i$, the amount of traffic splitting factor $\alpha_{h,k}^{s}$ for flow $s$ over hyperarc $h$ and code $k$ follows; $\overset{.}{\alpha}_{h,k}^{s} = \kappa_{i} [E_{i}[Q] - H_{h,k}^{s}Q_{h,k}^{s}]_{\alpha_{h,k}^{s}}^{+}$, where $\kappa_{i}$ is a positive constant, and $[b]_{z}^{+} = b$ if $z \geq 0$ and $[b]_{z}^{+} = 0$ if $b \leq 0$ and $z = 0$.
It can be seen that $\sum_{h(J)|h \in \Aset} \sum_{k \in K_{h} | s \in \Sset_{k}} \overset{.}{\alpha}_{h,k}^{s} = 0$ and $\sum_{h(J)|h \in \Aset} \sum_{k \in K_{h} | s \in \Sset_{k}} \overset{.}{\alpha}_{h,k}^{s}H_{h,k}^{s}Q_{h,k}^{s} \leq 0$. Also, $\sum_{h(J)|h \in \Aset} \sum_{k \in K_{h} | s \in \Sset_{k}} \overset{.}{\alpha}_{h,k}^{s}H_{h,k}^{s}Q_{h,k}^{s} = 0$ only if $\overset{.}{\alpha}_{h,k}^{s}=0$ which is possible only if $H_{h,k}^{s}Q_{h,k}^{s} \geq \overset{-}{Q}_{i}$, and $\alpha_{h,k}^{s}(H_{h,k}^{s}Q_{h,k}^{s} - \overset{-}{Q}_{i})=0$.

The structure of the optimal solution of Eq.~(\ref{opt:eq1_trafficSplit}) (\ie $\overset{.}{\alpha}_{h,k}^{s} = \kappa_{i} [E_{i}[Q] - H_{h,k}^{s}Q_{h,k}^{s}]_{\alpha_{h,k}^{s}}^{+}$) has the following interpretation: the higher the loss rate of antidotes on overhearing links $\rho_{h,k}^{s',s}$, the higher $Q_{h,k}^{s}$, and the smaller $\alpha_{h,k}^{s}$. This means that flow $s$ should code fewer packets with packets from flow(s) $s'$ in code $k$, when antidotes from $s'$ are likely to be lost.

{\em Example 1 - continued:} In Fig.~\ref{fig:one-hop}, this means that $I$ should combine fewer packets from the two flows if there is loss on the overhearing link $A_1-B_2$. In the extreme case where loss rate is 1 over the link $A_1-B_2$, inter-session coding should be turned off. At the other extreme, where there is no loss, the two flows should always be combined. \hfill $\blacksquare$

\textbf{Scheduling.} Third, we solve the Lagrangian for $\tau_{h,k}$. This problem is solved for every hyperarc and every clique for the conflict graphs in the hypergraph. \begin{align} \label{opt:eq1_scheduling}
\max_{\boldsymbol \tau} & \textstyle \sum_{h \in \Aset} \sum_{k \in \Kset_{h}}  \sum_{s \in \Sset_{k}} q_{h,k}^{s} R_{h} \tau_{h,k} \nonumber \\
\mbox{ s.t.} & \textstyle \sum_{h \in \Cset_{q}} \sum_{k \in \Kset_{h}} \tau_{h,k} \leq \tau, \mbox{  } \forall \Cset_{q} \subseteq A
\end{align} Let us assume that $Q_{h,k} = R_{h} \sum_{s \in \Sset_{k}}q_{h,k}^{s}$, and $E_{\Cset_{q}}[Q(t)]$ is the minimal $\overset{-}{Q}_{\Cset_{q}}(t)$ at time $t$ such that; $\overset{-}{Q}_{\Cset_{q}}(t) = \frac{1}{|\Aset_{\Cset_{q}}^{'}(t)|} \sum_{\phi \in \Aset_{\Cset_{q}}^{'}} Q_{h,k}(t)$ with $\Aset_{\Cset_{q}}^{'} := \{ \phi = \{h,k\} | \tau_{h,k} > 0 \mbox{  or  } Q_{h,k}(t) \geq \overset{-}{Q}_{\Cset_{q}}(t), h \in \Aset, k \in \Kset_{h}  \}$. At each clique $\Cset_{q}$, the fraction of the time $\tau_{h,k}$ that is allocated to hyperarc $h$, and code $k$ is as follows; $\overset{.}{\tau}_{h,k} = \varepsilon_{\Cset_{q}} [Q_{h,k} - E_{\Cset_{q}}[Q]]_{\tau_{h,k}}^{+}$, where $\varepsilon_{\Cset_{q}}$ is a positive constant and $[b]_{z}^{+} = b$ if $z \geq 0$ and $[b]_{z}^{+} = 0$ if $b \leq 0$ and $z = 0$. It can be seen that $\sum_{h \in \Cset_{q}} \sum_{k \in \Kset_{h}} \overset{.}{\tau}_{h,k} = 0$ and $\sum_{h \in \Cset_{q}} \sum_{k \in \Kset_{h}} \overset{.}{\tau}_{h,k} Q_{h,k} \geq 0$. Also, $\sum_{h \in \Cset_{q}} \sum_{k \in \Kset_{h}} \overset{.}{\tau}_{h,k} Q_{h,k} = 0$ only if $\overset{.}{\tau}_{h,k}=0$ which requires that $Q_{h,k}=\overset{-}{Q}_{\Cset_{q}}$ or $\tau_{h,k}=0$ and $Q_{h,k} < \overset{-}{Q}_{\Cset_{q}}$.

\textbf{Queue Update.} We find the Lagrange multipliers (queue sizes) $q_{h,k}^{s}$, using the gradient descent:
\begin{align} \label{opt:eq1_parameterUpdate}
& \textstyle q_{h,k}^{s}(t+1) =  \{ q_{h,k}^{s}(t) + c_t \{ \frac{H_{h,k}^{s}\alpha_{h,k}^{s} x_{s} }{1-\rho_{h}^{s}} + \nonumber \\
& \textstyle \sum _{s' \in \Sset_{k} - \{s\}} H_{h,k}^{s'} \alpha_{h,k}^{s'} x_{s'}\rho_{h,k}^{s,s'}  - R_{h} \tau_{h,k} \} \}^{+}
\end{align}
where $t$ is the iteration number, $c_t$ is a small constant, and the $~^+$ operator makes the Lagrange multipliers positive.
$q_{h,k}^{s}$ is interpreted as the queue for flow $s$ allocated for the $k$-th network code over hyperarc $\forall h \in \Aset$. Indeed, in Eq.~(\ref{opt:eq1_parameterUpdate}), $q_{h,k}^{s}$ is updated with the difference between the incoming $(H_{h,k}^{s}\alpha_{h,k}^{s} x_{s} )/(1-\rho_{h}^{s}) + \sum _{s' \in \Sset_{k} - \{s\}} H_{h,k}^{s'} \alpha_{h,k}^{s'} x_{s'}\rho_{h,k}^{s,s'}$ and outgoing $R_{h} \tau_{h,k}$ traffic rates at $h$.\footnote{Note that the queue update in Eq.~(\ref{opt:eq1_parameterUpdate}) can be re-written as; $\dot{q}_{h,k}^{s} = \gamma_{h} [ \frac{H_{h,k}^{s}\alpha_{h,k}^{s} x_{s} }{1-\rho_{h}^{s}} + \sum _{s' \in \Sset_{k} - \{s\}}$ $H_{h,k}^{s'} \alpha_{h,k}^{s'} x_{s'}\rho_{h,k}^{s,s'}  - R_{h} \tau_{h,k} ]_{q_{h,k}^{s}}^{+}$, where $\gamma_{h}$ is a positive constant.}

\subsection{I$^2$NC-stateless Scheme}
The second term in Eq.~(\ref{opt:eq1}) describes the redundancy added by node $i$ to protect flow $s$ from loss of antidotes on the overhearing link. An implicit assumption was that node $i$ knows what antidotes are available at the next hop and uses only those packets for inter-session coding.
However, this knowledge can be imperfect, especially in the presence of loss.
Here, we formulate a variation of the problem, where such knowledge is not necessary. Instead, node $i$ needs to know only the loss rate on all the links to the next hop for flow $s$ (\eg in Fig.~\ref{fig:one-hop} for flow 2 ($S_2$), these are links $I-B_2$ and $A_1-B_2$).

We replace the capacity constraint in Eq.~(\ref{opt:eq1}) with: %\sum_{s'\in \Sset_{k}-\{s\}
\begin{equation}\label{eq:P2}
\textstyle \frac{H_{h,k}^{s} \alpha_{h,k}^{s} x_{s}}{1-\rho_{h}^{s}} + \sum_{s' \in \Sset_{k}-\{s\}} \frac{ H_{h,k}^{s'}\alpha_{h,k}^{s'}x_{s'}\rho_{h,k}^{s,s'}}{1-\rho_{h}^{s}} \leq R_{h}\tau_{h,k}
\end{equation}
and this is $\forall  h  \in \Aset, k \in \Kset_{h}, s \in \Sset_{k}$. The other constraints remain the same as in Eq.~(\ref{opt:eq1}). The difference from Eq.~(\ref{opt:eq1}) is in the second term, related to the overheard packets at the next hop. Any fraction of flow $s'$ added as redundancy to flow $s$, as well as overheard packets from $s'$ in the next hop, help to decode inter-session coded packets of $s$ with flow $s'$. To protect transmissions of these ``helping'' fractions ($H_{h,k}^{s'}\alpha_{h,k}^{s'}x_{s'}\rho_{h,k}^{s,s'}$) from being lost on the direct link to the next hop of flow $s$ (\eg from $I$ to $B_2$), we add redundancy to match the loss rate of that direct link ($\rho_{h}^{s}$ in general, $\rho_{\{I,B_2\}}^{S_2}$ in the example). This is why the term $H_{h,k}^{s'}\alpha_{h,k}^{s'}x_{s'}\rho_{h,k}^{s,s'}$ is divided by ${1-\rho_{h}^{s}}$.

The solution of this optimization problem also decomposes into rate control, traffic splitting, and scheduling problems, which correspond to Eq.~(\ref{opt:eq1_rateControl1}), (\ref{opt:eq1_trafficSplit}), and (\ref{opt:eq1_scheduling}), respectively. $Q_{h,k}^{s}$ needs to be updated:
\begin{align} \label{opt:eq2_Q_h_k_s}
\textstyle Q_{h,k}^{s} = \frac{q_{h,k}^{s}}{1-\rho_{h}^{s}} + \sum _{s' \in \Sset_{k}-\{s\}} \frac{q_{h,k}^{s'}\rho_{h,k}^{s',s}}{1-\rho_{h}^{s'}}.
\end{align}
%\textcolor{blue}{
The Lagrange multiplier is updated as follows;
\begin{align} \label{opt:eq2_parameterUpdate}
& \textstyle q_{h,k}^{s}(t+1) =   \{ q_{h,k}^{s}(t) + c_t \{ \frac{H_{h,k}^{s}\alpha_{h,k}^{s} x_{s} }{1-\rho_{h}^{s}} + \nonumber \\
& \textstyle \sum _{s' \in \Sset_{k} - \{s\}} \frac{ H_{h,k}^{s'} \alpha_{h,k}^{s'} x_{s'}\rho_{h,k}^{s,s'} } {1-\rho_{h}^{s}} - R_{h} \tau_{h,k} \} \}^{+}
\end{align}
%}
We provide the convergence analysis of our solution in Appendix A.\footnote{We do not claim that the solution of our network utility maximization problem is the optimal solution to the general intra- and inter-session NC problem over wireless networks. This is well-known, open problem \cite{netinfflow}, \cite{algebraicNC}, \cite{hobook}. Even without an optimal, closed form solution, there is still value in using the structure of the solution to design mechanisms that perform well in practice, as we show through the numerical and simulations results in the next sections.} We first give the proof of convergence, then we verify the convergence through numerical calculations.

\section{\label{sec:algs}System Implementation}
We propose practical implementations of the {I$^2$NC-state} and {I$^2$NC-stateless} schemes (Fig.~\ref{fig:protocol_stack}), following the NUM formulation structure.

\subsection{Operation of End-Nodes}
At the end nodes, there is an adaptation layer between transport and intra-session NC  which has two tasks: (i) to interface between application and intra-session NC; and (ii) to optimize the reliability mechanism at the transport layer.

{\em Task (i):} At the source, the adaptation layer sets the generation (block) size $G^s$. $G^s$ is set according to application; \eg media transmission requirements for UDP, or set equal to TCP congestion window for TCP applications and changes over time. The adaptation layer receives $G^s$ original packets $p_1, p_2, ..., p_{G^{s}}$ from the transport layer of flow $s$ and generates $G^s$ intra-session coded packets; $a_1 = p_1$, $a_2 = p_1 + p_2$, $...$, $a_{G^{s}} = p_1 + ... + p_{G^{s}}$. We call this coding ``incremental additive coding''. We chose the incremental additive coding to avoid introducing coding delays (\ie our algorithm does not need to wait $G^{s}$ packets to encode packets) as proposed in \cite{NC_meets_TCP}. The intra-session header includes the block id, packet id, block size, and coding coefficients. At the receiver side, the reverse operations are performed.

{\em Task (ii):} To further optimize the interaction between I$^2$NC and transport, particularly TCP, we keep track of and acknowledge the number of received packets in a generation, rather than their sequence numbers (note that this part is not needed for UDP protocol). This idea is similar to the use of end-to-end FEC and intra-session NC that make TCP sequence agnostic \cite{LT_TCP,complement_TCP,NC_meets_TCP}. {\em E.g.}, if a receiver receives the first packet labeled with block id $g^{s}=1$, then it generates an ACK with block id $g^{s}=1$ and packet id $\eta^{s} = 1$. The uncoded packets,  $p_1, p_2, ..., p_{G^{s}}$, are stored in a buffer at the source for TCP ACK adaptation. {\em E.g.}, if an ACK for block id $g^{s}=1$ and packet id $\eta^{s} = 1$ is received by the source, then the TCP adapter matches this ACK to packet $p_1$ and informs TCP that packet $p_1$ is ACK-ed. As long as the TCP receiver transmits ACKs, the TCP clock moves, thus improving TCP goodput. After the ACK with the block and packet ids is transmitted by the TCP receiver, the packet is stored at the receiving buffer. When the last packet from a generation is received, then packets are decoded and passed to the application.

\subsection{Operation of Intermediate Nodes}
An intermediate node needs to take a number of actions when it receives (Alg.~\ref{alg:packet_insert}) or transmits (Alg.~\ref{alg:packet_transmit}) a packet.

\subsubsection{Receiving a packet and intra-session network coding}~\\
{\bf Buffer packets.}
A node $i$ may receive a packet from higher layers or from previous hops. In the latter case, if the received packet is inter-coded, it is decoded and the packet with destination to this node is stored (or is passed to transport if it is the last hop). If it is not the last hop, a packet $a_l \in \{a_1, a_2, ..., a_{G^{s}}\}$ is stored in the output queue $\Qset_{i}$. In addition to the physical output queue $\Qset_{i}$, the node $i$ keeps track of several virtual queues; $Q_{h,k}^{s}$ per (flow, hyperarc, code). The packet $a_l$ is labeled with $(h^{*}, k^{*}, s)$, which essentially indicates whether and how to code this packet according to the traffic splitting in Eq.~(\ref{opt:eq1_trafficSplit}): we pick  $\{h^{*},k^{*}\} = \arg\min _{h,k} \{H_{h,k}^{s}Q_{h,k}^{s}\}$, randomly breaking ties. Note that this labeling is local at the node, and does not introduce any transmission overhead.

Note that $H_{h,k}^{s}$ is the indicator whether flow $s$ is transmitted over hyperarc $h$ with code $k$. This indicator is determined at each node using a routing table which has a data structure to determine the next hops (note that paths do not need to be known by the sources or any node in the system). Basically, if a packet from flow $s$ is able to reach to the next hop determined by the routing table when it transmitted over hyperarc $h$ and with code $k$, then the indicator is set to $1$, otherwise $0$. We also note that in this system, as long as paths remain fixed for longer (at least longer than a time required to transmit a packet) time periods, we can see more benefit from NC, because each node will learn which flows can be network coded and estimate the loss rates better as time gets longer. However, even in the extreme case in which paths change very fast (say for example at every packet transmission), our system works well, but it does not fully exploit NC opportunities, since it cannot estimate whether NC is possible or not. However, it works not worse than a system without NC. Therefore, I$^2$NC is designed to adapt to path changes and to exploit NC benefit if possible.

\begin{algorithm}[t!]
\caption{Node $i$ processes packet $a_l$ from flow $s$.\label{alg:packet_insert}}
\begin{algorithmic}[1]
%\begin{footnotesize}
\begin{scriptsize}
\STATE Read the information: packet $a_l$, from flow $s$ (generation size $G^{s}$)
\STATE Insert $a_l$ into the physical output queue $\Qset_{i}$.
\STATE Determine $\{h^{*},k^{*}\}$ and label $a_l$ with $\{h^{*},k^{*}\}$ pair and $s$
\STATE Update $q_{h^{*},k^{*}}^{s}$ (using Eqs.~(\ref{opt:eq1_parameterUpdate}) and (\ref{opt:eq2_parameterUpdate})) and $q_{h^{*},k^{*}}^{s'}$
\STATE Calculate $Q_{h^{*},k^{*}}^{s}$ (using Eqs.~(\ref{opt:eq1_Q_h_k_s}) and (\ref{opt:eq2_Q_h_k_s})) and $Q_{i}^{s}$ (using Eq.~(\ref{opt:eq1_Q_i_s}))
\STATE $G_{h^{*},k^{*}}^{s} = G_{h^{*},k^{*}}^{s} + 1$
\IF {$G^{s}$ packets from flow $s$ are received at node $i$}
\STATE Calculate the number of parities $P_{h,k}^{s,s}$, $P_{h,k}^{s',s}$
\STATE Create $P_{h,k}^{s,s}$ parities from $s$ and $P_{h,k}^{s',s}$ parities from $s'$
\STATE Label all generated parities with $\{h,k\}$ pair and $s$
\ENDIF
\end{scriptsize}
%\end{footnotesize}
\end{algorithmic}
\end{algorithm}

{\bf Update Virtual Queue Sizes.}
When packet $a_l$ is selected to be transmitted with the $k^{*}$-th network code over hyperarc $h^{*}$, the virtual queues; $Q_{h^{*},k^{*}}^{s}$ and $q_{h^{*},k^{*}}^{s}$ should be updated. $q_{h^{*},k^{*}}^{s}$ is updated according to Eqs.~(\ref{opt:eq1_parameterUpdate}) and (\ref{opt:eq2_parameterUpdate}). $Q_{h^{*},k^{*}}^{s}$ is calculated according to Eq.~(\ref{opt:eq1_Q_h_k_s}) for {I$^2$NC-state} and Eq.~(\ref{opt:eq2_Q_h_k_s}) for {I$^2$NC-stateless}. $Q_{i}^{s}$ is calculated according to Eq.~(\ref{opt:eq1_Q_i_s}). Then, the number of packets $G_{h^{*},k^{*}}^{s}$ from the same generation that are allocated to $h^{*},k^{*}$ pair is incremented: $G_{h^{*},k^{*}}^{s} = G_{h^{*},k^{*}}^{s} + 1$. $G_{h,k}^{s}$ is set to 0 for each new generation.

{\bf Generate Parities.}
After $G^{s}$ packets from a generation of flow $s$ are received at node $i$, $P^s$ parity packets are generated via intra-session NC (which is performed according to random linear NC \cite{random_nc}) and labeled with information $(s, h, k)$. There are two types of parities.
\begin{itemize}
\item $P_{h,k}^{s,s} = \lceil{G_{h,k}^{s}\rho_{h}^{s}}/(1-\rho_{h}^{s})\rceil$ parities are added on flow $s$'s virtual queue to correct for loss during direct transmission to the next hop over hyperarc $h$.
\item $P_{h,k}^{s',s} = \lceil{G_{h,k}^{s} \rho_{h,k}^{s',s}}\rceil$, $\forall s' \in \Sset_{k}$ parities are added on the virtual queues of other flows $s'$ that are inter-session coded together with $s$. This is to help the next hop for $s'$ to decode despite losses on the overhearing link.
\end{itemize}
These parity packets are for {I$^2$NC-state}. For {I$^2$NC-stateless} $P_{h,k}^{s,s}$ is the same, but $P_{h,k}^{s',s}=\lceil{G_{h,k}^{s} \rho_{h,k}^{s',s}}/(1-\rho_{h}^{s'})\rceil$, \ie additional redundancy is used to protect parity packets from loss on the direct link.

\begin{algorithm}[t!]
 \caption{Node $i$ transmits a packet.} \label{alg:packet_transmit}
\begin{algorithmic}[1]
%\begin{footnotesize}
\begin{scriptsize}
\STATE Select $\{h^{\dag}, k^{\dag}\}$ pair that maximizes $Q_{h,k} = R_{h}(\sum _{s \in \Sset_k} q_{h,k}^{s})$
\STATE Initialize: $\xi = \emptyset$
\FOR {$a_l \in \Qset_{i}$}
\IF {$a_l$ is labeled with $\{h^{\dag}, k^{\dag}\}$ AND flow id label of $a_l$ is different from $\forall a_{l'} \in \xi$}
\IF {I$^2$NC-state AND $\xi \cup a_l$ is decodable OR I$^2$NC-stateless}
\STATE Insert packet to $\xi$
\ENDIF
\ENDIF
\ENDFOR
\STATE Network code (XOR) all packets in $\xi$
\STATE Broadcast the network coded packet over hyperarc $h^{\dag}$
\STATE Update $q_{h^{\dag}, k^{\dag}}^{s}$, $\forall s \in \Sset_{k}$
\STATE Re-calculate $Q_{h,k} = R_{h}(\sum _{s \in \Sset_k} q_{h,k}^{s})$ and $Q_{i}^{s}$ (using Eq.~(\ref{opt:eq1_Q_i_s}))
\end{scriptsize}
%\end{footnotesize}
\end{algorithmic}
\end{algorithm}

\subsubsection{Transmitting a packet and inter-session network coding}
We consider the 802.11 MAC. When a node $i$ accesses a channel, $\{h^{\dag}, k^{\dag}\}$ is chosen to maximize $Q_{h,k} = R_{h}(\sum _{s \in \Sset_k} q_{h,k}^{s})$ according to Eq.~(\ref{opt:eq1_scheduling}), randomly breaking ties. Although the pair $\{h^{\dag}, k^{\dag}\}$ determines the hyperarc, code and flows to be coded together in the next transmission, the specific packets from those flows still need to be selected and coded. We call these packets the set $\xi$, and select them using the procedure specified in Alg.~\ref{alg:packet_transmit}.\footnote{The inter-session NC header includes the number of coded packets together, next hop address, and the packet id's. Note that this header as well as the IP header of each packet are not network coded.}

To achieve this, we first initialize the set of network coded packets $\xi = \emptyset$. For each packet $a_l \in \Qset_{i}$, check whether $a_l$ is labeled with $\{h^{\dag}, k^{\dag}\}$. If it is, then we check whether its flow id label already exists in one of the packets in $\xi$, \ie another packet from the same flow has already been put in $\xi$. If not, there is one more check for I$^2$NC-state for decodability at the next hops of all packets in the network code, based on reports or estimates of overheard packets in the next hops, similarly to \cite{cope}. If the packet is decodable with some probability larger than a threshold (default value is 0.20) then, $a_l$ is inserted to $\xi$. In I$^2$NC-stateless, the packet $a_l$ is inserted to $\xi$ without checking the decodability, which is ensured through the additional redundancy packets. This is the strength of I$^2$NC-stateless: it eliminates the need to exchange detailed state, which is costly and unreliable at high loss rates. After all packets in $\Qset_{i}$ are checked, the labels ($h,k,s$) of the packets in $\xi$, inter-session NC header is added, and coded (XORed) and broadcast over $h$.

After a coded packet is transmitted, the virtual queues are updated according to Eqs.~(\ref{opt:eq1_parameterUpdate}), (\ref{opt:eq2_parameterUpdate}). The queues $Q_{h^{\dag},k^{\dag}}$  and $Q_{i}^{s}$ are calculated according to Eqs.~(\ref{opt:eq1_Q_h_k_s}), (\ref{opt:eq2_Q_h_k_s}), (\ref{opt:eq1_Q_i_s}).\footnote{Note that I$^2$NC may cause re-ordering at the receiver, but since we already implemented intra-session NC, and made TCP receiver sequence agnostic in this term, out of packet delivery is not a problem for TCP.}

We note that in both I$^2$NC-state and stateless, packets are network coded if some conditions are satisfied. However, if these conditions are not met, a packet without NC is still transmitted, because at least one packet is inserted in $\xi$ (Alg.~\ref{alg:packet_transmit}). Thus, we do not delay any packets in our schemes. Yet, delaying packets may create more NC opportunities and there is a tradeoff between delay and throughput. These issues have been considered in some previous work \cite{towsley_secon}, \cite{hs_icc_2010}. However, this is an aspect orthogonal to the focus of I$^2$NC (which is the synergy between inter- and intra-session NC) and can potentially be combined with it.

\subsubsection{Keeping Track of and Exchanging State Information}
For {I$^2$NC-state}, intermediate nodes need also to keep track of and exchange information with each other, so as to enable the intra- and inter-session NC modules to make their redundancy and coding decisions and to provide reliability. An approach similar to COPE is used: ACKs are sent after the reception and successful decoding of a packet. Information about overheard packets is piggy-backed on the ACKs. With I$^2$NC-stateless, we only need neighbors to exchange information about the loss rates at the neighboring nodes. Information about the loss rates as well as the number of received packets at a generation is reported through control packets for every generation.\footnote{In our implementation, the loss probabilities are calculated as weighted average of the loss rates. The weighted average is calculated over a window of 10 samples. The last 10 samples are ordered such that the newest sample is the first sample, and the oldest sample is the $10^{th}$ sample. Each sample is given a weight inversely proportional to its sample number.}
In order to provide reliability, we consider re-transmissions. In I$^2$NC-state, a packet is removed from the output queue only after an ACK related to the packets is received. Otherwise, the packet is re-transmitted after a round trip time. In I$^2$NC-stateless, packets are removed from the output queue when a control packets is received and confirms the successful transmission of all packets of the corresponding generation. Otherwise, a number of intra-session coded packets from the generation which are missing at the receiver are generated from the packets kept in the queue and transmitted.

\subsubsection{Congestion Control and Queue Management}
End-to-end congestion control (\ie rate control) is given by Eq.~(\ref{opt:eq1_rateControl1}) in which if $U_s(x_s)= \log(x_s)$, then $x_s = 1 / \left( \sum _{i \in \Pset_{s}} Q_{i}^{s} \right)$. This means that flow rate $x_s$ is inversely proportional with increasing queue size over the path of flow $s$. This behavior is similar to TCP's end-to-end congestion control algorithm, where congestion at a node may result in one or
more packets may be dropped from the buffer at this node. TCP reacts to packet drops by reducing its rate. Thus, TCP reduces its flow rate when queue size increases. This gives us intuition that TCP mimics the rate control part of the decomposed solution. This intuition has been validated in \cite{tutorial_doyle}, \cite{book_srikant}, \cite{low_1}, \cite{low_2}. %(by reverse engineering TCP).

Similarly, we consider that TCP already mimics the structure of the rate control part in Eq.~(\ref{opt:eq1_rateControl1}).  Therefore, upon congestion at node $i$, the per-flow queue sizes $Q_i^s$ are compared and the last packet from flow $s$ having the largest $Q_i^s$ is dropped from the queue; in case of a tie, an incoming packet is dropped. We do not make any additional updates to TCP's end-to-end congestion control algorithm. Also, we do not implement any end-to-end congestion control mechanism for UDP. Our goal is to keep UDP as it is (without any end-to-end control) and show the effectiveness of I$^2$NC-state and I$^2$NC-stateless when there is no end-to-end control.

\begin{example}\label{ex4}
Let us re-visit the X-topology from Fig.~\ref{fig:one-hop}, shown again for convenience in Fig.~\ref{fig:example_intra_inter}, and illustrate  how we perform intra- and inter-session NC under scheme {I$^2$NC-stateless}. The loss probabilities over the direct ($I-B_2$) and overhearing ($A_1-B_2$) links are assumed $0.5$ and $0.25$.

\begin{figure}
%\vspace{-5pt}
\centering
\subfigure[Intra-session coding]{{\includegraphics[width=0.45\columnwidth, angle=-90]{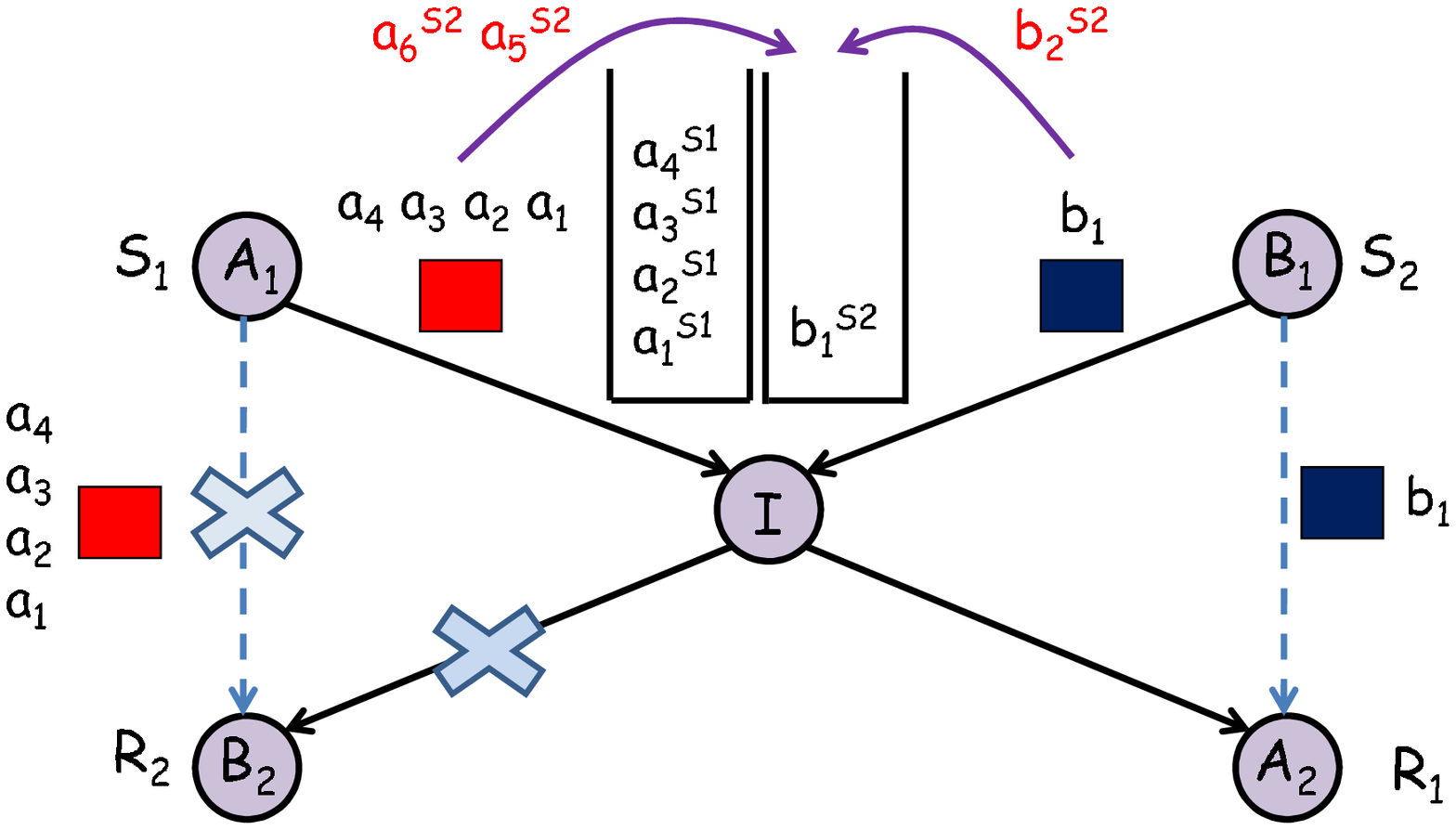}\vspace{-45pt}}} \\
\subfigure[Inter-session coding]{{\includegraphics[width=0.45\columnwidth, angle=-90]{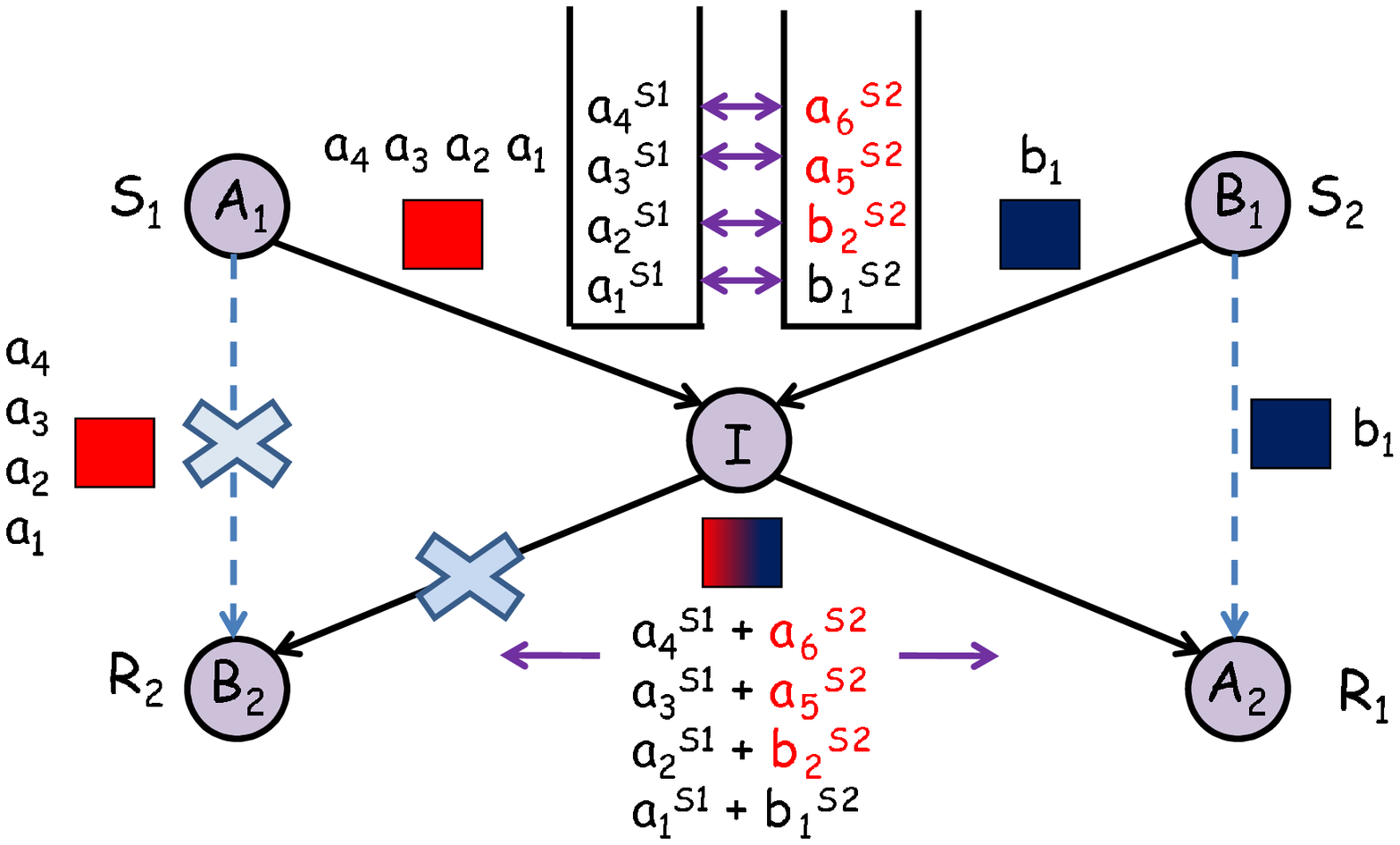}\vspace{-45pt}}}
\caption{Example of coding (under scheme {I$^2$NC-stateless}) at node $I$ in the X-topology. There is loss only on two links: the direct link $I-B_2$ (with probability 0.5) and the overhearing link $A_1-B_2$ (with probability $0.25$).}
\vspace{-15pt}
\label{fig:example_intra_inter}
\end{figure}

In Fig.~\ref{fig:example_intra_inter}(a), we describe intra-session NC. Let us assume the generation size of $S_1$ is $G^{S_1}=4$ and $S_2$ is $G^{S_2}=1$. The packets transmitted by $A_1$, $B_1$ are $a_1, a_2, a_3, a_4$ and $b_1$, respectively. Note that there is only one option for inter-session NC, \ie to XOR packets from the two flows, thus there exists only one possible network code $k=1$ over hyperarc $h=(I,\{B_2,A_2\})$.
All packets are labeled with this information and their flow ids. The labeled packets are $a_1^{S_1}, a_2^{S_1}, a_3^{S_1}, a_4^{S_1}$ and $b_{1}^{S_2}$. Parities are generated as follows. Since $G_{I,\{B_2,A_2\}}^{S_1} = 4$ and $G_{I,\{B_2,A_2\}}^{S_2} = 1$, the number of parities is  $P_{I,\{B_2,A_2\}}^{S_1,S_1} = 0$, $P_{I,\{B_2,A_2\}}^{S_1,S_2} = 0$, $P_{I,\{B_2,A_2\}}^{S_2,S_2} = 1$ (thus generating one parity from flow $S_2$ and labeling it with $S_2$, {\em i.e.}, $b_2^{S_2}$), and $P_{I,\{B_2,A_2\}}^{S_2,S_1} = 2$ (thus generating two parities from flow $S_1$ and labeling them with $S_2$, {\em i.e.}, $a_5^{S_2}, a_6^{S_2}$).

In Fig.~\ref{fig:example_intra_inter}(b), we describe inter-session NC. Node $I$ performs inter-session NC and transmits packets according to Alg.~\ref{alg:packet_transmit}: it XORs packets from the two queues, for $S_1, S_2$, and broadcasts over the hyperarc $(I,\{B_2,A_2\})$. In particular, it transmits the following packets: $a_1^{S_1} \oplus b_1^{S_2}$, $a_2^{S_1} \oplus b_2^{S_2}$, $a_3^{S_1} \oplus a_5^{S_2}$, and $a_4^{S_1} \oplus a_6^{S_2}$. $A_2$ receives and decodes all the packets. $B_2$ receives $3$ packets on the average over overhearing link $A_1-B_2$ and receives $2$ packets over transmission link $I-B_2$. Five received packets allows $B_2$ to decode all five packets $a_1, a_2, a_3, a_4, b_1$, so $b_1$ is successfully decoded.
\hfill $\blacksquare$
\end{example}

\section{\label{sec:performance}Performance Evaluation}
\subsection{Simulation Setup}
\begin{figure*}[t!]
\vspace{-15pt}
\begin{center}
\subfigure[X topology]{{\includegraphics[width=5.5cm]{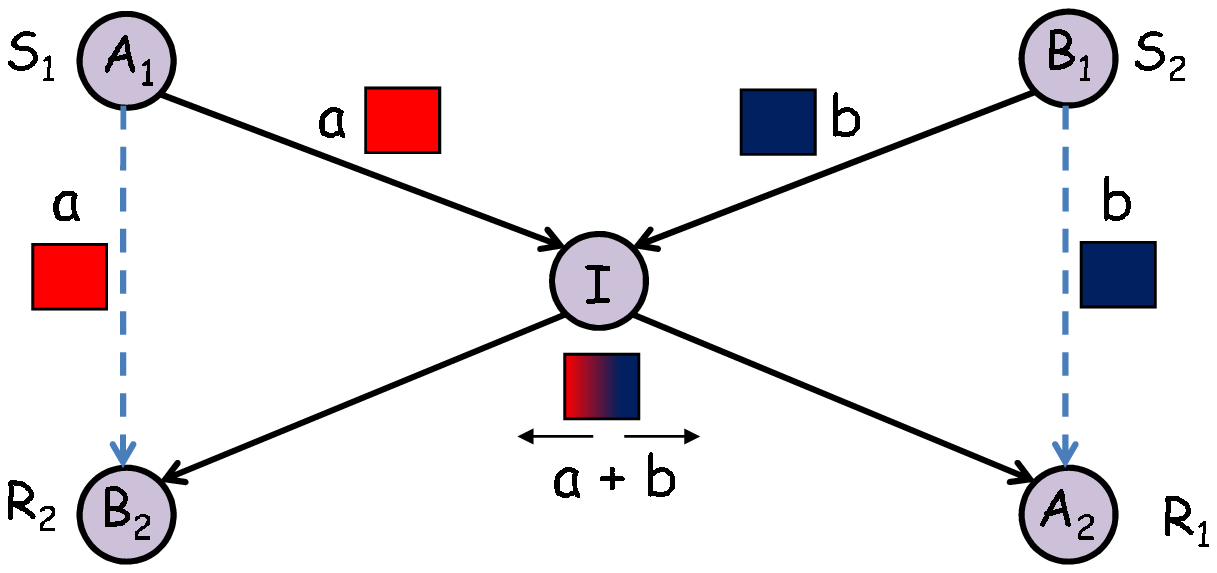}}} %\hspace{+40pt}
\subfigure[Cross topology]{{\includegraphics[width=5.5cm]{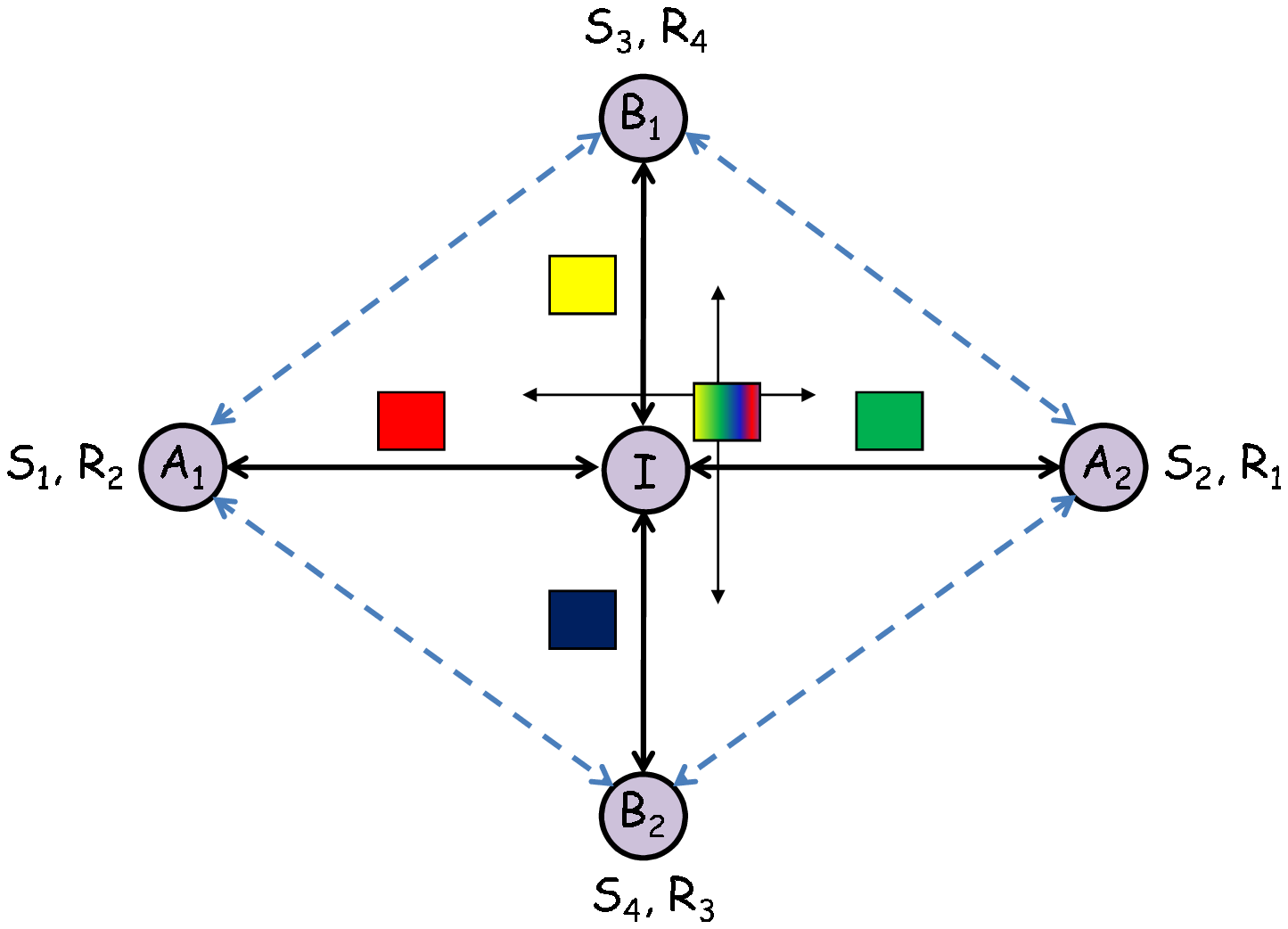}}} %\hspace{+40pt}
\subfigure[Wheel topology]{{\includegraphics[width=5.5cm]{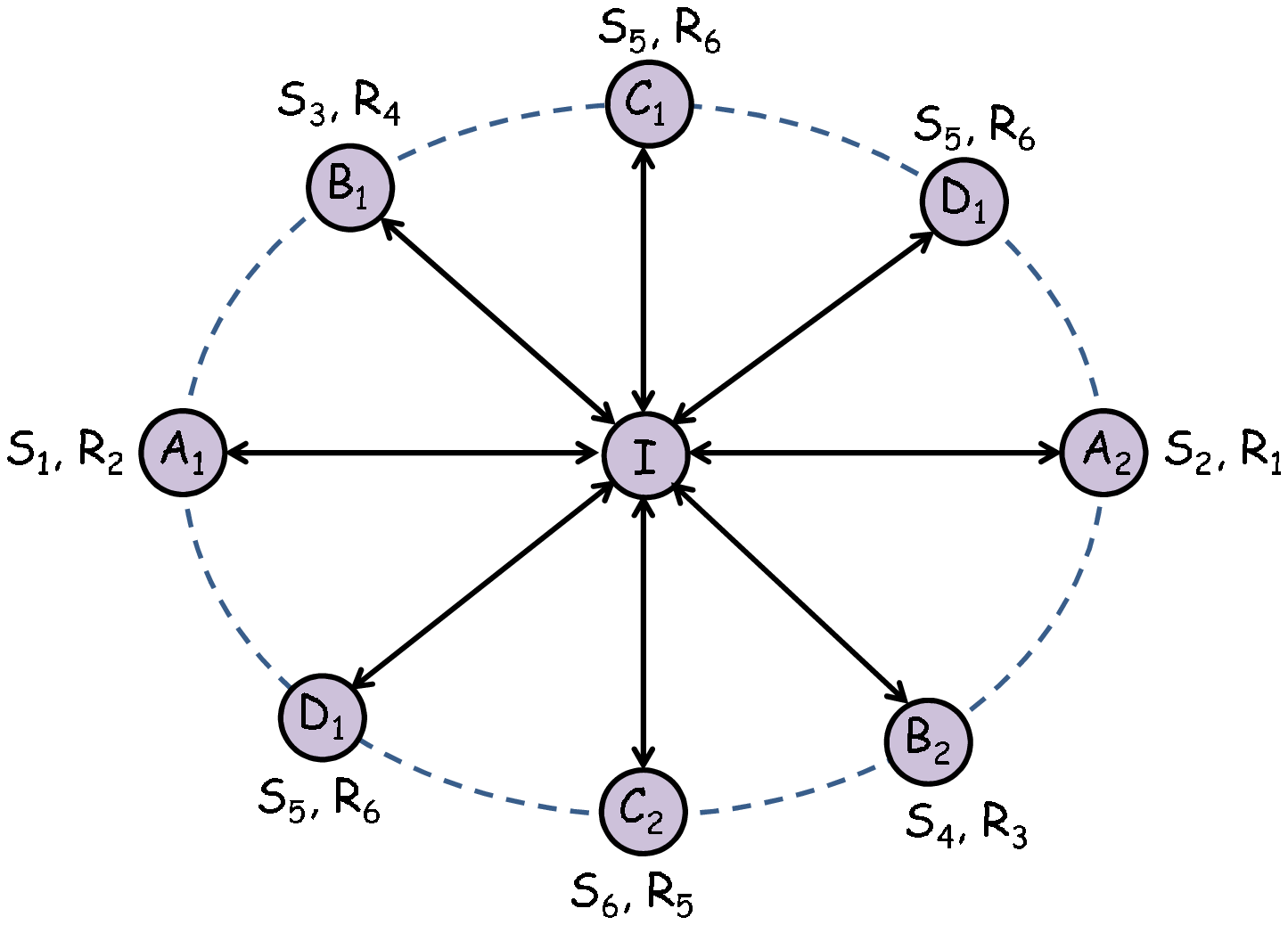}}} %\hspace{+40pt}
\end{center}
\begin{center}
\caption{\label{fig:all_topologies}  Topologies under consideration. (a) X topology. Two unicast flows, $S_1,R_1$, and $S_2,R_2$, meeting at intermediate node $I$. (b) Cross topology. Four unicast flows, $S_1,R_1$, $S_2,R_2$, $S_3,R_3$, and $S_4,R_4$, meeting at intermediate node $I$. (c) Wheel topology. Multiple unicast flows $S_1,R_1$, $S_2,R_2$, etc., meeting at intermediate node $I$. In all three topologies, $I$ opportunistically combine the packets and broadcast.}
\vspace{-10pt}
\end{center}
\end{figure*}

We used the {\tt GloMoSim} simulator \cite{glomosim}, which is well suited for simulating wireless environments. We considered various {\em topologies}: X topology, shown in part of Fig.~\ref{fig:one-hop} and repeated in Fig.~\ref{fig:all_topologies}(a); the cross-topology with four end-nodes generating bi-directional traffic, with one relay shown in Fig.~\ref{fig:all_topologies}(b); the wheel topology shown in Fig.~\ref{fig:all_topologies}(c); and the multi-hop topology shown in Fig.~\ref{fig:one-hop}. In X, cross, and wheel topologies, the intermediate node $I$ is placed in a center of of circle with radius $90m$ over $200m \times 200m$ terrain and all other nodes $A_1$, $B_1$ and etc. are placed around the circle. In the multi-hop topology of Fig.~\ref{fig:one-hop}, two X topologies are cascaded and the distance between consecutive nodes is set to $90m$. The topology is over a $800m \times 300m$ terrain.

%% Topologies

We also considered various {\em traffic scenarios}: FTP/TCP and CBR/UDP. TCP and CBR flows start at random times within the first $5sec$ and are on until the end of the simulation which is $60sec$. The CBR flow generates data packets at every $0.1ms$. IEEE 802.11b is used in the {\em MAC layer}, with the addition of the pseudo-broadcasting mechanism, as in  COPE \cite{cope}. In terms of {\em wireless channel}, we simulated the two-ray path loss model and a Rayleigh fading channel with average channel loss rates $0, 20, 30, 40, 50$ \%.
We have repeated each $60sec$ simulation for $10$ seeds. Channel capacity is $1Mbps$, the buffer size at each node is set to $100$ packets, packet sizes are set to $500B$, the generation size is set to 15 packets for UDP flows and to the TCP window size for TCP flows.

We compare our schemes (I$^2$NC-state and I$^2$NC-stateless) to no network coding ({\em noNC}), and {\em COPE} \cite{cope}, in terms of total transport-level throughput (added over all flows).

\subsection{Simulation Results}
\begin{figure}[t!]
\centering
\subfigure[Loss only on overhearing link]{{\includegraphics[width=6.4cm]{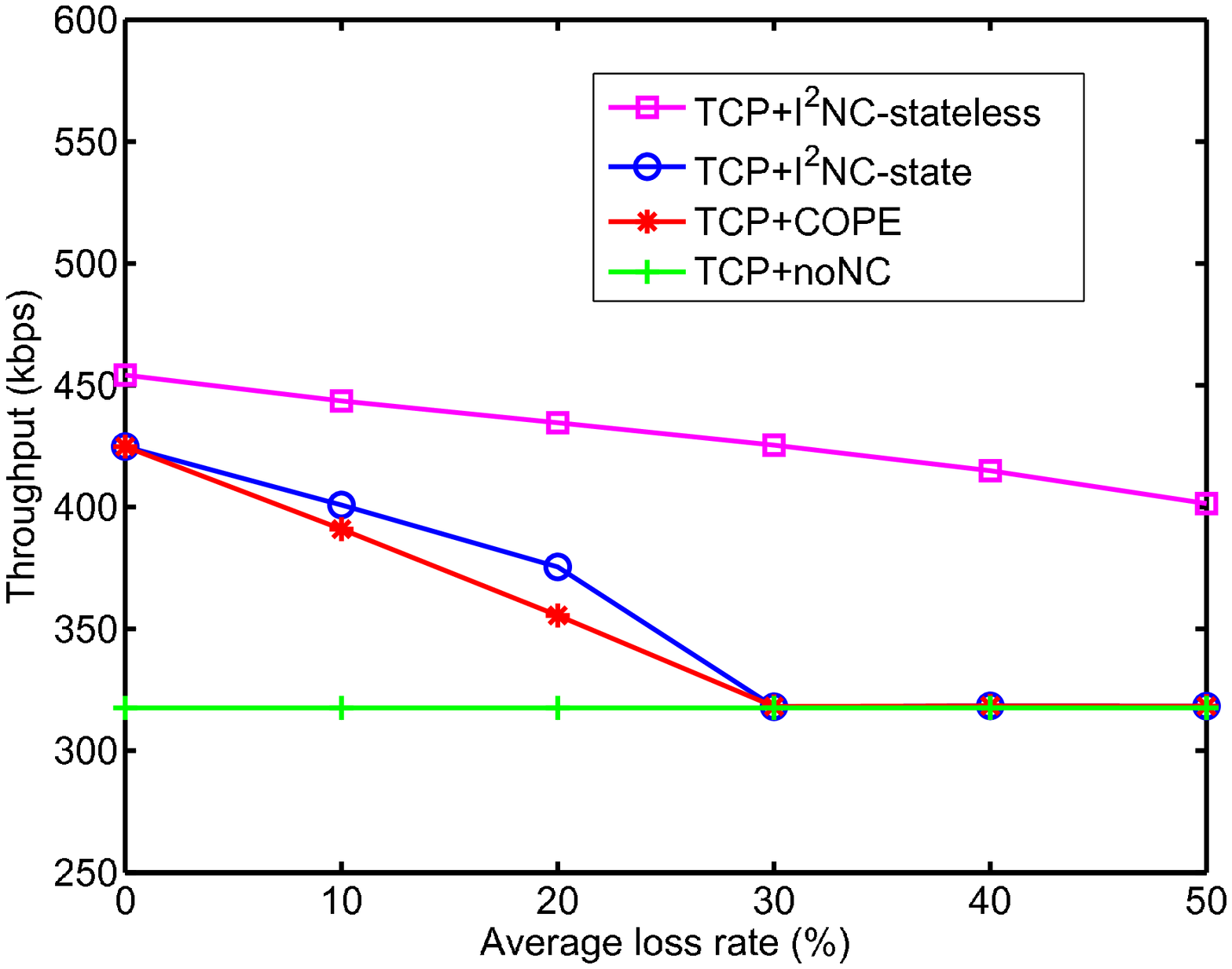}}}  \\
\subfigure[Loss only on direct link]{{\includegraphics[width=6cm]{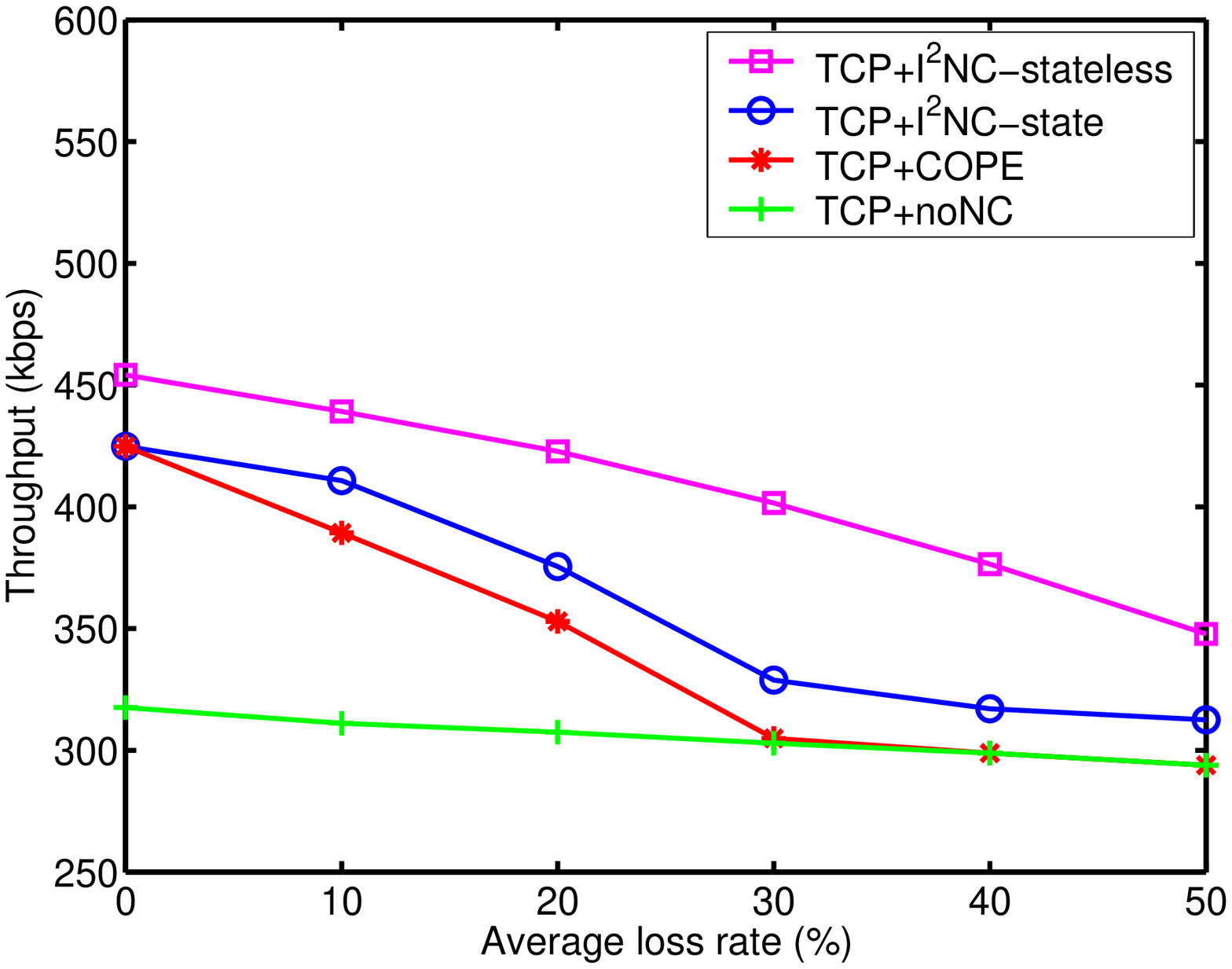}}}
\vspace{-5pt}
\caption{\label{fig:tcp_results_vs_specific_loss}  X topology in Fig.~\ref{fig:all_topologies}(a). We show the total TCP throughput (added over two flows) vs. link loss rate, for two specific loss patterns. Loss happens only on one link, either: (a) the overhearing link $A_1-B_2$ or (b) the direct link $I-B_2$. All other links are lossless.}
\vspace{-15pt}
\end{figure}

{\em TCP Traffic.} In Fig.~\ref{fig:tcp_results_vs_specific_loss}, we present simulation results for two TCP flows in X topology shown in Fig.~\ref{fig:all_topologies}(a) to illustrate the key intuition of our approach. Consider, for the moment, that loss occurs only on one link, either (a) the overhearing link $A_1-B_2$ or (b) the direct link $I-B_2$.

%% Losses in some specific patterns - TCP

The first case is depicted in Fig.~\ref{fig:tcp_results_vs_specific_loss}(a). Loss on the overhearing link does not affect the uncoded streams, thus the throughput of TCP+noNC does not change with loss rate. When NC is employed, reports carrying information about overheard packets may be delivered late to intermediate node $I$. Thus, there are some instances that intermediate node should make a decision even if it does not have the exact knowledge. In this case, $I$ makes a decision probabilistically. Specifically, if decoding probability exceeds some threshold (20\% in our simulations), $I$ codes packets. However, some of these packets may not be decodable at the receiver. It is why the performance of TCP+COPE and TCP+I$^2$NC-state reduce with increasing loss rate and equals to the throughput of TCP+noNC after 20\% loss rate (NC is turned off after 20\% loss rate). However, TCP+I$^2$NC-state is still better than TCP+COPE, because when it makes probabilistic NC decision (when loss rate is less than 20\%), it adds redundancy considering the loss rate over the overhearing link. This improves throughput, because adding redundancy using intra-session NC makes all packets equally beneficial to the receiver and the probability of decoding inter-session network coded packets increases. TCP+\stateless outperforms other schemes over the entire loss range. For example, if there is no loss, \stateless still brings the benefit due to eliminating ACK packets and using less overhead to communicate information (\ie COPE and \state exchanges the information about the overheard packets, while \stateless exchanges the information about the loss rates), thus using the medium more efficiently. When the loss rate increases, the improvement of \stateless becomes significant, reaching up to 30\%. The reason is that at high loss rates, \state and \cope do not have reliable knowledge of the decoding buffers of their neighbors and cannot do NC efficiently. In contrast, \stateless does not rely on this information, but on the loss rate of the overhearing link to make NC decision. In the discussion of Example 1, we mentioned that at 50\% loss rate, 16.6\% improvement can be achieved via NC. Here, we see this improvement (13\%) as well as the the additional benefit of eliminating ACK packets (12\%). Note that the total improvement is 25\%.

The second case is depicted in Fig.~\ref{fig:tcp_results_vs_specific_loss}(b). The throughput of TCP+noNC decreases with increasing loss rate because, the loss is over the direct link and some packets whether they are coded or not are lost on the direct link ($I-B_2$). This leads to decrease in throughput level. TCP+\state outperforms TCP+\cope in this scenario, because \state corrects errors on the direct link thanks to the added redundancy which reduces the number of re-transmissions. Thus, \state uses the channel more efficiently than \cope and improves the throughput. Note that TCP+\state outperforms TCP+\cope even after 20\% loss rate, although inter-session NC is turned off after this level. The reason is that although \state does not do inter-session NC after 20\% loss rate, it keeps doing intra-session NC which adds redundancy to correct errors. Due to this property, TCP+\state outperforms TCP+\cope even at high loss rates. TCP+\stateless significantly outperforms all alternatives again due to performing NC at all loss rates and eliminating ACK packets.

\begin{figure*}[t!]
\begin{center}
\subfigure[X topology (shown in Fig.~\ref{fig:all_topologies}(a))]{{\includegraphics[width=5cm]{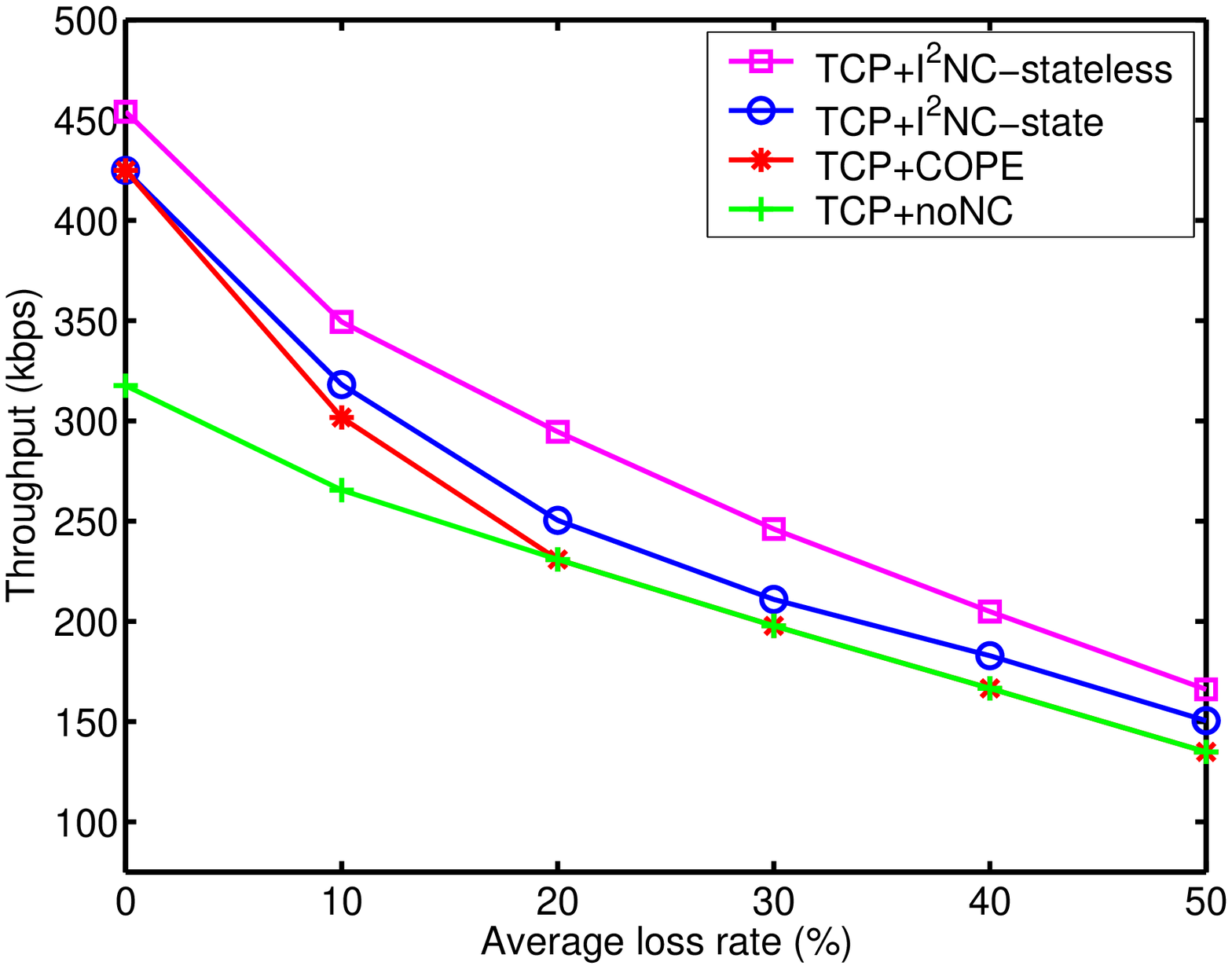}}} \hspace{-0pt}
\subfigure[Cross topology (shown in Fig.~\ref{fig:all_topologies}(b))]{{\includegraphics[width=5cm]{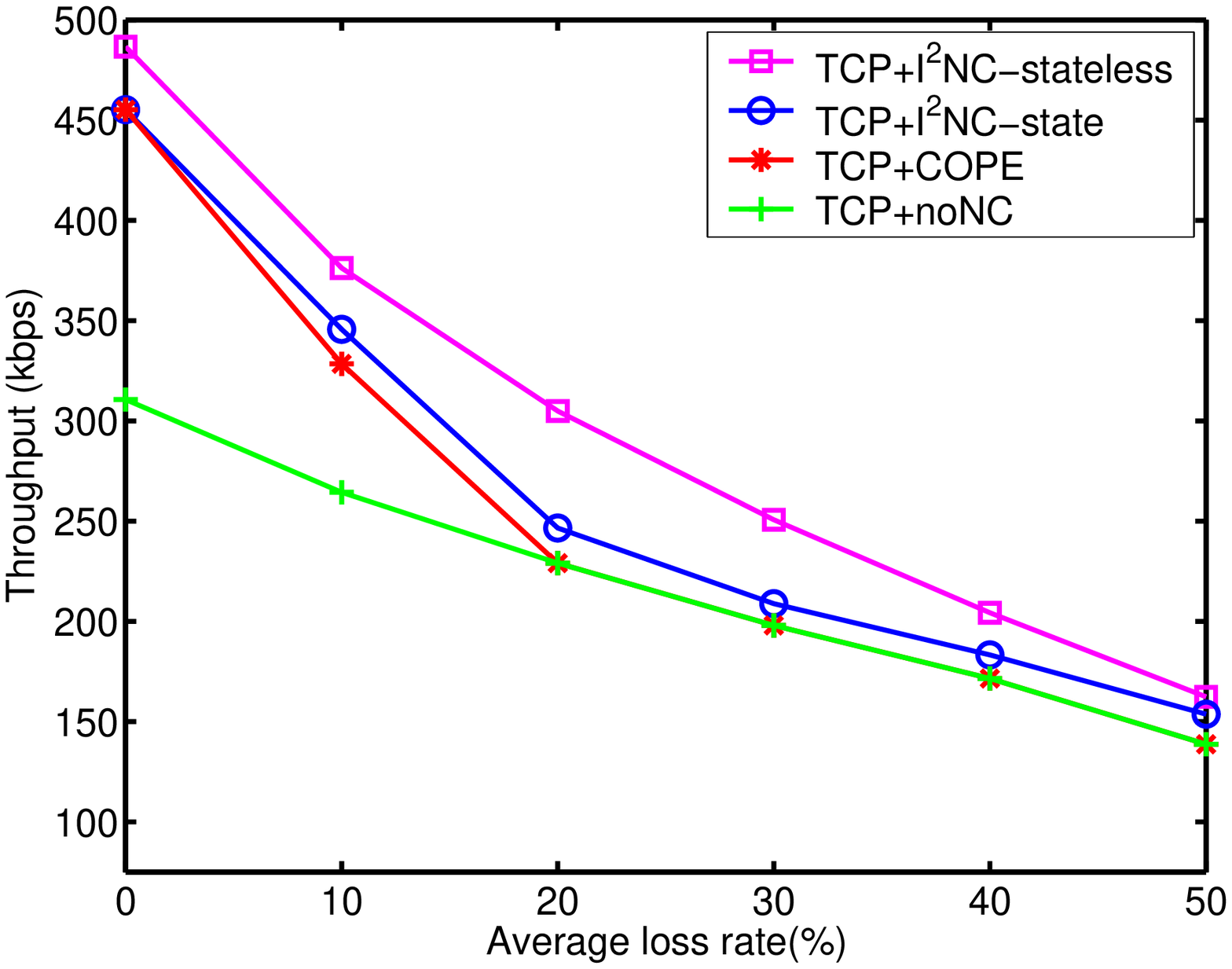}}} \hspace{-0pt}
\subfigure[Multi-hop topology (shown in Fig.~\ref{fig:one-hop})]{{\includegraphics[width=5cm]{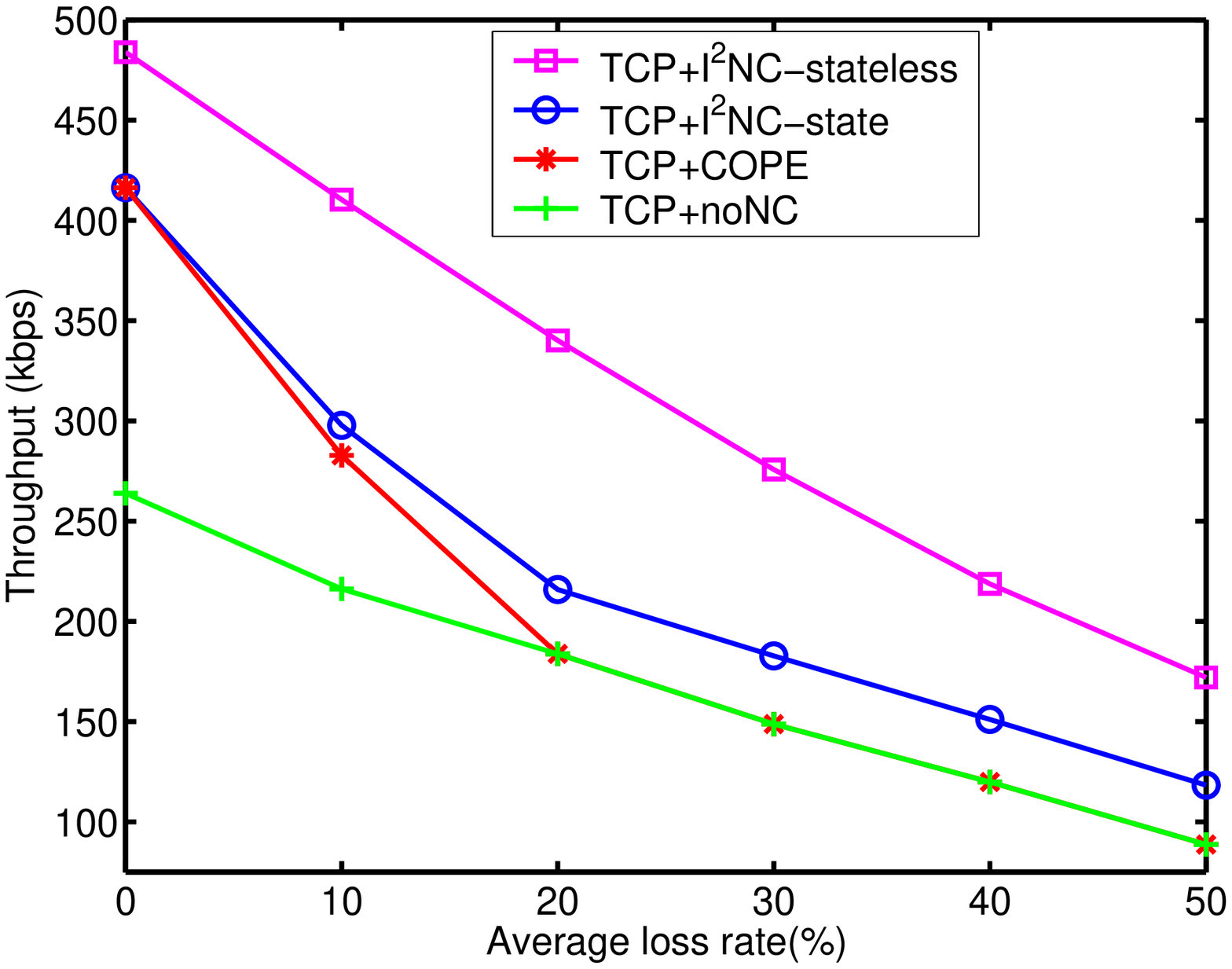}}} \hspace{-0pt}
\end{center}
\begin{center}
\vspace{-5pt}
\caption{\label{fig:tcp_results_vs_loss} Total TCP throughput vs. average loss rate (for ease of presentation, the same loss rate is assumed on all links) in three different topologies.}
\vspace{-15pt}
\end{center}
\end{figure*}

%{\em TCP Traffic.}
Fig.~\ref{fig:tcp_results_vs_loss} presents simulation results for TCP traffic over X, cross, and the multi-hop topologies assuming loss on all links. For ease of presentation, here, we report only the results when all links have the same loss probability.
%% Losss over all links - TCP

Fig.~\ref{fig:tcp_results_vs_loss}(a) shows the results for the X topology. At low-medium loss rates (10\% - 30\%), \state and \cope are still able to do NC, so TCP+I$^2$NC-state and TCP+COPE improve throughput significantly as compared to TCP+noNC. At higher loss rates, I$^2$NC-state and COPE do not have reliable knowledge of the decoding buffers of their neighbors and cannot do NC efficiently. As a result, the improvement of TCP+I$^2$NC-state and TCP+COPE as compared to TCP+noNC reduce with increasing loss rate. TCP+I$^2$NC-state is better than TCP+COPE at higher loss rates thanks to its error correction mechanism. TCP+I$^2$NCstateless outperforms other schemes over the entire loss range thanks to combining NC and error correction as well as eliminating ACKs. For example, if there is no loss, TCP+I$^2$NC-stateless still brings the benefit by eliminating ACK packets, thus using the medium more efficiently. When the loss rate increases, the improvement of I$^2$NC-stateless becomes significant, because I$^2$NC-stateless does not rely on the knowledge of the decoding buffers of their neighbors, but only on the link loss rates for inter-session NC.

Fig.~\ref{fig:tcp_results_vs_loss}(b) shows the results for the cross topology. The improvement of TCP+I$^2$NC-stateless is higher as compared to the X topology, because there are more NC opportunities here for I$^2$NC-stateless to exploit. We also performed simulations with increasing number of flows, ({\em i.e.}, nodes in this topology); the details are provided later in this section.

Fig.~\ref{fig:tcp_results_vs_loss}(c) presents the results for the multi-hop topology in Fig.~\ref{fig:one-hop}. The improvement of TCP+I$^2$NC-state is higher than in the X and cross topologies, especially at higher loss rates. This is because intra-session coding, employed by I$^2$NC-state, reduces the dependency on link level ARQ. More specifically, in this multi-hop topology, the end-to-end residual loss rate increases with the number of hops. Intra-session NC overcomes this, thus increasing TCP throughput. The improvement of I$^2$NC-stateless is even more significant for this topology, because the benefit of eliminating ACKs is more pronounced with larger number of hops.

\begin{figure}[t!]
\centering %\vspace{-5pt}
\includegraphics[width=6cm]{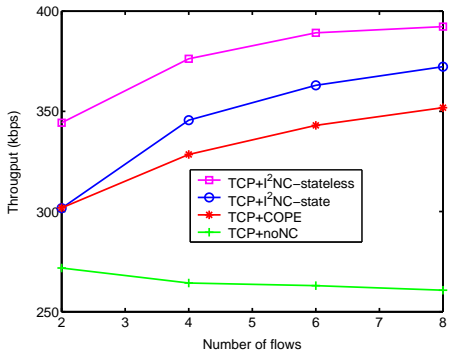}
\vspace{-5pt}
\caption{\label{fig:cross_number_of_flows} Wheel topology shown in Fig.~\ref{fig:all_topologies}(c) with increasing number of flows. Loss rate on all links is set to 10\%.}
\vspace{-15pt}
\end{figure}

We also performed simulations with increasing number of flows, \ie nodes in wheel topology in Fig.~\ref{fig:all_topologies}(c). It is seen in Fig.~\ref{fig:cross_number_of_flows} that the total throughput achieved by NC schemes increases with the increasing number of flows. When the number of flows increases, the probability of NC at the intermediate node $I$ increases. More NC opportunities leads to higher throughput.
%% Wheel topology

\begin{figure*}[t!]
\vspace{-5pt}
\begin{center}
\subfigure[X topology (shown in Fig.~\ref{fig:all_topologies}(a))]{{\includegraphics[width=5cm]{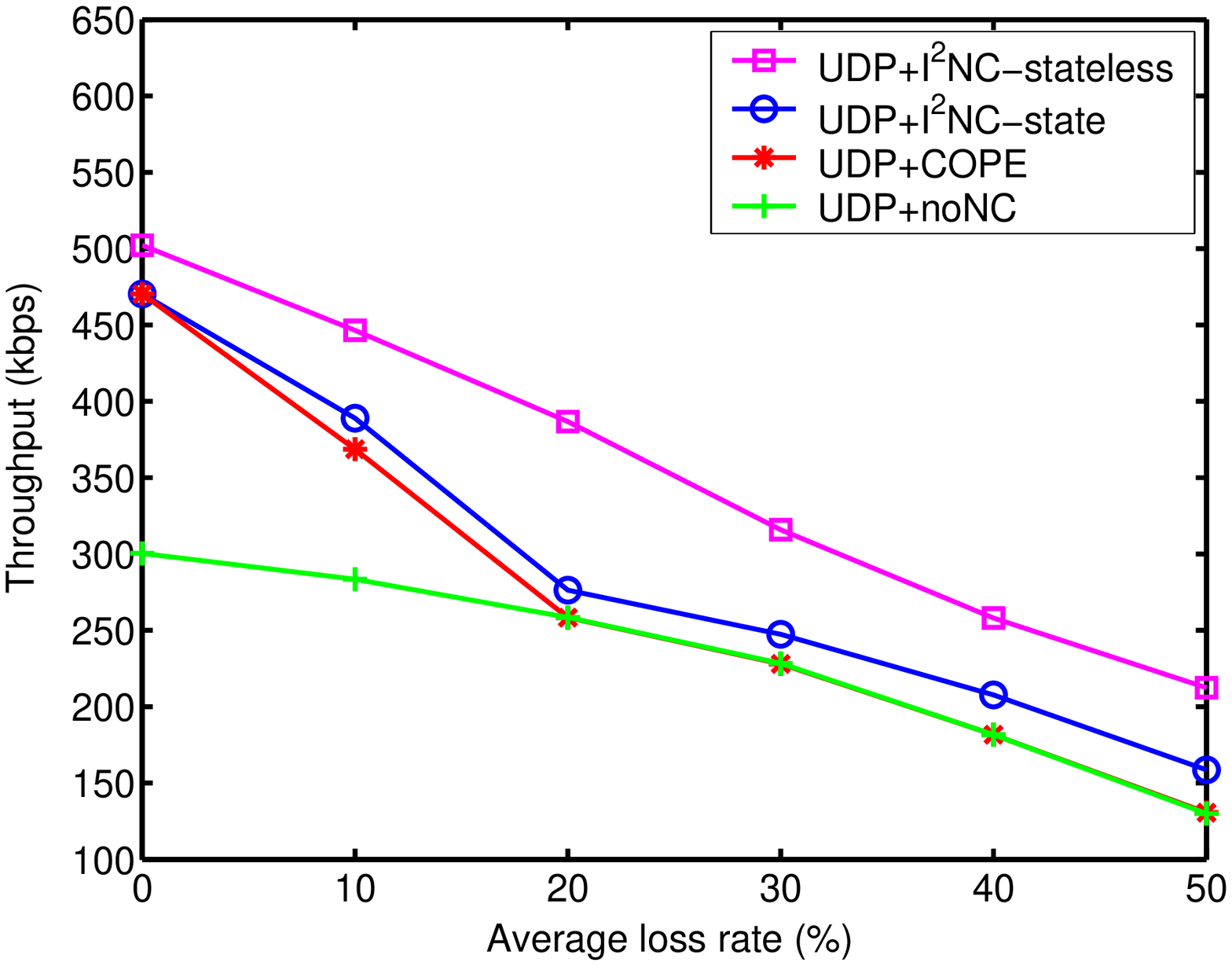}}} \hspace{-0pt}
\subfigure[Cross topology (shown in Fig.~\ref{fig:all_topologies}(b))]{{\includegraphics[width=5cm]{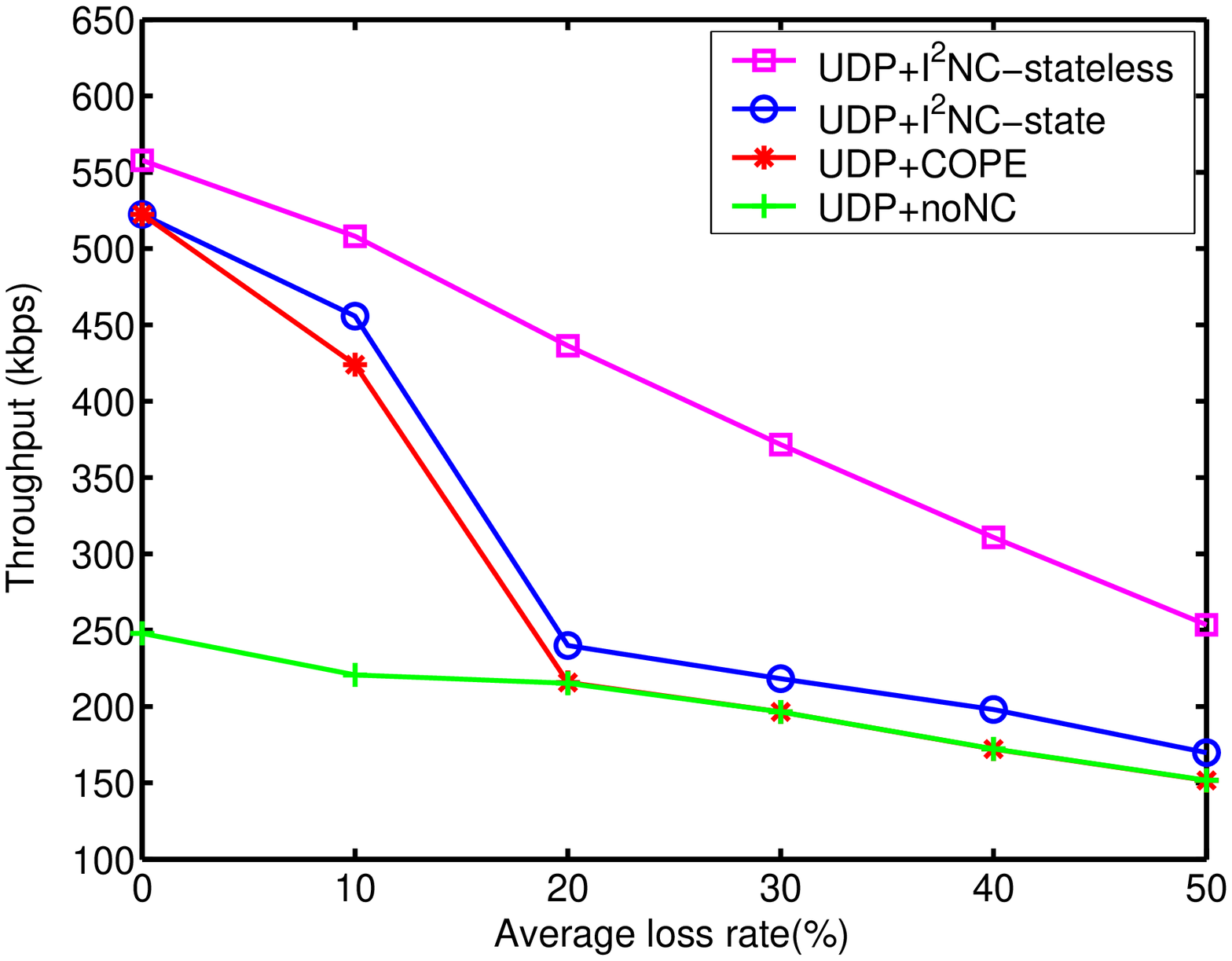}}} \hspace{-0pt}
\subfigure[Multi-hop topology (shown in Fig.~\ref{fig:one-hop})]{{\includegraphics[width=5cm]{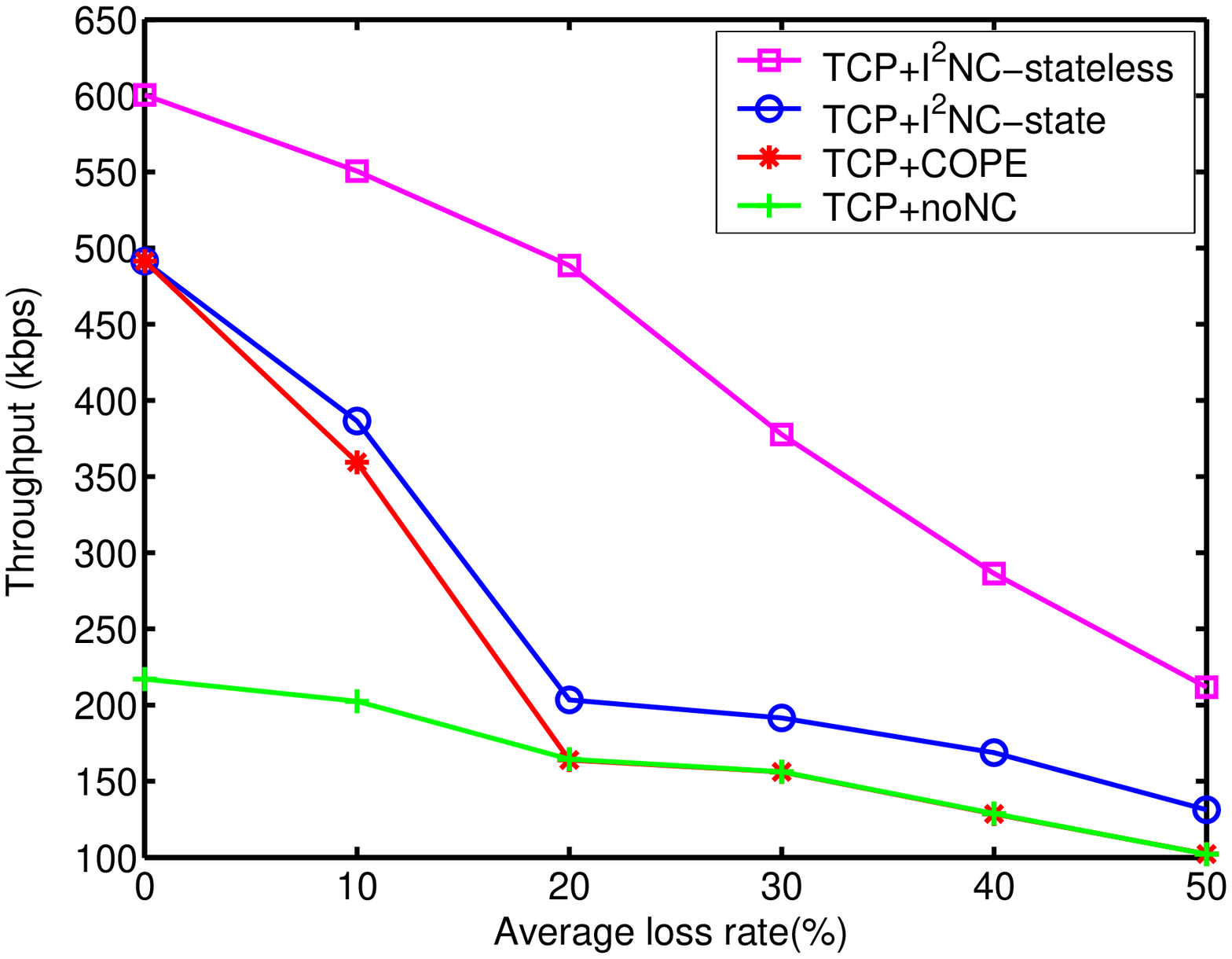}}} \hspace{-0pt}
%\vspace{-15pt}
\end{center}
\begin{center}
\vspace{-5pt}
\caption{\label{fig:udp_results_vs_loss} Total UDP throughput vs. average loss rate (the same loss rate is assumed on all links) in three different topologies.}
\vspace{-15pt}
\end{center}
\end{figure*}

{\em UDP traffic.} We repeated the simulations for the three topologies for the case that there is loss over all links. The results are presented in Fig.~\ref{fig:udp_results_vs_loss}.

%% Losss over all links - UDP

Fig.~\ref{fig:udp_results_vs_loss}(a) presents the results for the X topology. The improvement of UDP+\stateless is up to 60\% as compared to UDP+noNC. This is significantly higher than the improvement of TCP+\stateless and the optimal scheme (in which the improvement is 33.3\%). The reason is the MAC gain as explained in \cite{cope}.\footnote{The MAC gain observed with UDP flows when NC is used can be summarized as follows. When NC is employed, the coded wireless network can handle larger amount of load as compared to its uncoded counterpart. Therefore, when coded system saturates at some load level, uncoded system can not handle this level of load. Thus, several packets are dropped from output queues at each node in the system. Some of these packets may be dropped from intermediate packets. In this case, resources (bandwidth in our case) to transmit these packets (which will be eventually dropped) is wasted. Therefore, the gap between the achieved throughput level of coded and uncoded systems becomes significant.} We present the results for the load at which the system saturates. At this load, UDP+noNC is already saturated, several packets are dropped from the buffers, and they do not arrive to their receivers. This reduces the throughput of noNC, while NC schemes still handle the traffic created by the load. Notice that even at 50\% loss rate, UDP+\stateless improves over UDP+noNC by 40\%, which is significant.

Fig.~\ref{fig:udp_results_vs_loss}(b) presents the results for the cross topology. In this topology, the improvement of NC is very large. When there is no loss, the improvement is around 250\%. The effectiveness of UDP+\stateless is also significant in this topology: at 50\% loss rate the improvement of UDP+\stateless over UDP+noNC is 70\%.

Fig.~\ref{fig:udp_results_vs_loss}(c) presents the results for multi-hop topology. We see similar behavior as observed by Figs.~\ref{fig:udp_results_vs_loss}(b) and (c). However, the improvement of UDP+\stateless is larger in this topology, because the benefit of eliminating ACKs is more pronounced with larger number of hops.

\subsection{Numerical Results}
We consider the X and cross topologies shown in Figs.~\ref{fig:all_topologies}(a) and \ref{fig:all_topologies}(b). In the X topology, $A_1$ transmits packets to $A_2$ via $I$ with rate $x_1$, and $B_1$ transmits packets to $B_2$ via $I$ with rate $x_2$. In the cross topology, $A_1$ transmits packets to $A_2$ with rate $x_1$, $A_2$ transmits packets to $A_1$ with rate $x_2$, $B_1$ transmits packets to $B_2$ with rate $x_3$, and $B_2$ transmits packets to $B_1$ with rate $x_4$. All transmissions are via $I$. In both topologies, the data rate of each link is set to $1$ packet/transmission. We compare our schemes \state and \stateless with noNC which is also formulated in a network utility maximization framework without any NC constraints.

\begin{figure}[t!]
\begin{center}
\subfigure[Loss on $A_1-B_2$]{{\includegraphics[width=4.3cm]{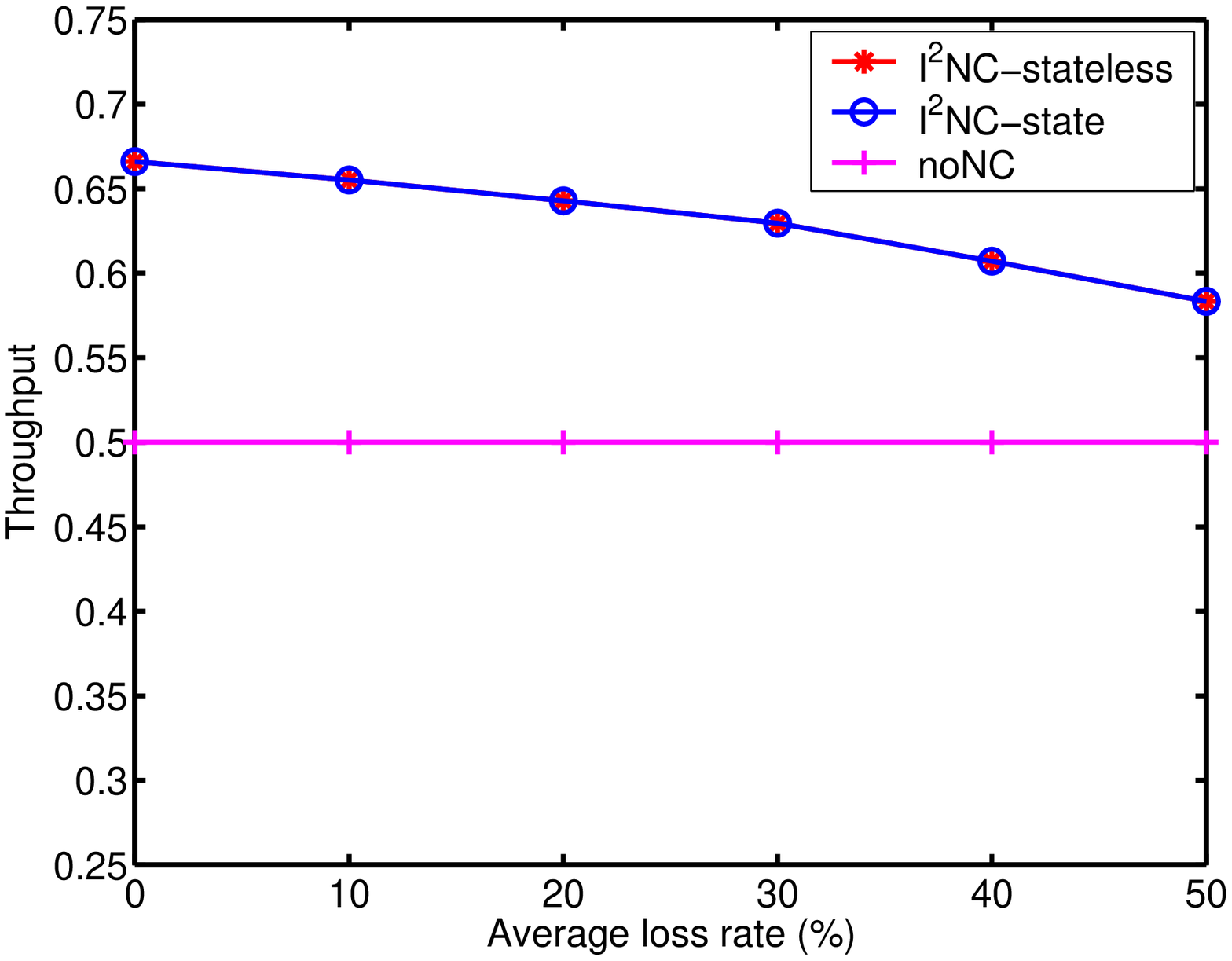}}}
\subfigure[Loss on $I-B_2$]{{\includegraphics[width=4.3cm]{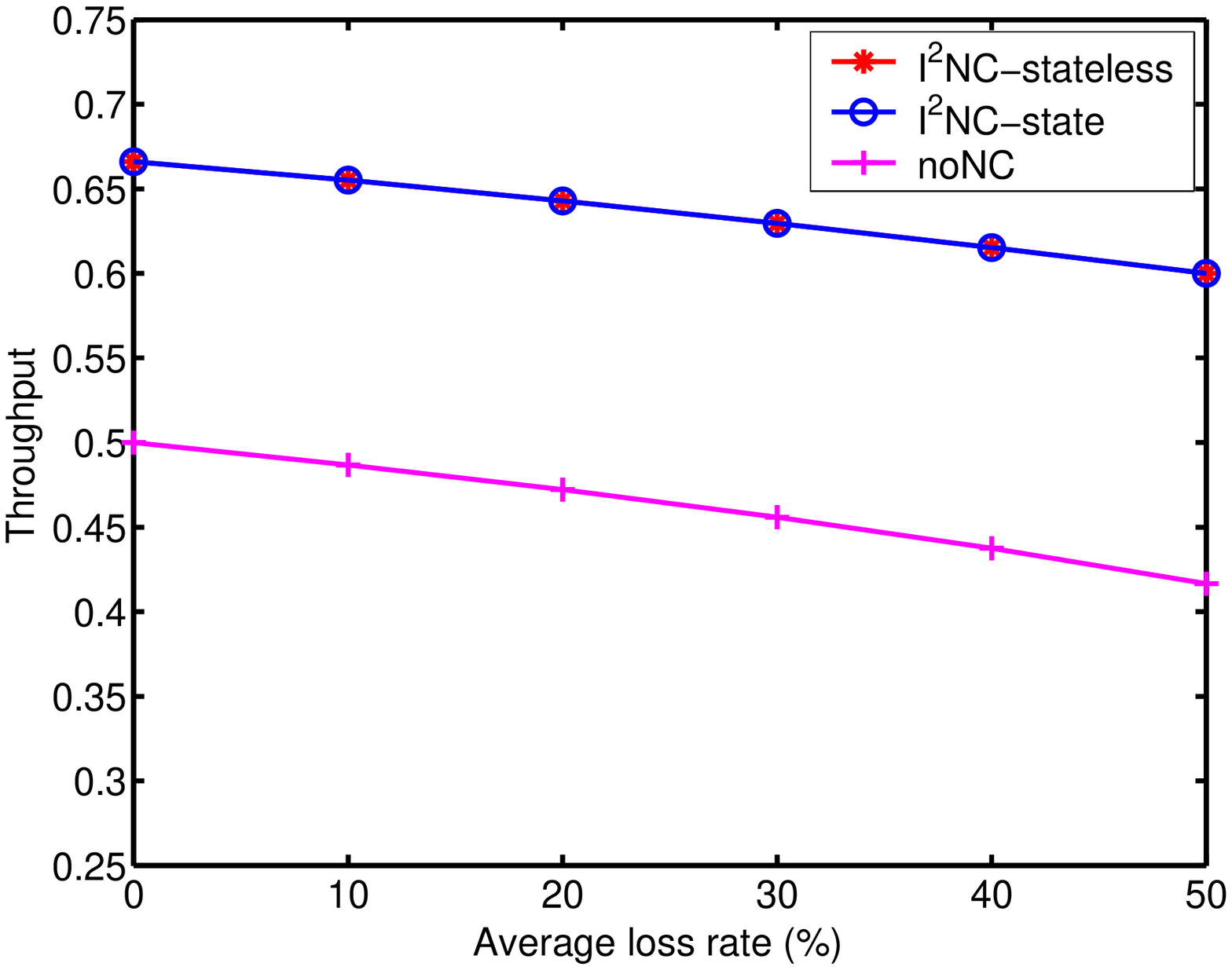}}} \\
\subfigure[Loss on $A_1-B_2$ and $I-B_2$]{{\includegraphics[width=4.3cm]{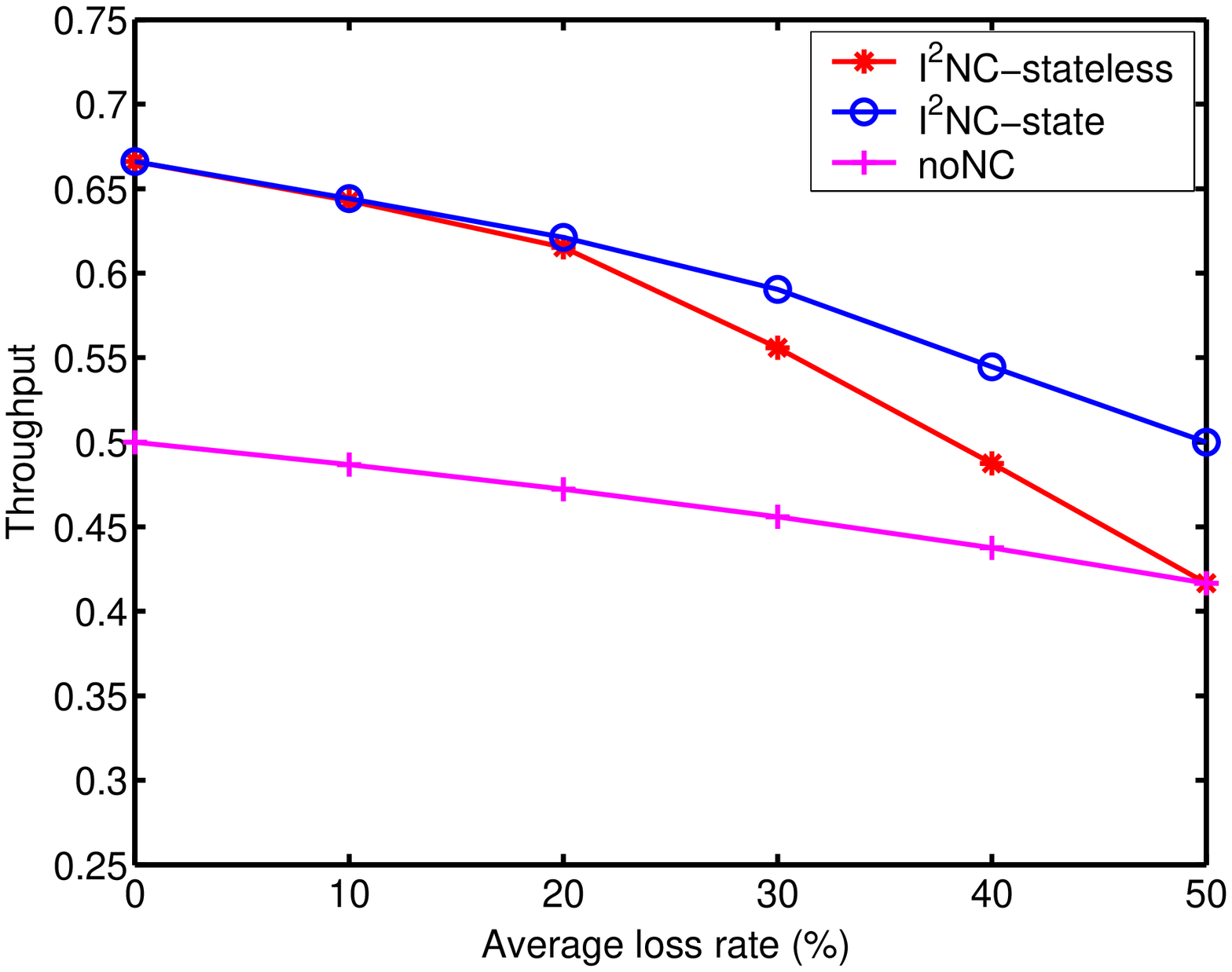}}}
\subfigure[Loss on all links]{{\includegraphics[width=4.3cm]{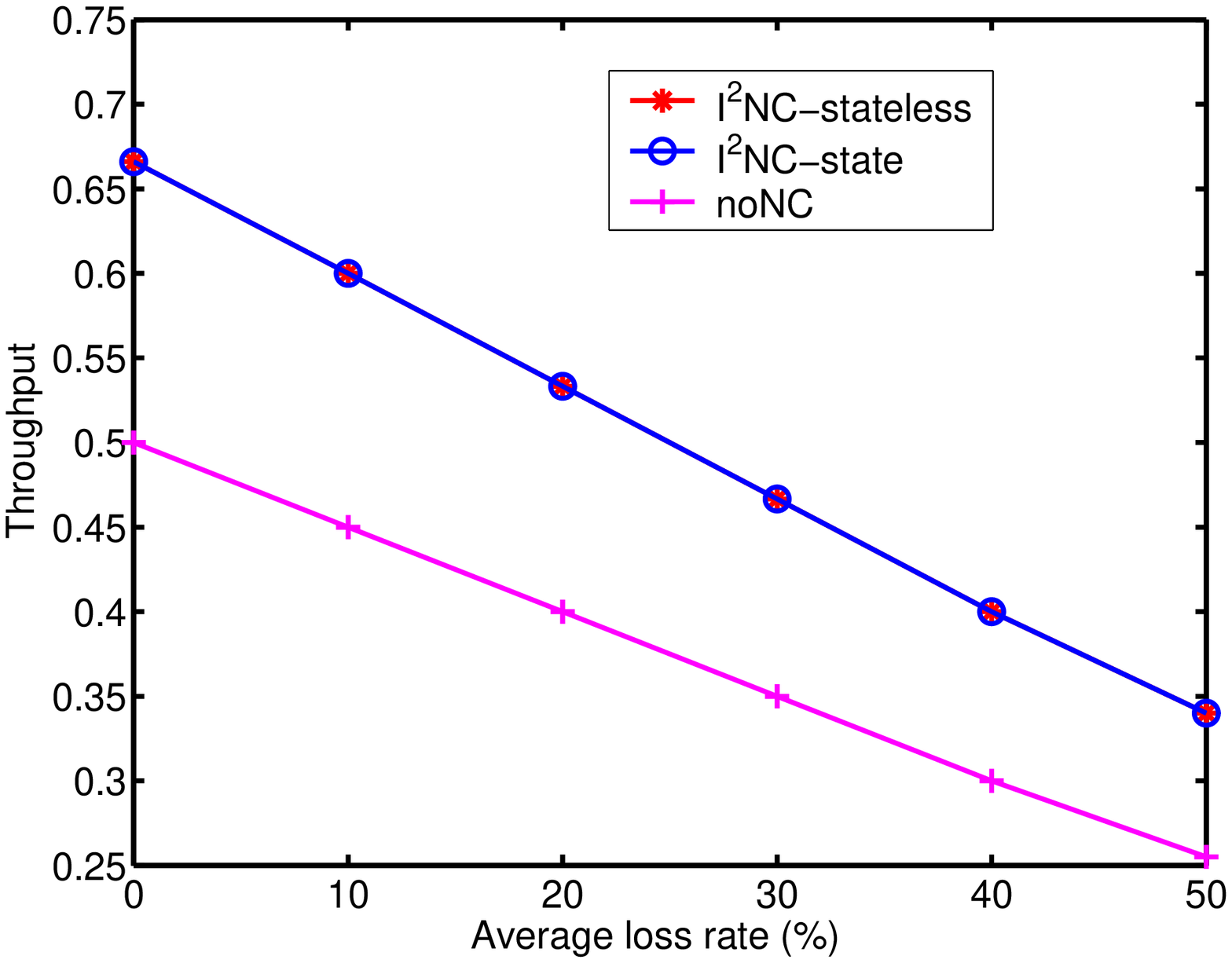}}}
\end{center}
\begin{center}
\caption{\label{fig:x_nc_results_vs_loss} X topology. Throughput vs. loss rate (the same loss rate is assumed on $A_1-B_2$ and $I-B_2$ in (c), and the same loss rate is assumed on all links in (d)).}
\vspace{-15pt}
\end{center}
\end{figure}

Fig.~\ref{fig:x_nc_results_vs_loss} shows the total throughput; $x_1+x_2$ for X topology. Fig.~\ref{fig:x_nc_results_vs_loss}(a) shows the results when there is loss on $A_1-B_2$. It is seen that the throughput of noNC is flat with increasing loss rate, because it is not affected by the loss rate on the overhearing link. \state and \stateless improve over noNC, because they exploit NC benefit. When the loss rate increases, the improvement reduces, because $B_2$ overhears only part of the data transmitted by $A_1$. Although the improvement decreases with increasing loss rate, it is still significant, \eg 16.6\% at 50\% loss rate. Note that
Fig.~\ref{fig:x_nc_results_vs_loss}(a) is the counterpart of the simulation results presented in Fig.~\ref{fig:tcp_results_vs_specific_loss}(a). It is seen that TCP+\stateless in Fig.~\ref{fig:tcp_results_vs_specific_loss}(a) shows similar performance as \stateless in Fig.~\ref{fig:x_nc_results_vs_loss}(a). This shows the effectiveness of \stateless in a realistic simulation environment. 

% Total achieved rate vs loss rate - X topology

Fig.~\ref{fig:x_nc_results_vs_loss}(b) shows the results when there is loss on $I-B_2$. It is seen that \state and \stateless improve over noNC significantly at all loss rates. It is also interesting to note that at 50\% loss rate, \state and \stateless improve over noNC by 44\% which is even higher than in the no loss case (33\%). The reason is in the following. In the optimal solution, the throughput values are $x_1 = 0.4$ and $x_2 = 0.2$. In this case, in the downlink $I-B_2$, data part of $x_2$ with rate $0.2$ and the parity part with rate $0.2$ (considering loss rate 50\%) are combined with $x_1$. This means that our schemes combine both parity and data parts of a flow with other flows and this improves the throughput significantly. This is one of the important contributions of I$^2$NC.

Fig.~\ref{fig:x_nc_results_vs_loss}(c) shows the results when there is loss on links $A_1-B_2$ and $I-B_2$. It is seen that \state improves the throughput significantly while the improvement of \stateless reduces to $0$ with increasing loss rate. The reason is that, \stateless is a more conservative scheme as compared to \state in the sense that it eliminates the perfect knowledge on antidotes. Yet, it still improves the throughput significantly, \eg it improves over noNC by 22\% at 30\% loss rate.

Fig.~\ref{fig:x_nc_results_vs_loss}(d) shows the results when there is loss on all links. It is seen that \state and \stateless improve over noNC significantly at all loss rates. Note that throughput of \stateless reduces to that of noNC at 50\% loss rate in Fig.~\ref{fig:x_nc_results_vs_loss}(c). The reader might wonder why we do not see such behavior in Fig.~\ref{fig:x_nc_results_vs_loss}(d). The reason is that since there is loss over link $A_1-I$ as well as $A_1-B_2$, the number of parities added by $A_1$ to correct losses over link $A_1-I$ also increases the number of overheard packets at $B_2$. Therefore, \stateless does not add redundancy at node $I$ for both $A_1-B_2$ and $I-B_2$ as in Fig.~\ref{fig:x_nc_results_vs_loss}(c), but adds redundancy only for loss on link $I-B_2$. This improves the performance of \stateless. Note that the counterpart of these results are presented in Fig.~\ref{fig:tcp_results_vs_loss}(a). It is seen that the throughput improvement of \stateless over noNC at 50\% loss rate is around 30\% in Fig.~\ref{fig:x_nc_results_vs_loss}(d). As compared to this, the improvement of TCP+\stateless over noNC is limited in Fig.~\ref{fig:tcp_results_vs_loss}(a), because, in simulations, the block size is limited and fixed, and the scheduling is not perfect (we consider IEEE 802.11). Yet, the throughput improvement of TCP+\stateless over noNC is around 20\% in Fig.~\ref{fig:tcp_results_vs_loss}(a), which is significant.

\begin{figure}[t!]
\begin{center}
\subfigure[Loss on $A_1-B_2$]{{\includegraphics[width=4.3cm]{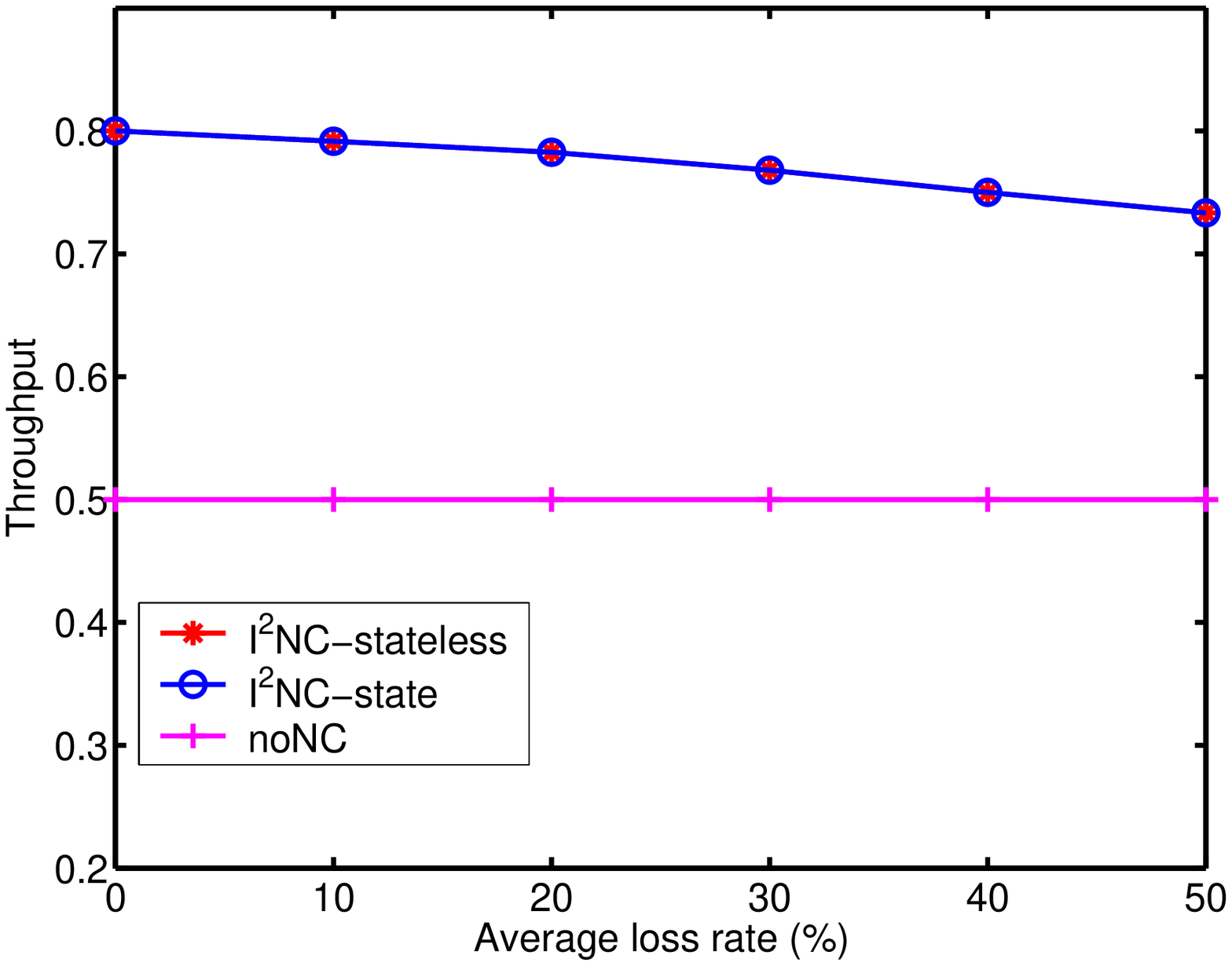}}}
\subfigure[Loss on $I-B_2$]{{\includegraphics[width=4.3cm]{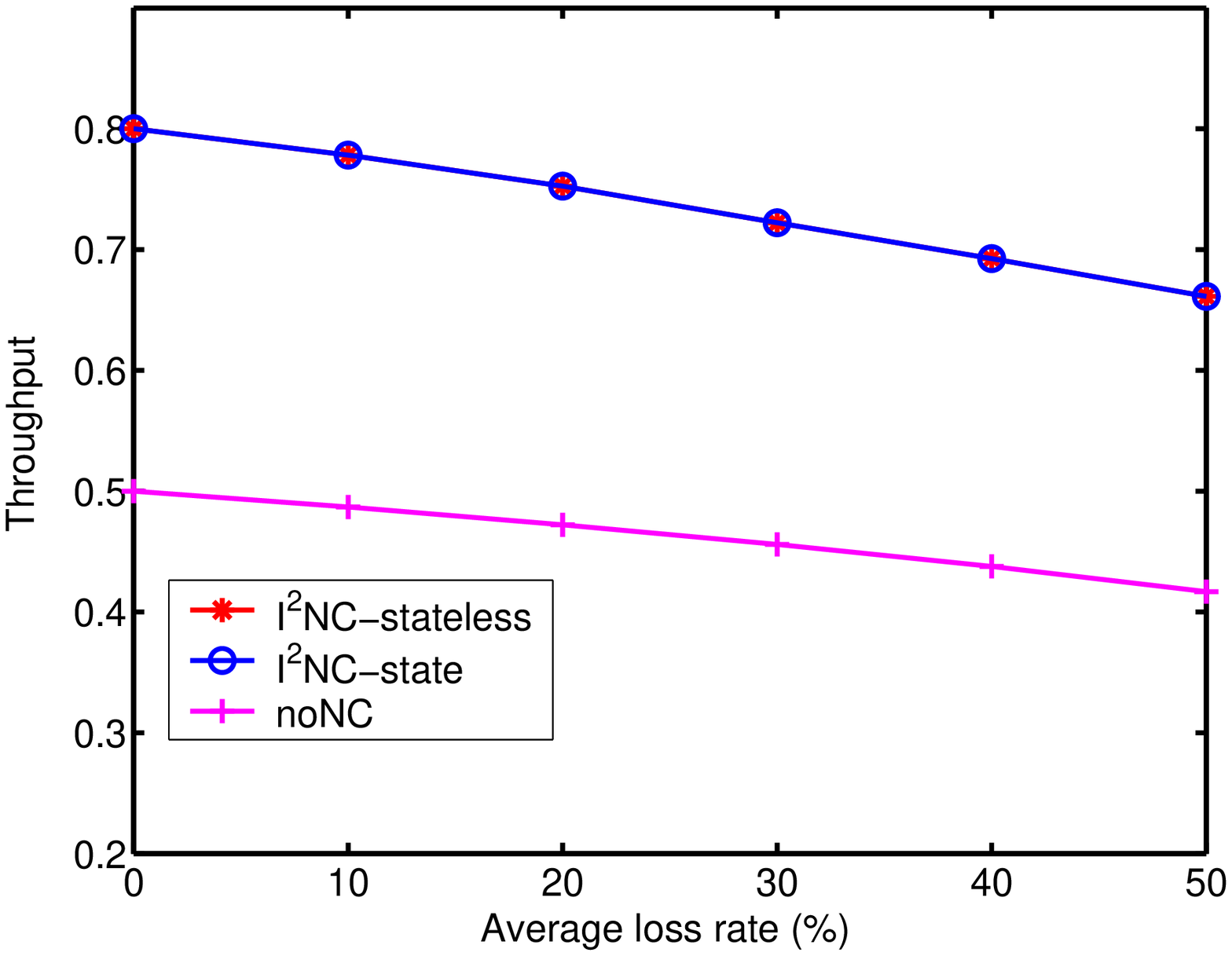}}} \\
\subfigure[Loss on $A_1-B_2$ and $I-B_2$]{{\includegraphics[width=4.3cm]{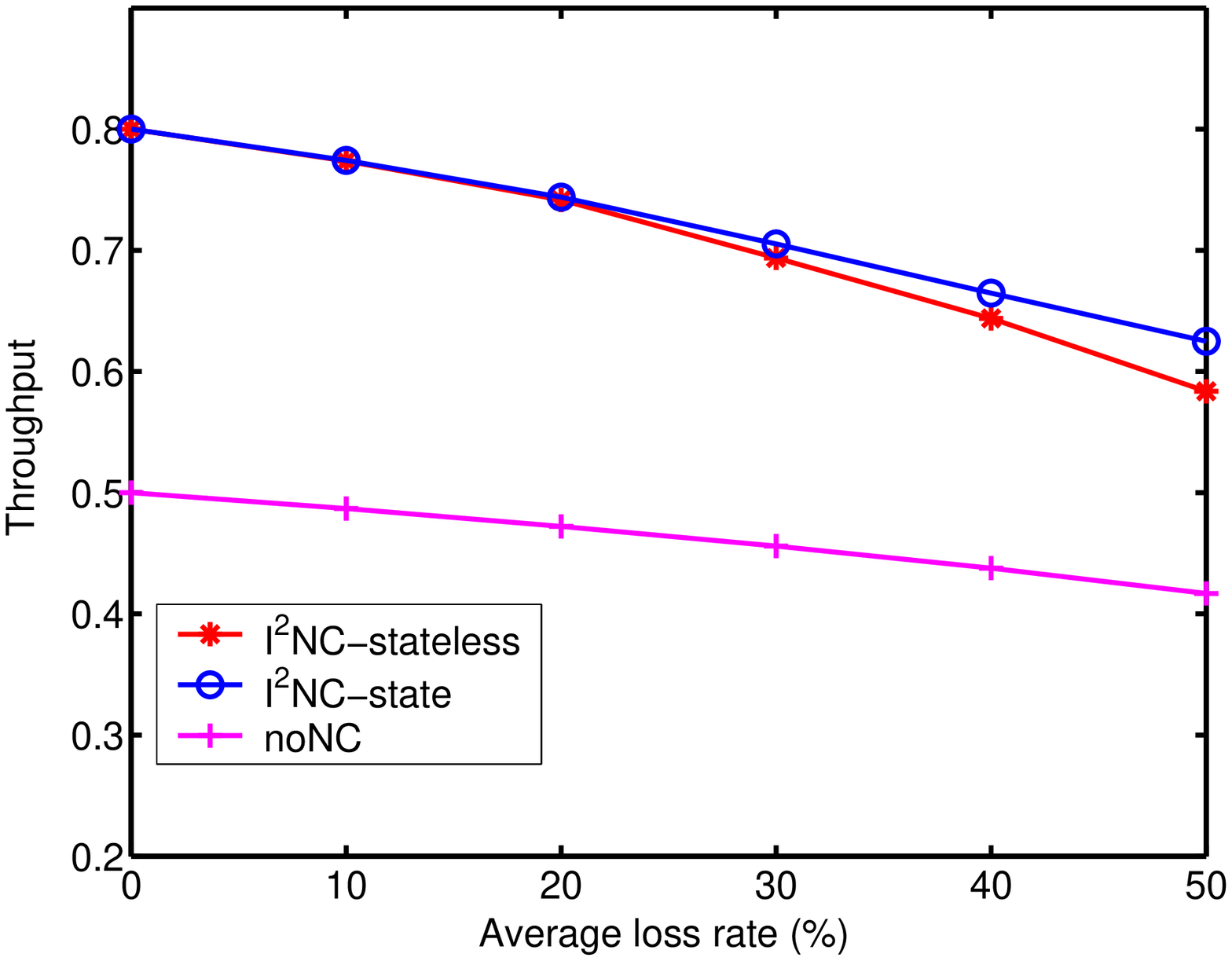}}}
\subfigure[Loss on all links]{{\includegraphics[width=4.3cm]{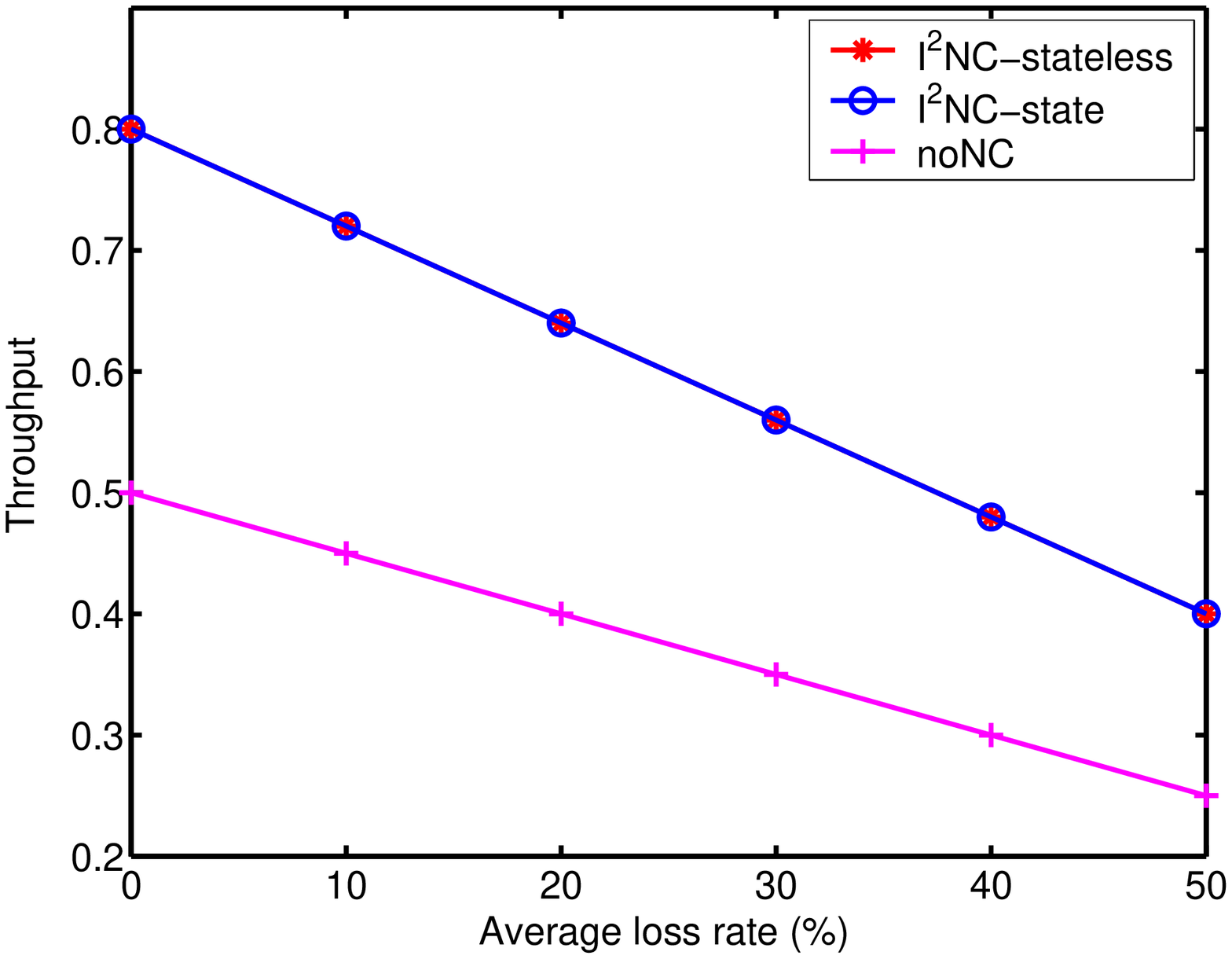}}}
\end{center}
\begin{center}
\caption{\label{fig:cross_nc_results_vs_loss} Cross topology. Throughput vs. loss rate (the same loss rate is assumed on $A_1-B_2$ and $I-B_2$ in (c), and the same loss rate is assumed on all links in (d)).}
\vspace{-15pt}
\end{center}
\end{figure}

Fig.~\ref{fig:cross_nc_results_vs_loss} shows the total throughput; $x_1+x_2+x_3+x_4$ for {I$^2$NC-state}, \stateless and noNC for the cross topology shown in Fig.~\ref{fig:all_topologies}(b) for different loss patterns. It is seen that the results are similar to the ones in Fig.~\ref{fig:x_nc_results_vs_loss}. One difference is that the throughput improvement of NC schemes is higher, {\em i.e.}, up to 80\%, because there are more NC opportunities in the cross topology.

% Total achieved rate vs loss rate - Cross topology

\section{\label{sec:conclusion}Conclusion}

In this paper, we proposed {I$^2$NC}: a one-hop intra- and inter-session network coding approach for wireless networks. {I$^2$NC} builds on and improves COPE in two aspects: it is resilient to loss and it does not need to rely on the exact knowledge of the state of the neighbors. Our design is grounded on a NUM formulation and its solution. Simulations in {\tt GloMoSim} demonstrate significant throughput gain of our approach compared to no network coding and COPE.

\bibliographystyle{IEEEtran}

\section*{\label{sec:num_results}Appendix A: Convergence Analysis}
In this section, we analyze the convergence of the distributed solution of the NUM problem, given in Section~\ref{sec:opt2}. First, we provide a proof of convergence, and then present some numerical calculations to verify the convergence.

\subsection{Proof of Convergence}
%\textbf{Optimality Conditions:}
Let us first consider the optimality conditions below. Note that $\overset{*}{x}_s$, $\overset{*}{\alpha}_{h,k}^{s}$, $\overset{*}{\tau}_{h,k}^{s}$, and $\overset{*}{q}_{h,k}^{s}$ are the optimal values.
\begin{align} \label{appendix:opt_cond_eq1}
& \textstyle \overset{*}{x}_s = ({U'_{s}})^{-1} \Bigl( \sum _{i \in \Pset_{s}} \sum _{\substack{h(J) \mid h \in \Aset}} \sum _{k \in \Kset_{h} \mid s \in \Sset_{k}} H_{h,k}^{s} \overset{*}{\alpha}_{h,k}^{s} \bigl( \frac{\overset{*}{q}_{h,k}^{s}}{1-\rho_{h}^{s}} \nonumber \\
& \textstyle + \sum _{s' \in  \Sset_{k}-\{s\}} \overset{*}{q}_{h,k}^{s'}\rho_{h,k}^{s',s} \bigr) \Bigr),
\end{align}
\begin{align} \label{appendix:opt_cond_eq2}
\textstyle \overset{*}{\tau}_{h,k}^{s} = \arg \max_{\sum_{h \in C_{q}} \sum_{k \in \Kset_{h}} \tau_{h,k} \leq \gamma} \sum_{h \in \Aset} \sum_{k \in \Kset_{h}} \overset{*}{Q}_{h,k} \tau_{h,k},
\end{align}
\begin{align} \label{appendix:opt_cond_eq3}
& \textstyle  \overset{*}{q}_{h,k}^{s} \bigl( \frac{H_{h,k}^{s} \overset{*}{\alpha}_{h,k}^{s} \overset{*}{x}_{s} }{1-\rho_{h}^{s}} + \sum _{\substack{s' \in  \Sset_{k} - \{s\}}} H_{h,k}^{s'} \overset{*}{\alpha}_{h,k}^{s'} \overset{*}{x}_{s'}\rho_{h,k}^{s,s'} \nonumber \\
& \textstyle - R_{h} \overset{*}{\tau}_{h,k} \bigr) = 0, \mbox{  } \overset{*}{q}_{h,k}^{s} \geq 0,
\end{align}
\begin{align} \label{appendix:opt_cond_eq4}
& \textstyle \frac{H_{h,k}^{s} \overset{*}{\alpha}_{h,k}^{s} \overset{*}{x}_{s} }{1-\rho_{h}^{s}} + \sum _{s' \in \Sset_{k} - \{s\}} H_{h,k}^{s'} \overset{*}{\alpha}_{h,k}^{s'} \overset{*}{x}_{s'}\rho_{h,k}^{s,s'} \leq  R_{h} \overset{*}{\tau}_{h,k}, \nonumber \\
& \textstyle \sum_{h \in \Cset_{q}} \sum_{k \in \Kset_{h}} \overset{*}{\tau}_{h,k} \leq \gamma,
\end{align}
\begin{align} \label{appendix:opt_cond_eq5}
& \overset{*}{\alpha}_{h,k}^{s} = \arg \min_{ \sum_{\substack {h(\Jset) | \\ h \in \Aset}} \sum_{\substack{k \in \Kset_{h}| \\ s \in \Sset_{k}}} \alpha_{h,k}^{s} = 1 } \overset{*}{x}_s  \sum_{h \in \Aset} \sum_{\substack{k \in \Kset_{h}  | \\ s \in \Sset_{k}}} \alpha_{h,k}^{s}H_{h,k}^{s}\overset{*}{Q}_{h,k}^{s}, \nonumber \\
& \textstyle \sum_{h(\Jset)|h \in \Aset} \sum_{k \in \Kset_{h}|s \in \Sset_{k}} \overset{*}{\alpha}_{h,k}^{s} = 1.
\end{align}

We consider a similar Lyapunov function considered in \cite{cc_multicast_nc}; $V(q,\tau,\alpha) = \sum_{h \in \Aset} \sum_{k \in \Kset_{h}} \sum_{s \in \Sset_{k}}$ $\frac{(q_{h,k}^{s} - \overset{*}{q}_{h,k}^{s})^{2}}{2\gamma_h} + \sum_{h \in \Aset} \sum_{k \in \Kset_{h}} \frac{(\tau_{h,k} - \overset{*}{\tau}_{h,k})^{2}}{2\varepsilon_{\Cset_{q}}} + \sum_{h \in \Aset} \sum_{k \in \Kset_{h}} \sum_{s \in \Sset_{k}} \overset{*}{x_{s}} \frac{(\alpha_{h,k}^{s} - \overset{*}{\alpha}_{h,k}^{s})^2}{2\kappa_i}$.

The derivative of the Lyapunov function with respect to the Lagrange multipliers is expressed as; $\overset{.}{V}(q,\tau,\alpha)  =  \sum_{h \in \Aset} \sum_{k \in \Kset_{h}} \sum_{s \in \Sset_{k}} (q_{h,k}^{s} - \overset{*}{q}_{h,k}^{s}) \bigr[ \frac{H_{h,k}^{s} \alpha_{h,k}^{s} x_{s} }{1-\rho_{h}^{s}} + \sum _{s' \in \Sset_{k} - \{s\}} H_{h,k}^{s'} \alpha_{h,k}^{s'} x_{s'}\rho_{h,k}^{s,s'}  - R_{h} \tau_{h,k} \bigl]_{q_{h,k}^{s}}^{+} + \sum_{h \in \Aset} \sum_{k \in \Kset_{h}} (\tau_{h,k} - \overset{*}{\tau}_{h,k}) [Q_{h,k} - E_{\Cset_q}[Q]]_{\tau_{h,k}}^{+} + \sum_{h \in \Aset} \sum_{k \in \Kset_{h}} \sum_{s \in \Sset_{k}} \overset{*}{x_{s}} (\alpha_{h,k}^{s} - \overset{*}{\alpha}_{h,k}^{s}) [E_{i}[Q] - H_{h,k}^{s}Q_{h,k}^{s}]_{\alpha_{h,k}^{s}}^{+}$.

By using the definition of the function $[b]_{z}^{+}$, and considering that Since $E_{\Cset_{q}}[Q]$ is the minimal $\overset{-}{Q}_{\Cset_{q}}(t)$ and $E_{i}[Q]$ is the maximal $\overset{-}{Q}_{i}(t)$, the following holds;
$\overset{.}{V}(q,\tau,\alpha) \leq  \sum_{h \in \Aset} \sum_{k \in \Kset_{h}} \sum_{s \in \Sset_{k}} (q_{h,k}^{s} - \overset{*}{q}_{h,k}^{s}) \bigr( \frac{H_{h,k}^{s} \alpha_{h,k}^{s} x_{s} }{1-\rho_{h}^{s}} + \sum _{s' \in \Sset_{k} - \{s\}} H_{h,k}^{s'} \alpha_{h,k}^{s'} x_{s'}\rho_{h,k}^{s,s'}  - R_{h} \tau_{h,k} \bigl) + \sum_{h \in \Aset} \sum_{k \in \Kset_{h}} (\tau_{h,k} - \overset{*}{\tau}_{h,k}) (Q_{h,k} - E_{\Cset_q}[Q]) + \sum_{h \in \Aset} \sum_{k \in \Kset_{h}} \sum_{s \in \Sset_{k}} \overset{*}{x_{s}} (\alpha_{h,k}^{s} - \overset{*}{\alpha}_{h,k}^{s}) (E_{i}[Q] - H_{h,k}^{s}Q_{h,k}^{s})$.

Since $Q_{h,k} \geq Q_{h,k} - E_{\Cset_{q}}[Q]$ and $H_{h,k}^{s} Q_{h,k}^{s} \geq H_{h,k}^{s} Q_{h,k}^{s} - E_{i}[Q]$,
%$E_{\Cset_{q}}[Q]$ is the minimal $\overset{-}{Q}_{\Cset_{q}}(t)$ and $E_{i}[Q]$ is the maximal $\overset{-}{Q}_{i}(t)$,
the following holds; $\overset{.}{V}(q,\tau,\alpha)  \leq  \sum_{h \in \Aset} \sum_{k \in \Kset_{h}} \sum_{s \in \Sset_{k}} (q_{h,k}^{s} - \overset{*}{q}_{h,k}^{s}) ( \frac{H_{h,k}^{s} \alpha_{h,k}^{s} x_{s} }{1-\rho_{h}^{s}} + \sum _{s' \in \Sset_{k} - \{s\}} H_{h,k}^{s'} \alpha_{h,k}^{s'} x_{s'}\rho_{h,k}^{s,s'}  - R_{h} \tau_{h,k} )  + \sum_{h \in \Aset} \sum_{k \in \Kset_{h}} (\tau_{h,k} - \overset{*}{\tau}_{h,k}) Q_{h,k} + \sum_{h \in \Aset} \sum_{k \in \Kset_{h}} \sum_{s \in \Sset_{k}} \overset{*}{x_{s}} (\overset{*}{\alpha}_{h,k}^{s} - \alpha_{h,k}^{s}) H_{h,k}^{s}Q_{h,k}^{s}$.

Substituting $Q_{h,k} = R_{h} \sum_{s \in \Sset_{k}}q_{h,k}^{s}$ and $Q_{h,k}^{s} = \frac{q_{h,k}^{s}}{1-\rho_{h}^{s}} + \sum _{s' \in \Sset_{k}-\{s\}} q_{h,k}^{s'}\rho_{h,k}^{s',s}$ to the above inequality; $\overset{.}{V}(q,\tau,\alpha) \leq \sum_{h \in \Aset} \sum_{k \in \Kset_{h}} \sum_{s \in \Sset_{k}} (q_{h,k}^{s} - \overset{*}{q}_{h,k}^{s}) ( \frac{H_{h,k}^{s} \alpha_{h,k}^{s} x_{s} }{1-\rho_{h}^{s}} + \sum _{s' \in \Sset_{k} - \{s\}} H_{h,k}^{s'} \alpha_{h,k}^{s'} x_{s'}\rho_{h,k}^{s,s'}  - R_{h} \tau_{h,k} ) + \sum_{h \in \Aset} \sum_{k \in \Kset_{h}} (\tau_{h,k} - \overset{*}{\tau}_{h,k})  R_{h} \sum_{s \in \Sset_{k}} q_{h,k}^{s} + \sum_{h \in \Aset} \sum_{k \in \Kset_{h}} \sum_{s \in \Sset_{k}} \overset{*}{x_{s}} (\overset{*}{\alpha}_{h,k}^{s} - \alpha_{h,k}^{s}) H_{h,k}^{s}(\frac{q_{h,k}^{s}}{1-\rho_{h}^{s}} + \sum _{s' \in \Sset_{k}-\{s\}} q_{h,k}^{s'}\rho_{h,k}^{s',s})$.

When we arrange the terms in the above inequality by adding and removing terms, we have;
%\iffalse
\iftrue
\begin{align*}
& \overset{.}{V}(q,\tau,\alpha)   \leq \sum_{h \in \Aset} \sum_{k \in \Kset_{h}} \sum_{s \in \Sset_{k}} (q_{h,k}^{s} - \overset{*}{q}_{h,k}^{s})
( \frac{H_{h,k}^{s} \alpha_{h,k}^{s} x_{s} }{1-\rho_{h}^{s}} \nonumber \\
& + \sum _{s' \in \Sset_{k} - \{s\}} H_{h,k}^{s'} \alpha_{h,k}^{s'} x_{s'}\rho_{h,k}^{s,s'}  -  \frac{H_{h,k}^{s} \alpha_{h,k}^{s} \overset{*}{x}_{s} }{1-\rho_{h}^{s}} \nonumber \\
& - \sum _{s' \in \Sset_{k} - \{s\}} H_{h,k}^{s'} \alpha_{h,k}^{s'} \overset{*}{x}_{s'}\rho_{h,k}^{s,s'} ) + \sum_{h \in \Aset} \sum_{k \in \Kset_{h}} \sum_{s \in \Sset_{k}} (q_{h,k}^{s} - \overset{*}{q}_{h,k}^{s}) \nonumber \\
& ( \frac{H_{h,k}^{s} \alpha_{h,k}^{s} \overset{*}{x}_{s} }{1-\rho_{h}^{s}} + \sum _{s' \in \Sset_{k} - \{s\}} H_{h,k}^{s'} \alpha_{h,k}^{s'} \overset{*}{x}_{s'}\rho_{h,k}^{s,s'}  - R_{h} \tau_{h,k} ) \nonumber \\
&  + \sum_{h \in \Aset} \sum_{k \in \Kset_{h}} \sum_{s \in \Sset_{k}} R_{h} ( \tau_{h,k} - \overset{*}{\tau_{h,k}}) q_{h,k}^{s} \nonumber \\
& + \sum_{h \in \Aset} \sum_{k \in \Kset_{h}} \sum_{s \in \Sset_{k}} \overset{*}{x}_{s} ( \overset{*}{\alpha}_{h,k}^{s} - \alpha_{h,k}^{s} ) (\frac{H_{h,k}^{s}q_{h,k}^{s}}{1-\rho_{h}^{s}} \nonumber \\
& + \sum _{s' \in \Sset_{k}-\{s\}} H_{h,k}^{s}q_{h,k}^{s'}\rho_{h,k}^{s',s})
\end{align*}
\begin{align*}
\textstyle & = \sum_{h \in \Aset} \sum_{k \in \Kset_{h}} \sum_{s \in \Sset_{k}} (q_{h,k}^{s} - \overset{*}{q}_{h,k}^{s}) ( \frac{H_{h,k}^{s} \alpha_{h,k}^{s} (x_{s} - \overset{*}{x_{s}}) }{1-\rho_{h}^{s}} \nonumber \\
&  + \sum _{s' \in \Sset_{k} - \{s\}} H_{h,k}^{s'} \alpha_{h,k}^{s'} \rho_{h,k}^{s,s'} (x_{s'} - \overset{*}{x}_{s'}) ) \nonumber \\
& + \sum_{h \in \Aset} \sum_{k \in \Kset_{h}} \sum_{s \in \Sset_{k}} (q_{h,k}^{s} - \overset{*}{q}_{h,k}^{s})
( \frac{H_{h,k}^{s} \alpha_{h,k}^{s} \overset{*}{x}_{s} }{1-\rho_{h}^{s}}  \nonumber \\
& + \sum _{s' \in \Sset_{k} - \{s\}} H_{h,k}^{s'} \alpha_{h,k}^{s'} \overset{*}{x}_{s'}\rho_{h,k}^{s,s'}  -  \frac{H_{h,k}^{s} \overset{*}{\alpha}_{h,k}^{s} \overset{*}{x}_{s} }{1-\rho_{h}^{s}} \nonumber \\
& - \sum _{s' \in \Sset_{k} - \{s\}} H_{h,k}^{s'} \overset{*}{\alpha}_{h,k}^{s'} \overset{*}{x}_{s'}\rho_{h,k}^{s,s'} ) \nonumber \\
&  + \sum_{h \in \Aset} \sum_{k \in \Kset_{h}} \sum_{s \in \Sset_{k}} (q_{h,k}^{s} - \overset{*}{q}_{h,k}^{s}) ( \frac{H_{h,k}^{s} \overset{*}{\alpha}_{h,k}^{s} \overset{*}{x}_{s} }{1-\rho_{h}^{s}} \nonumber \\
& + \sum _{s' \in \Sset_{k} - \{s\}} H_{h,k}^{s'} \overset{*}{\alpha}_{h,k}^{s'} \overset{*}{x}_{s'}\rho_{h,k}^{s,s'}  - R_{h} \tau_{h,k} ) \nonumber \\
&  + \sum_{h \in \Aset} \sum_{k \in \Kset_{h}} \sum_{s \in \Sset_{k}} R_{h} ( \tau_{h,k} - \overset{*}{\tau_{h,k}} )q_{h,k}^{s} \nonumber \\
& +  \sum_{h \in \Aset} \sum_{k \in \Kset_{h}} \sum_{s \in \Sset_{k}} \overset{*}{x}_{s} ( \overset{*}{\alpha}_{h,k}^{s} - \alpha_{h,k}^{s}  ) (\frac{H_{h,k}^{s}q_{h,k}^{s}}{1-\rho_{h}^{s}} \nonumber \\
& + \sum _{s' \in \Sset_{k}-\{s\}} H_{h,k}^{s}q_{h,k}^{s'}\rho_{h,k}^{s',s})
\end{align*}
\begin{align*}
\textstyle & = \sum_{h \in \Aset} \sum_{k \in \Kset_{h}} \sum_{s \in \Sset_{k}} (q_{h,k}^{s} - \overset{*}{q}_{h,k}^{s}) ( \frac{H_{h,k}^{s} \alpha_{h,k}^{s} (x_{s} - \overset{*}{x_{s}}) }{1-\rho_{h}^{s}} \nonumber \\
& + \sum _{s' \in \Sset_{k} - \{s\}} H_{h,k}^{s'} \alpha_{h,k}^{s'} \rho_{h,k}^{s,s'} (x_{s'} - \overset{*}{x}_{s'}) ) \nonumber \\
\textstyle & + \sum_{h \in \Aset} \sum_{k \in \Kset_{h}} \sum_{s \in \Sset_{k}} (q_{h,k}^{s} - \overset{*}{q}_{h,k}^{s})
( \frac{H_{h,k}^{s} \alpha_{h,k}^{s} \overset{*}{x}_{s} }{1-\rho_{h}^{s}} \nonumber \\
& + \sum _{s' \in \Sset_{k} - \{s\}} H_{h,k}^{s'} \alpha_{h,k}^{s'} \overset{*}{x}_{s'}\rho_{h,k}^{s,s'}  -  \frac{H_{h,k}^{s} \overset{*}{\alpha}_{h,k}^{s} \overset{*}{x}_{s} }{1-\rho_{h}^{s}} \nonumber \\
& - \sum _{s' \in \Sset_{k} - \{s\}} H_{h,k}^{s'} \overset{*}{\alpha}_{h,k}^{s'} \overset{*}{x}_{s'}\rho_{h,k}^{s,s'} ) \nonumber \\
\textstyle &  + \sum_{h \in \Aset} \sum_{k \in \Kset_{h}} \sum_{s \in \Sset_{k}} (q_{h,k}^{s} - \overset{*}{q}_{h,k}^{s}) ( \frac{H_{h,k}^{s} \overset{*}{\alpha}_{h,k}^{s} \overset{*}{x}_{s} }{1-\rho_{h}^{s}} \nonumber \\
& + \sum _{s' \in \Sset_{k} - \{s\}} H_{h,k}^{s'} \overset{*}{\alpha}_{h,k}^{s'} \overset{*}{x}_{s'}\rho_{h,k}^{s,s'}  - R_{h} \overset{*}{\tau}_{h,k} ) \nonumber \\
\textstyle & + \sum_{h \in \Aset} \sum_{k \in \Kset_{h}} \sum_{s \in \Sset_{k}} (q_{h,k}^{s} - \overset{*}{q}_{h,k}^{s}) R_{h}(\overset{*}{\tau}_{h,k} - \tau_{h,k} ) \nonumber \\
& + \sum_{h \in \Aset} \sum_{k \in \Kset_{h}} \sum_{s \in \Sset_{k}} R_{h} ( \tau_{h,k} - \overset{*}{\tau_{h,k}} )q_{h,k}^{s} \nonumber
\\
\textstyle &  + \sum_{h \in \Aset} \sum_{k \in \Kset_{h}} \sum_{s \in \Sset_{k}} \overset{*}{x}_{s} ( \overset{*}{\alpha}_{h,k}^{s} - \alpha_{h,k}^{s}) (\frac{H_{h,k}^{s}q_{h,k}^{s}}{1-\rho_{h}^{s}}\nonumber \\
& + \sum _{s' \in \Sset_{k}-\{s\}} H_{h,k}^{s}q_{h,k}^{s'}\rho_{h,k}^{s',s}) \end{align*}
\begin{align*}
\textstyle & = \sum_{h \in \Aset} \sum_{k \in \Kset_{h}} \sum_{s \in \Sset_{k}} (q_{h,k}^{s} - \overset{*}{q}_{h,k}^{s}) ( \frac{H_{h,k}^{s} \alpha_{h,k}^{s} (x_{s} - \overset{*}{x_{s}}) }{1-\rho_{h}^{s}} \nonumber \\
& + \sum _{s' \in \Sset_{k} - \{s\}} H_{h,k}^{s'} \alpha_{h,k}^{s'} \rho_{h,k}^{s,s'} (x_{s'} - \overset{*}{x}_{s'}) ) \nonumber \\
& + \sum_{h \in \Aset} \sum_{k \in \Kset_{h}} \sum_{s \in \Sset_{k}} (q_{h,k}^{s} - \overset{*}{q}_{h,k}^{s}) ( \frac{H_{h,k}^{s} \overset{*}{\alpha}_{h,k}^{s} \overset{*}{x}_{s} }{1-\rho_{h}^{s}} \nonumber \\
& + \sum _{s' \in \Sset_{k} - \{s\}} H_{h,k}^{s'} \overset{*}{\alpha}_{h,k}^{s'} \overset{*}{x}_{s'}\rho_{h,k}^{s,s'}  - R_{h} \overset{*}{\tau}_{h,k} ) \nonumber \\
& + \sum_{h \in \Aset} \sum_{k \in \Kset_{h}} \sum_{s \in \Sset_{k}} \overset{*}{q}_{h,k}^{s} R_{h} ( \tau_{h,k} - \overset{*}{\tau}_{h,k} ) \nonumber \\
& + \sum_{h \in \Aset} \sum_{k \in \Kset_{h}} \sum_{s \in \Sset_{k}} (q_{h,k}^{s} - \overset{*}{q}_{h,k}^{s})
( \frac{H_{h,k}^{s} \alpha_{h,k}^{s} \overset{*}{x}_{s} }{1-\rho_{h}^{s}} \nonumber \\
& + \sum _{s' \in \Sset_{k} - \{s\}} H_{h,k}^{s'} \alpha_{h,k}^{s'} \overset{*}{x}_{s'}\rho_{h,k}^{s,s'}  -  \frac{H_{h,k}^{s} \overset{*}{\alpha}_{h,k}^{s} \overset{*}{x}_{s} }{1-\rho_{h}^{s}}\nonumber \\
& - \sum _{s' \in \Sset_{k} - \{s\}} H_{h,k}^{s'} \overset{*}{\alpha}_{h,k}^{s'} \overset{*}{x}_{s'}\rho_{h,k}^{s,s'} ) \nonumber \\
&  + \sum_{h \in \Aset} \sum_{k \in \Kset_{h}} \sum_{s \in \Sset_{k}} \overset{*}{x}_{s} ( \overset{*}{\alpha}_{h,k}^{s} - \alpha_{h,k}^{s}  ) (\frac{H_{h,k}^{s}q_{h,k}^{s}}{1-\rho_{h}^{s}} \nonumber \\
& + \sum _{s' \in \Sset_{k}-\{s\}} H_{h,k}^{s}q_{h,k}^{s'}\rho_{h,k}^{s',s})
\end{align*}
\begin{align}
\label{appendix:ly_eq8_1}
& \overset{.}{V}(q,\tau,\alpha) \leq \sum_{h \in \Aset} \sum_{k \in \Kset_{h}} \sum_{s \in \Sset_{k}} (q_{h,k}^{s} - \overset{*}{q}_{h,k}^{s}) ( \frac{H_{h,k}^{s} \alpha_{h,k}^{s} (x_{s} - \overset{*}{x_{s}}) }{1-\rho_{h}^{s}} \nonumber \\
& + \sum _{s' \in \Sset_{k} - \{s\}} H_{h,k}^{s'} \alpha_{h,k}^{s'} \rho_{h,k}^{s,s'} (x_{s'} - \overset{*}{x}_{s'}) )  \\
\label{appendix:ly_eq8_2}
& + \sum_{h \in \Aset} \sum_{k \in \Kset_{h}} \sum_{s \in \Sset_{k}} (q_{h,k}^{s} - \overset{*}{q}_{h,k}^{s}) ( \frac{H_{h,k}^{s} \overset{*}{\alpha}_{h,k}^{s} \overset{*}{x}_{s} }{1-\rho_{h}^{s}} \nonumber \\
& + \sum _{s' \in \Sset_{k} - \{s\}} H_{h,k}^{s'} \overset{*}{\alpha}_{h,k}^{s'} \overset{*}{x}_{s'}\rho_{h,k}^{s,s'}  - R_{h} \overset{*}{\tau}_{h,k} )  \\
\label{appendix:ly_eq8_3}
& + \sum_{h \in \Aset} \sum_{k \in \Kset_{h}} \sum_{s \in \Sset_{k}} \overset{*}{q}_{h,k}^{s} R_{h} ( \tau_{h,k} - \overset{*}{\tau}_{h,k} )  \\
\label{appendix:ly_eq8_4}
%& + \overset{*}{x}_{s} \sum_{h \in \Aset} \sum_{k \in \Kset_{h}} \sum_{s \in \Sset_{k}} (\overset{*}{\alpha}_{h,k}^{s} H_{h,k}^{s} ( %\frac{\overset{*}{q}_{h,k}^{s}}{1-\rho_{h}^{s}} + \sum_{s' \in \Sset_{k} -\{s\}} \overset{*}{q}_{h,k}^{s'}\rho_{h,k}^{s',s} ) \nonumber \\
%& - \alpha_{h,k}^{s} H_{h,k}^{s} ( \frac{\overset{*}{q}_{h,k}^{s}}{1-\rho_{h}^{s}} + \sum_{s' \in \Sset_{k} -\{s\}} \overset{*}{q}_{h,k}^{s'}\rho_{h,k}^{s',s} ))
& + \sum_{h \in \Aset} \sum_{k \in \Kset_{h}} \sum_{s \in \Sset_{k}} \overset{*}{x}_{s} H_{h,k}^{s} (\frac{\overset{*}{q}_{h,k}^{s}}{1-\rho_{h}^{s}} \\
& + \sum_{s' \in \Sset_{k} -\{s\}} \overset{*}{q}_{h,k}^{s'}\rho_{h,k}^{s',s} ) (\overset{*}{\alpha}_{h,k}^{s} - \alpha_{h,k}^{s})
\end{align}
\fi
\begin{align}
\label{appendix:ly_eq8_1}
& \overset{.}{V}(q,\tau,\alpha) \leq \sum_{h \in \Aset} \sum_{k \in \Kset_{h}} \sum_{s \in \Sset_{k}} (q_{h,k}^{s} - \overset{*}{q}_{h,k}^{s}) ( \frac{H_{h,k}^{s} \alpha_{h,k}^{s} (x_{s} - \overset{*}{x_{s}}) }{1-\rho_{h}^{s}} \nonumber \\
& + \sum _{s' \in \Sset_{k} - \{s\}} H_{h,k}^{s'} \alpha_{h,k}^{s'} \rho_{h,k}^{s,s'} (x_{s'} - \overset{*}{x}_{s'}) )  \\
\label{appendix:ly_eq8_2}
& + \sum_{h \in \Aset} \sum_{k \in \Kset_{h}} \sum_{s \in \Sset_{k}} (q_{h,k}^{s} - \overset{*}{q}_{h,k}^{s}) ( \frac{H_{h,k}^{s} \overset{*}{\alpha}_{h,k}^{s} \overset{*}{x}_{s} }{1-\rho_{h}^{s}} \nonumber \\
& + \sum _{s' \in \Sset_{k} - \{s\}} H_{h,k}^{s'} \overset{*}{\alpha}_{h,k}^{s'} \overset{*}{x}_{s'}\rho_{h,k}^{s,s'}  - R_{h} \overset{*}{\tau}_{h,k} )  \\
\label{appendix:ly_eq8_3}
& + \sum_{h \in \Aset} \sum_{k \in \Kset_{h}} \sum_{s \in \Sset_{k}} \overset{*}{q}_{h,k}^{s} R_{h} ( \tau_{h,k} - \overset{*}{\tau}_{h,k} )  \\
\label{appendix:ly_eq8_4}
%& + \overset{*}{x}_{s} \sum_{h \in \Aset} \sum_{k \in \Kset_{h}} \sum_{s \in \Sset_{k}} (\overset{*}{\alpha}_{h,k}^{s} H_{h,k}^{s} ( %\frac{\overset{*}{q}_{h,k}^{s}}{1-\rho_{h}^{s}} + \sum_{s' \in \Sset_{k} -\{s\}} \overset{*}{q}_{h,k}^{s'}\rho_{h,k}^{s',s} ) \nonumber \\
%& - \alpha_{h,k}^{s} H_{h,k}^{s} ( \frac{\overset{*}{q}_{h,k}^{s}}{1-\rho_{h}^{s}} + \sum_{s' \in \Sset_{k} -\{s\}} \overset{*}{q}_{h,k}^{s'}\rho_{h,k}^{s',s} ))
& + \sum_{h \in \Aset} \sum_{k \in \Kset_{h}} \sum_{s \in \Sset_{k}} \overset{*}{x}_{s} H_{h,k}^{s} \overset{*}{Q}_{h,k}^{s} (\overset{*}{\alpha}_{h,k}^{s} - \alpha_{h,k}^{s})   \end{align}
Since the marginal utility $U_{s}^{'}(.)$ is a decreasing function, its inverse, \ie the Eq.~(\ref{appendix:ly_eq8_1}) is less than 0. Due to the optimality condition in Eq.~(\ref{appendix:opt_cond_eq2}) and Eq.~(\ref{appendix:opt_cond_eq4}), Eq.~(\ref{appendix:ly_eq8_2}) is less than 0. Due to the optimality condition in Eq.~(\ref{appendix:opt_cond_eq3}), Eq.~(\ref{appendix:ly_eq8_3}) is less than 0. Due to the optimality condition in Eq.~(\ref{appendix:opt_cond_eq5}), Eq.~(\ref{appendix:ly_eq8_4}) is less than 0. Thus, $\overset{.}{V}(q,\tau,\alpha) \leq 0$. This implies the convergence of our solutions, \cite{cc_multicast_nc}, \cite{nonlinearsystems}.

\iftrue
\subsection{Numerical Results}
We consider again the X and cross topologies shown in Figs.~\ref{fig:all_topologies}(a) and \ref{fig:all_topologies}(b). In the X topology, $A_1$ transmits packets to $A_2$ via $I$ with rate $x_1$, and $B_1$ transmits packets to $B_2$ via $I$ with rate $x_2$. In the cross topology, $A_1$ transmits packets to $A_2$ with rate $x_1$, $A_2$ transmits packets to $A_1$ with rate $x_2$, $B_1$ transmits packets to $B_2$ with rate $x_3$, and $B_2$ transmits packets to $B_1$ with rate $x_4$. All transmissions are via $I$. In both topologies, the data rate of each link is set to $1$ packet/transmission and the loss rate is set to 30\%.

In Figs.~\ref{fig:x_nc_state_convergence} and \ref{fig:x_nc_stateless_convergence}, we present the throughput vs. the iteration number for the X topology at different loss patterns for \state and {I$^2$NC-stateless}, respectively. Each figure shows the convergence of $x_1$, $x_2$, and $x_1+x_2$ to their optimum values. {\em E.g.}, $x_1+x_2$ converges to its optimum value $0.59$ in Fig.~\ref{fig:x_nc_state_convergence}(c) and $x_1+x_2$ converges to its optimum value $0.55$ in Fig.~\ref{fig:x_nc_stateless_convergence}(c).

Fig.~\ref{fig:cross_nc_state_convergence} and \ref{fig:cross_nc_stateless_convergence} present the throughput vs. the iteration number for the cross topology at different loss patterns for \state and {I$^2$NC-stateless}, respectively. We see similar convergence results. Specifically, each flow rate, $x_1$, $x_2$, $x_3$, $x_4$, and the total rate converge to their optimum values.

%% Convergence for state - X topology
\begin{figure}[t!]
\begin{center}
\subfigure[\scriptsize Loss only on overhearing link $A_1-B_2$]{{\includegraphics[width=4.3cm]{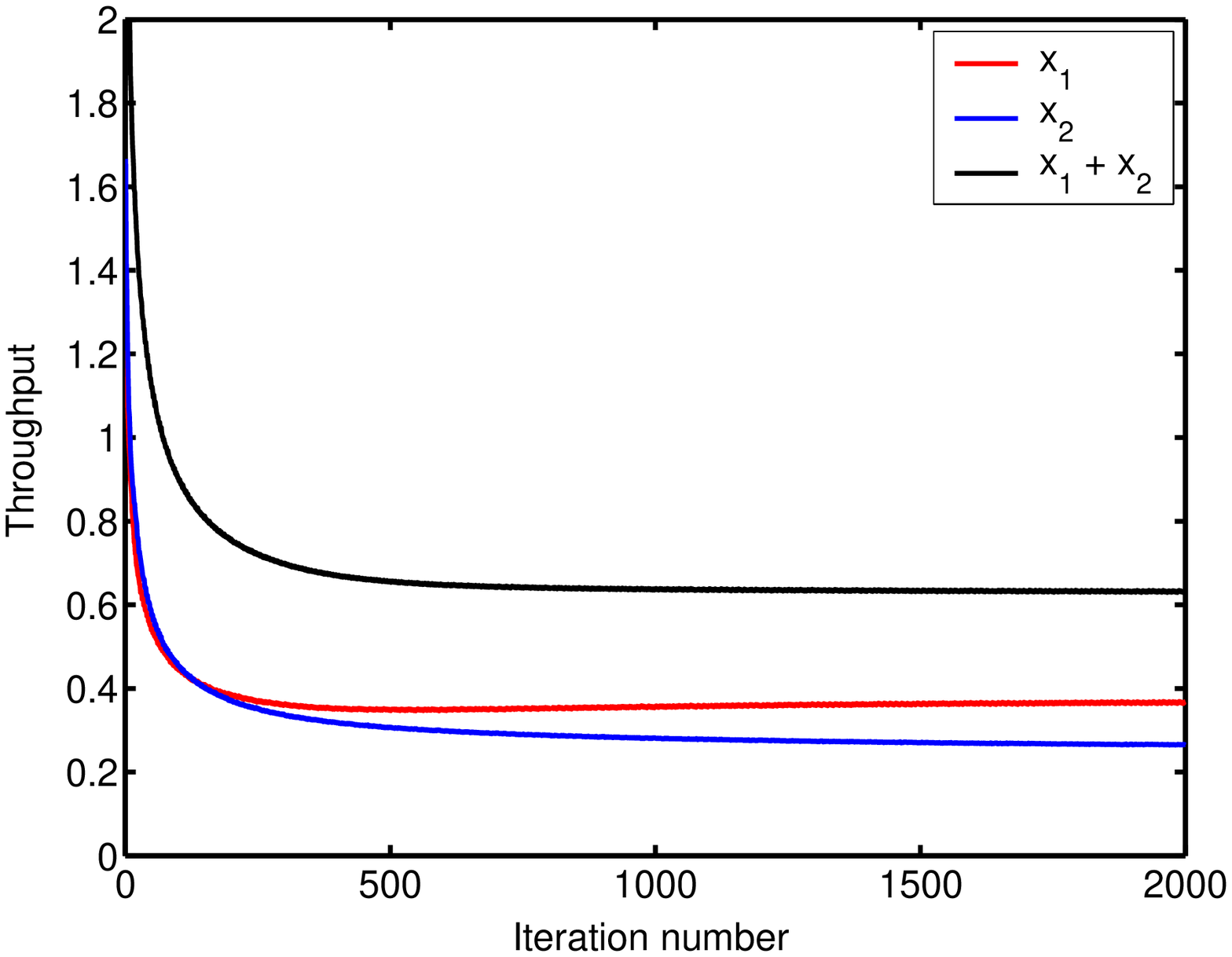}}}
\subfigure[\scriptsize Loss only on direct link $I-B_2$]{{\includegraphics[width=4.3cm]{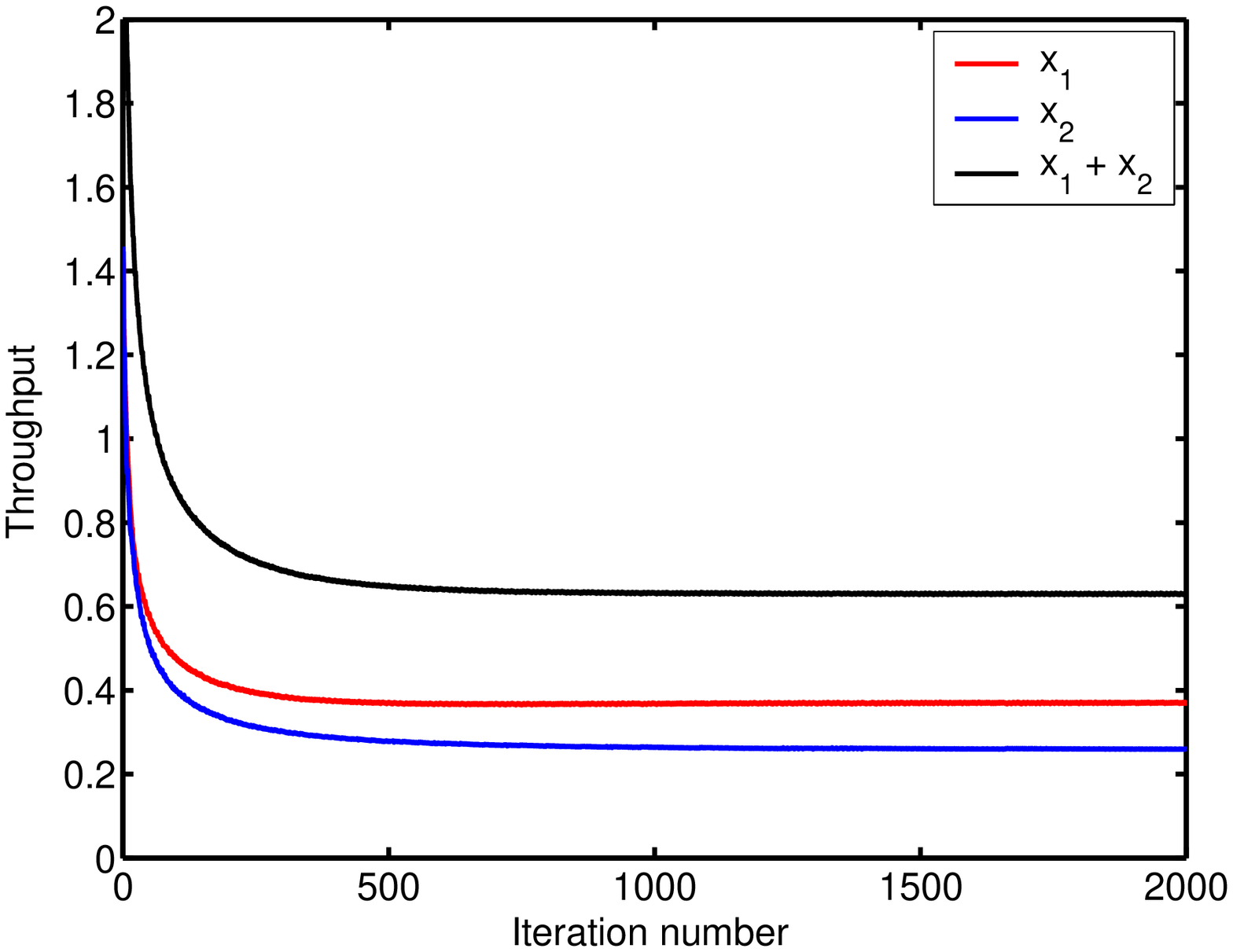}}} \\
\subfigure[\scriptsize Loss on links $A_1-B_2$ and $I-B_2$]{{\includegraphics[width=4.3cm]{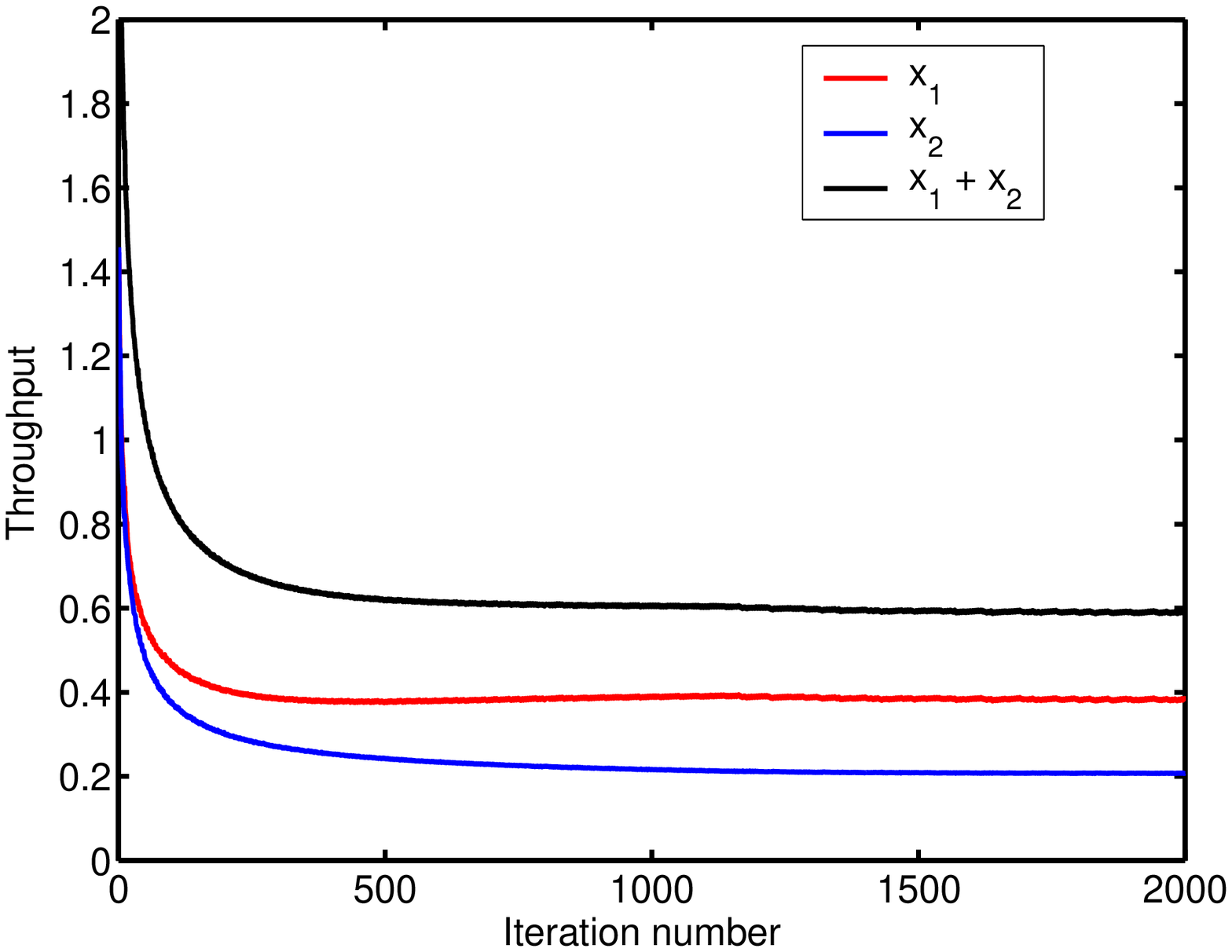}}}
\subfigure[\scriptsize Loss on all links]{{\includegraphics[width=4.3cm]{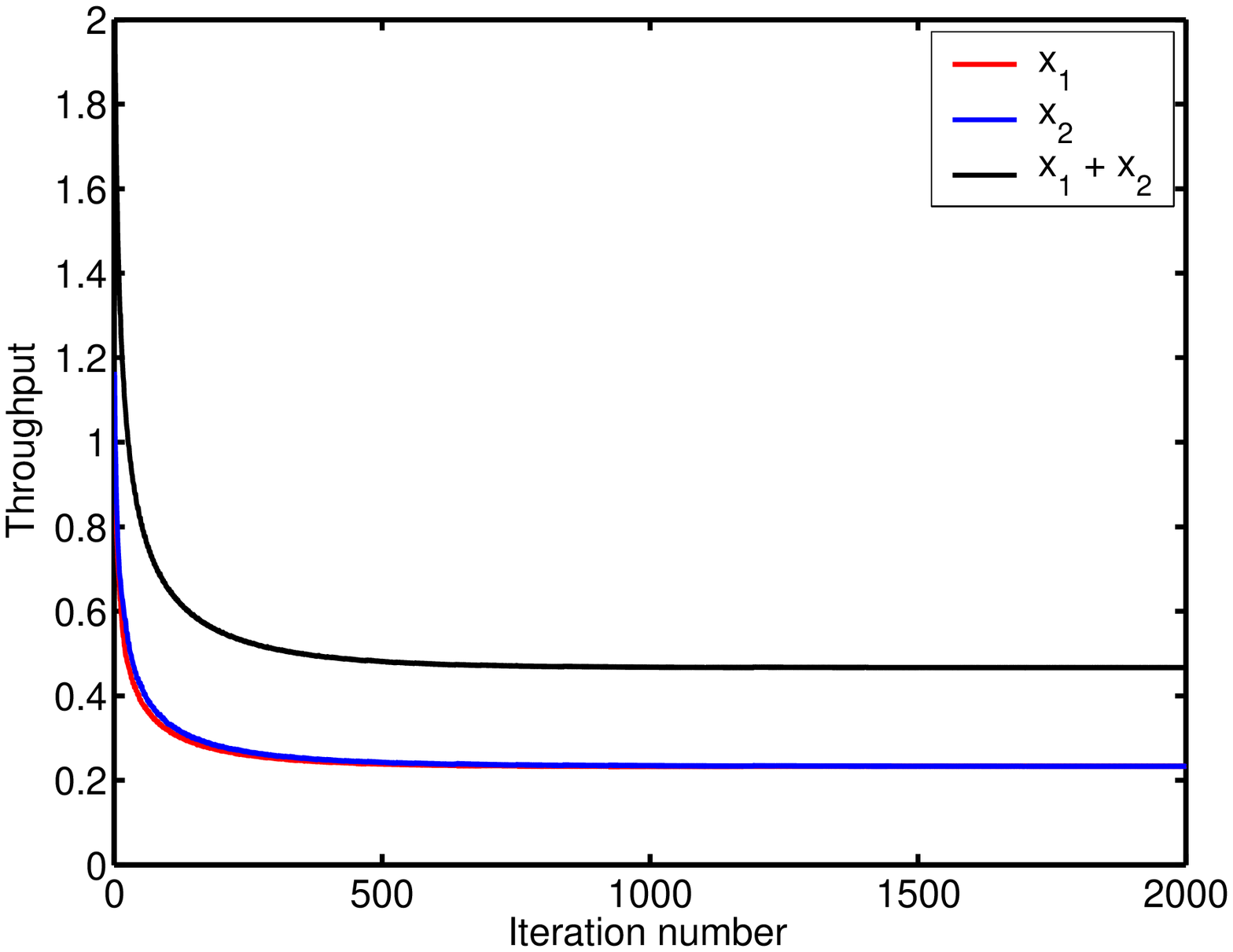}}}
%\vspace{-15pt}
\end{center}
\begin{center}
%\vspace{-10pt}
\caption{\label{fig:x_nc_state_convergence} \scriptsize  X topology. Convergence of $x_1$, $x_2$, and $x_1+x_2$ for {I$^2$NC-state}. Loss rate is 30\%.}
\vspace{-10pt}
\end{center}
\end{figure}

%% Convergence for stateless - X topology
\begin{figure}[t!]
\begin{center}
\subfigure[\scriptsize Loss only on overhearing link $A_1-B_2$]{{\includegraphics[width=4.3cm]{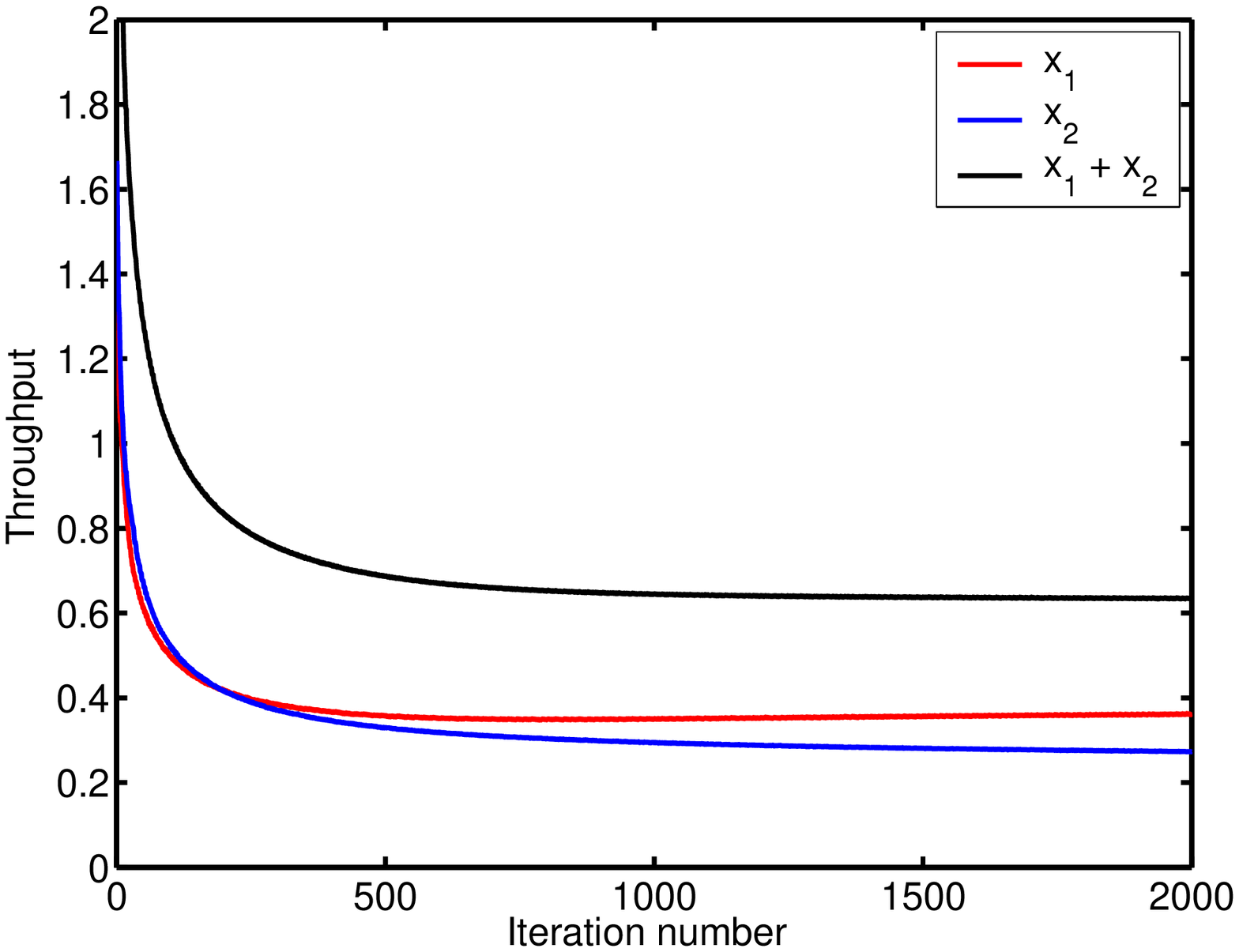}}}
\subfigure[\scriptsize Loss only on direct link $I-B_2$]{{\includegraphics[width=4.3cm]{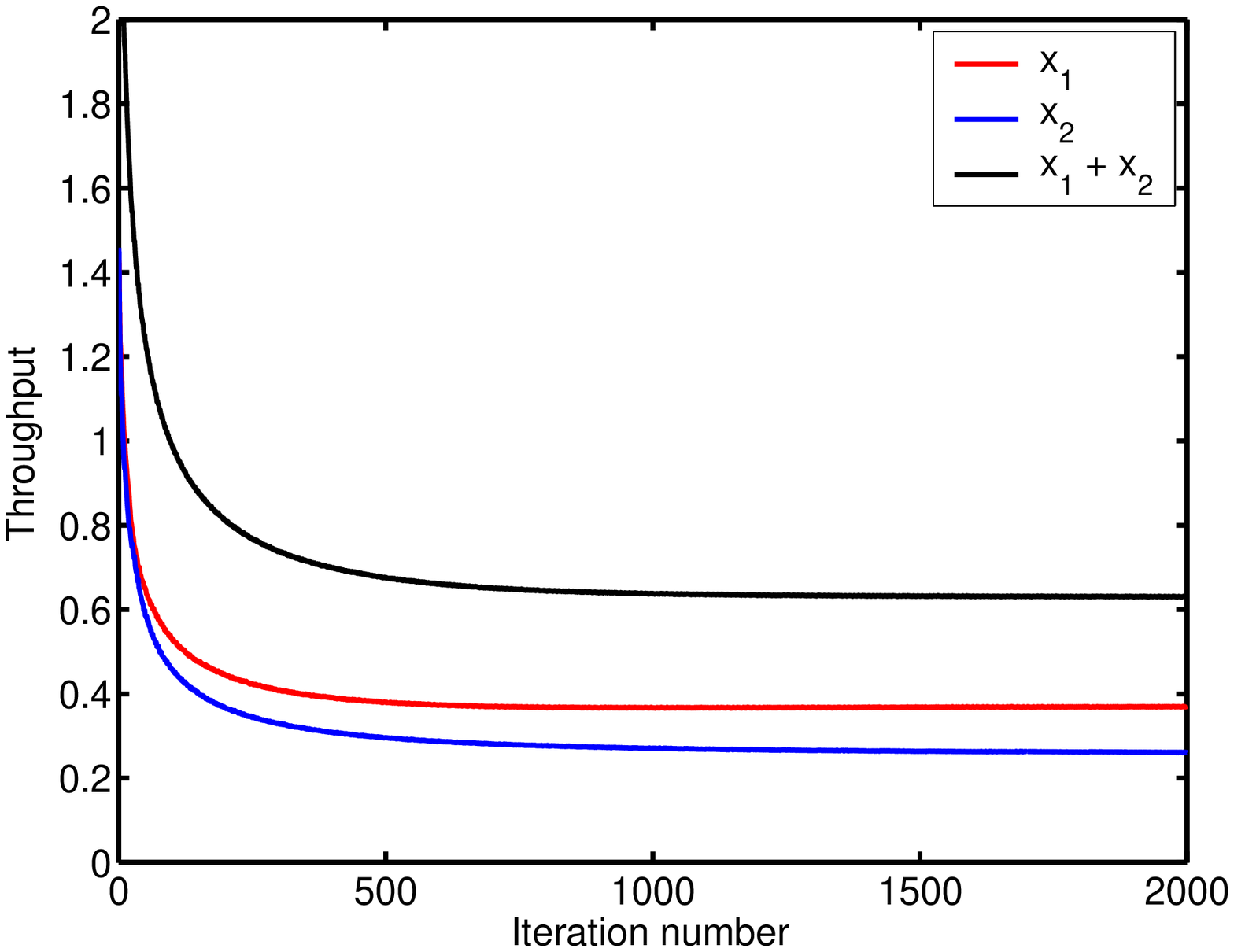}}} \\
\subfigure[\scriptsize Loss on links $A_1-B_2$ and $I-B_2$]{{\includegraphics[width=4.3cm]{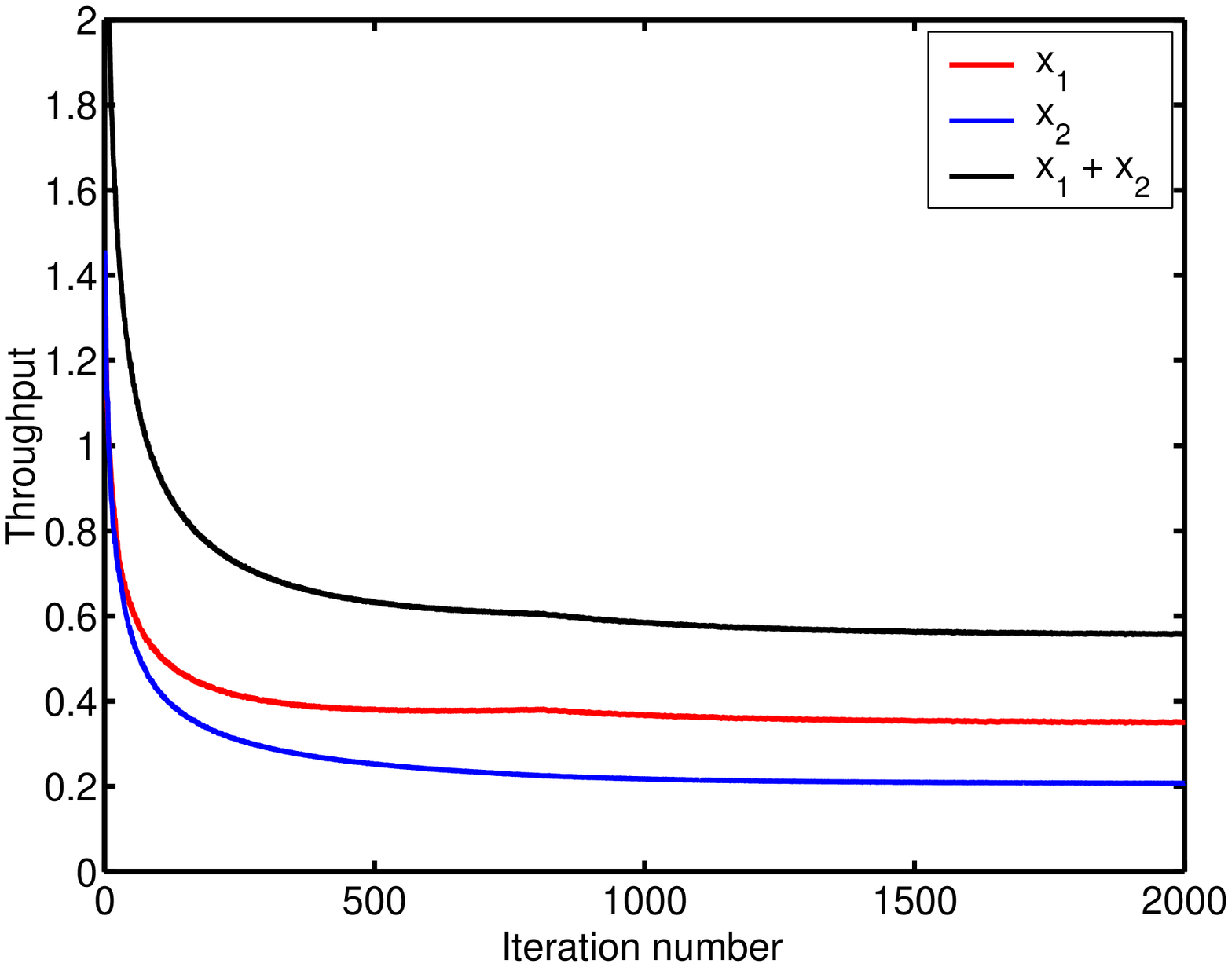}}}
\subfigure[\scriptsize Loss on all links]{{\includegraphics[width=4.3cm]{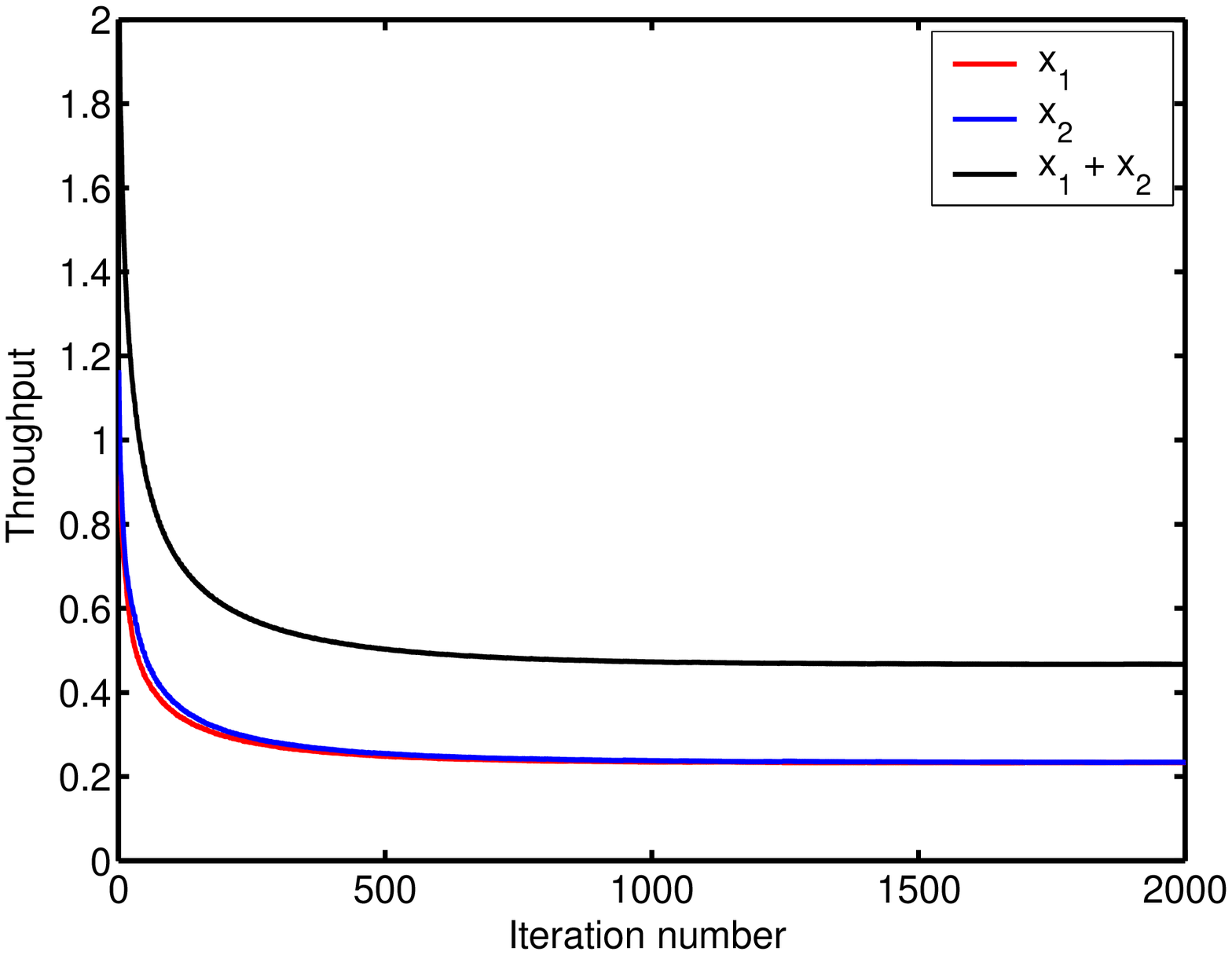}}}
%\vspace{-15pt}
\end{center}
\begin{center}
%\vspace{-10pt}
\caption{\label{fig:x_nc_stateless_convergence} \scriptsize  X topology. Convergence of $x_1$, $x_2$, and $x_1+x_2$ for {I$^2$NC-stateless}. Loss rate is 30\%.}
\vspace{-10pt}
\end{center}
\end{figure}

%% Convergence for state - Cross topology
\begin{figure}[t!]
\begin{center}
\subfigure[\scriptsize Loss only on overhearing link $A_1-B_2$]{{\includegraphics[width=4.3cm]{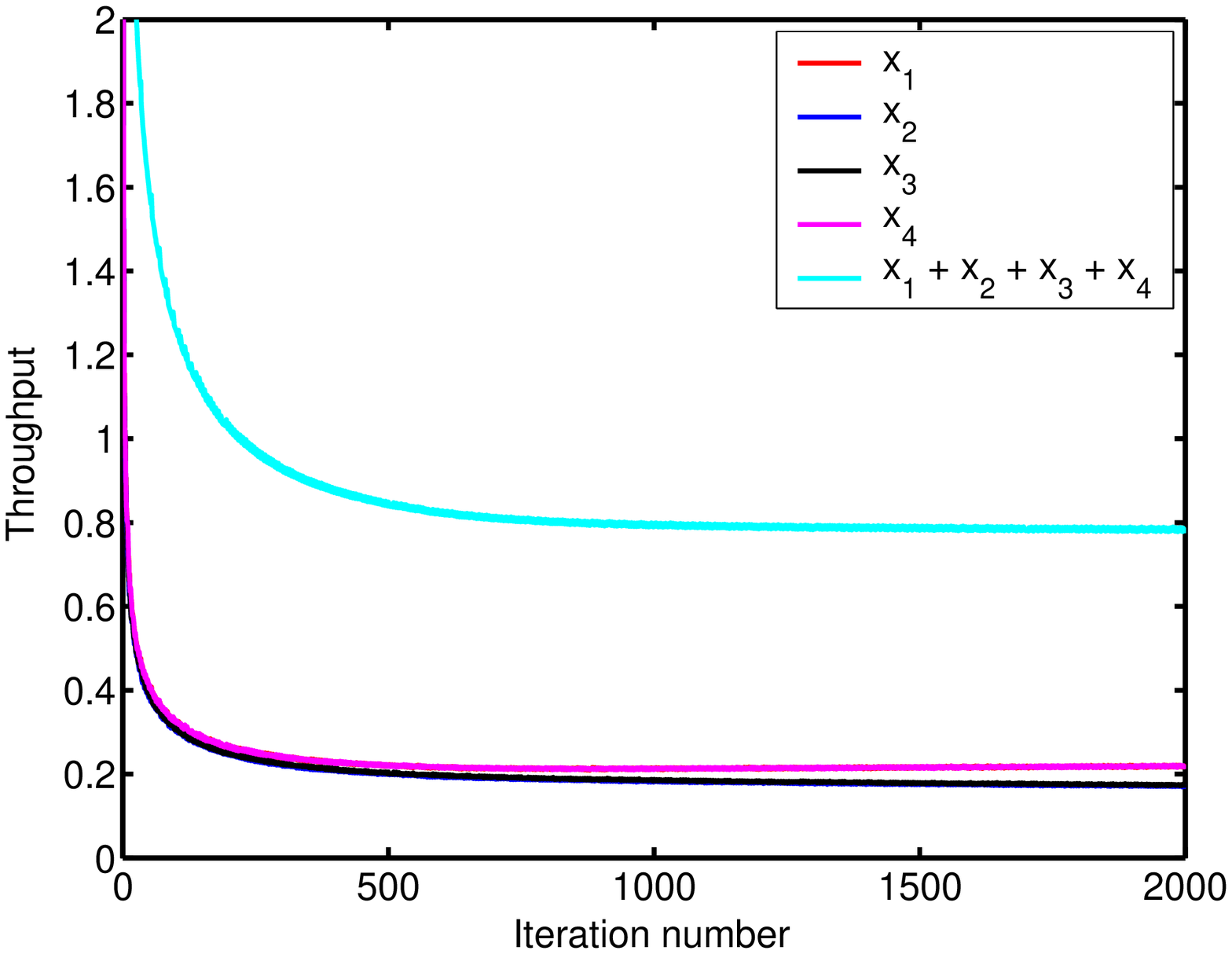}}}
\subfigure[\scriptsize Loss only on direct link $I-B_2$]{{\includegraphics[width=4.3cm]{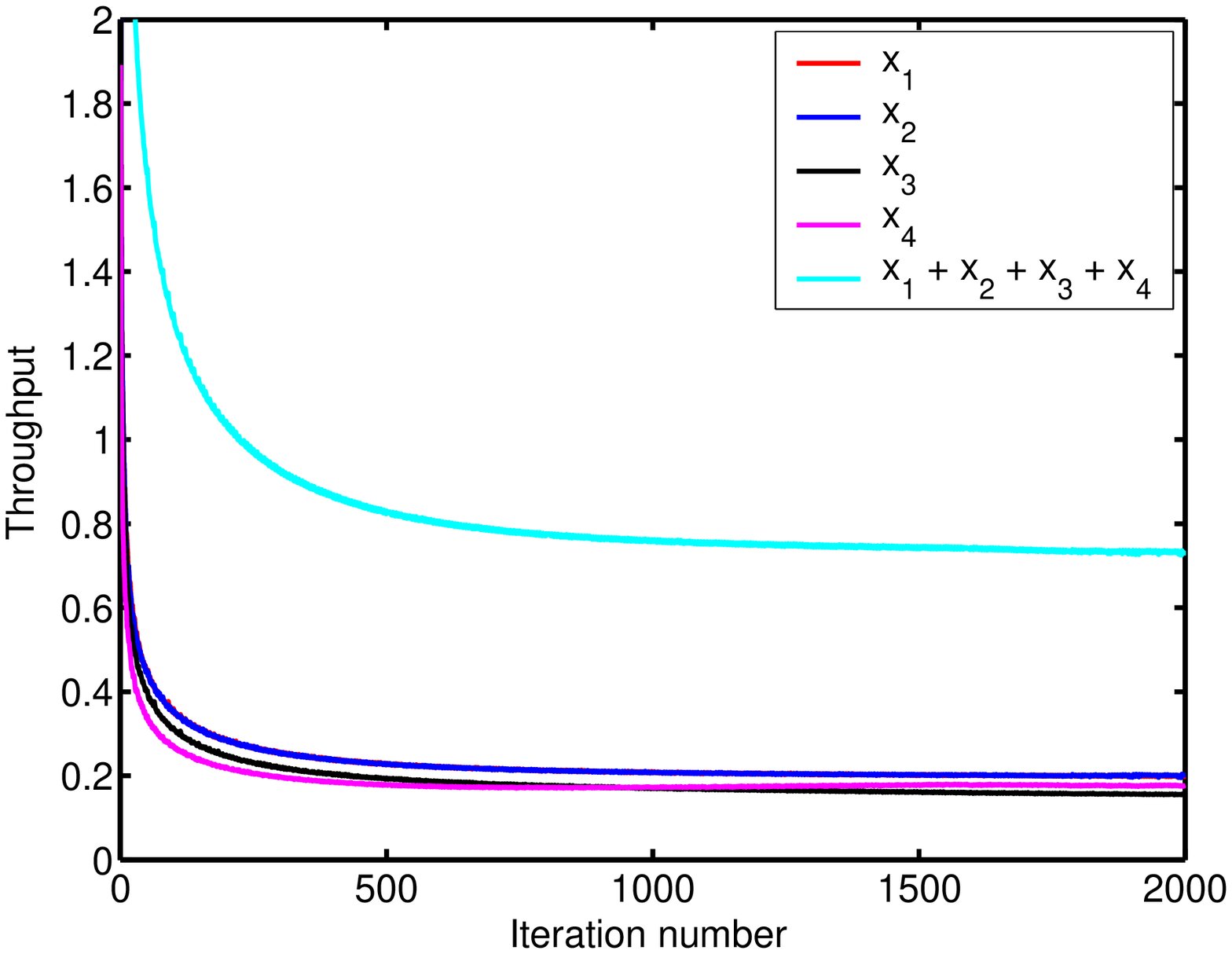}}} \\
\subfigure[\scriptsize Loss on links $A_1-B_2$ and $I-B_2$]{{\includegraphics[width=4.3cm]{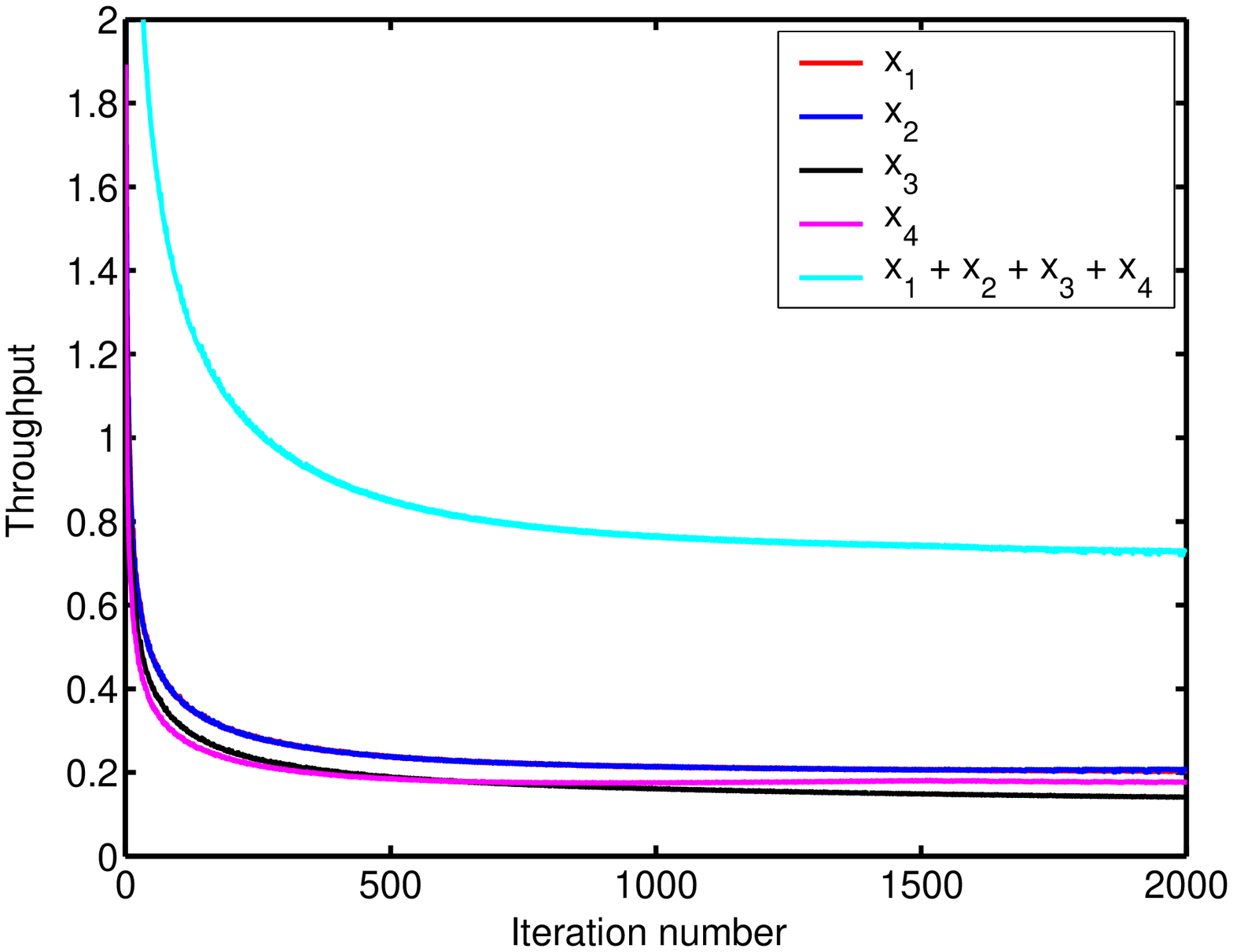}}}
\subfigure[\scriptsize Loss on all links]{{\includegraphics[width=4.3cm]{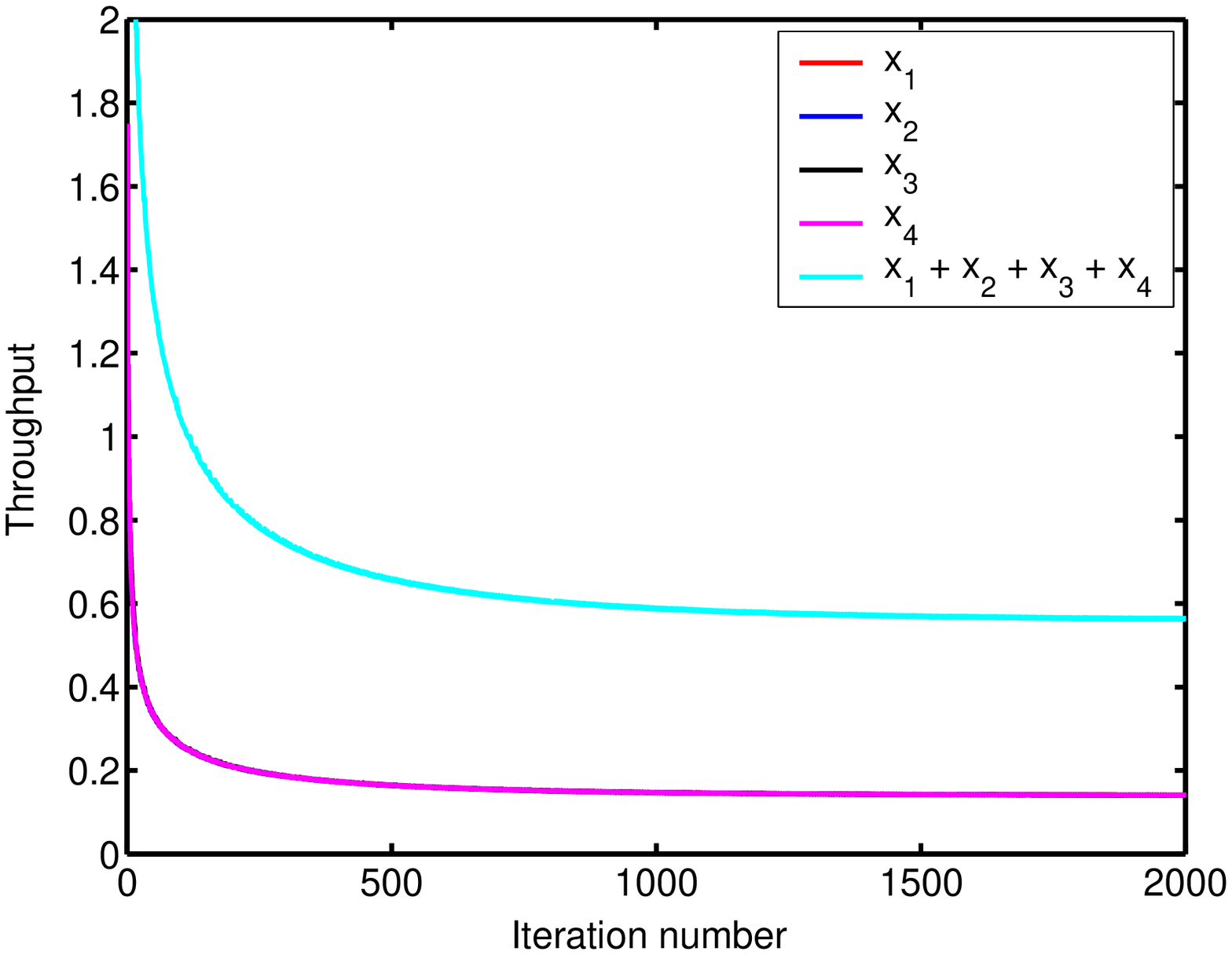}}}
%\vspace{-15pt}
\end{center}
\begin{center}
%\vspace{-10pt}
\caption{\label{fig:cross_nc_state_convergence} \scriptsize  Cross topology. Convergence of $x_1$, $x_2$, $x_3$, $x_4$, and $x_1+x_2+x_3+x_4$ for {I$^2$NC-state}. Loss rate is 30\%.}
\vspace{-10pt}
\end{center}
\end{figure}

%% Convergence for stateless - Cross topology
\begin{figure}[t!]
\begin{center}
\subfigure[\scriptsize Loss only on overhearing link $A_1-B_2$]{{\includegraphics[width=4.3cm]{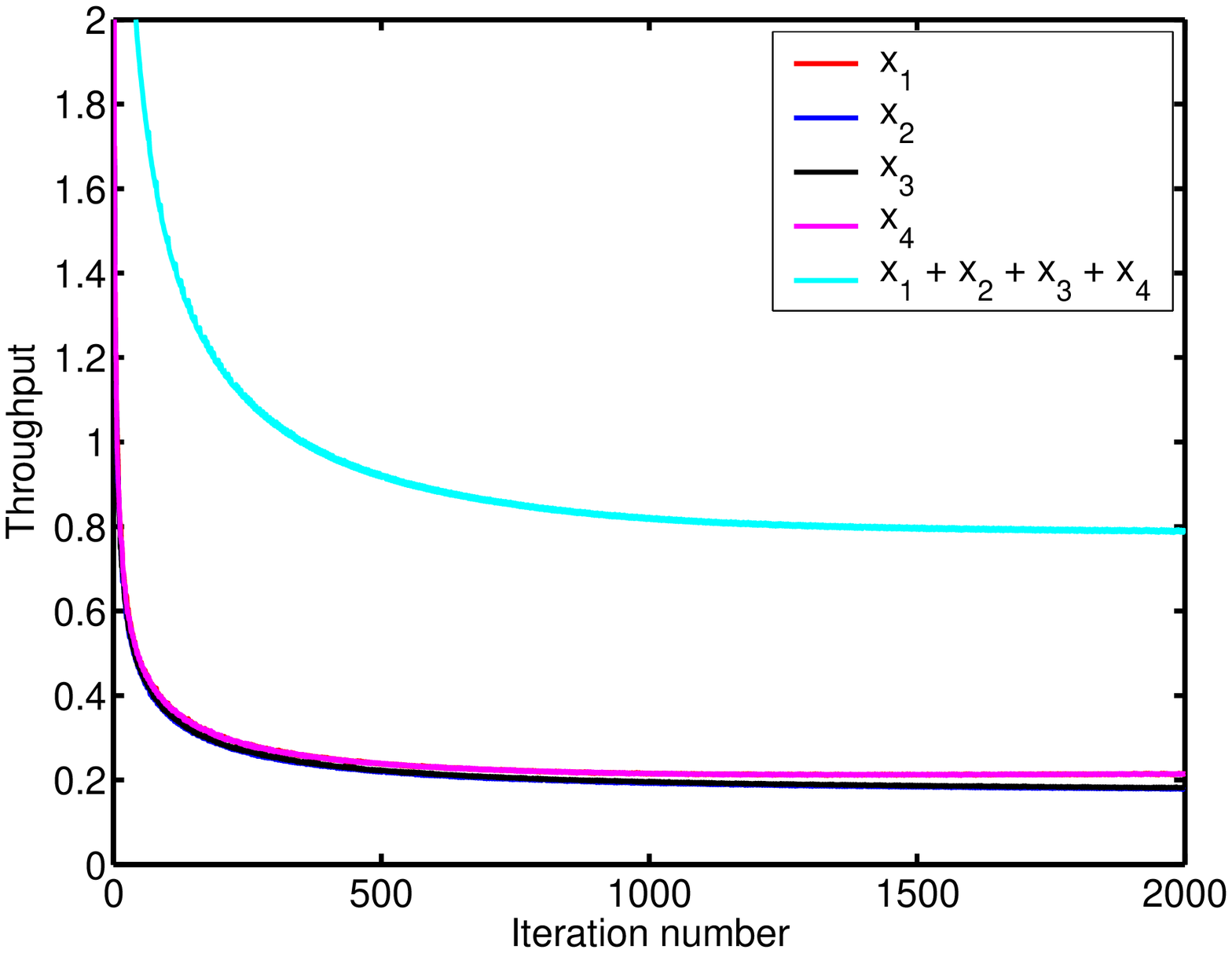}}}
\subfigure[\scriptsize Loss only on direct link $I-B_2$]{{\includegraphics[width=4.3cm]{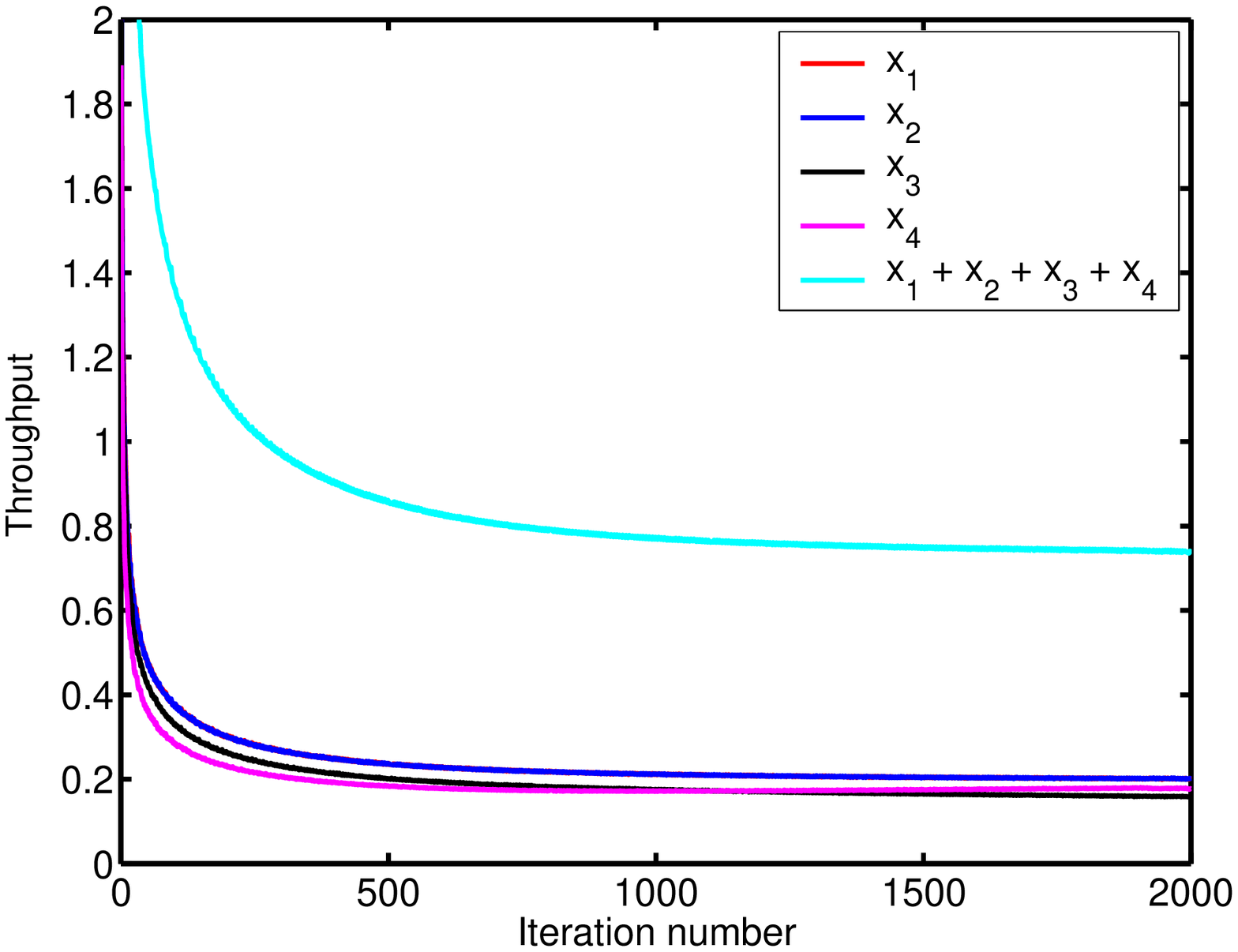}}} \\
\subfigure[\scriptsize Loss on links $A_1-B_2$ and $I-B_2$]{{\includegraphics[width=4.3cm]{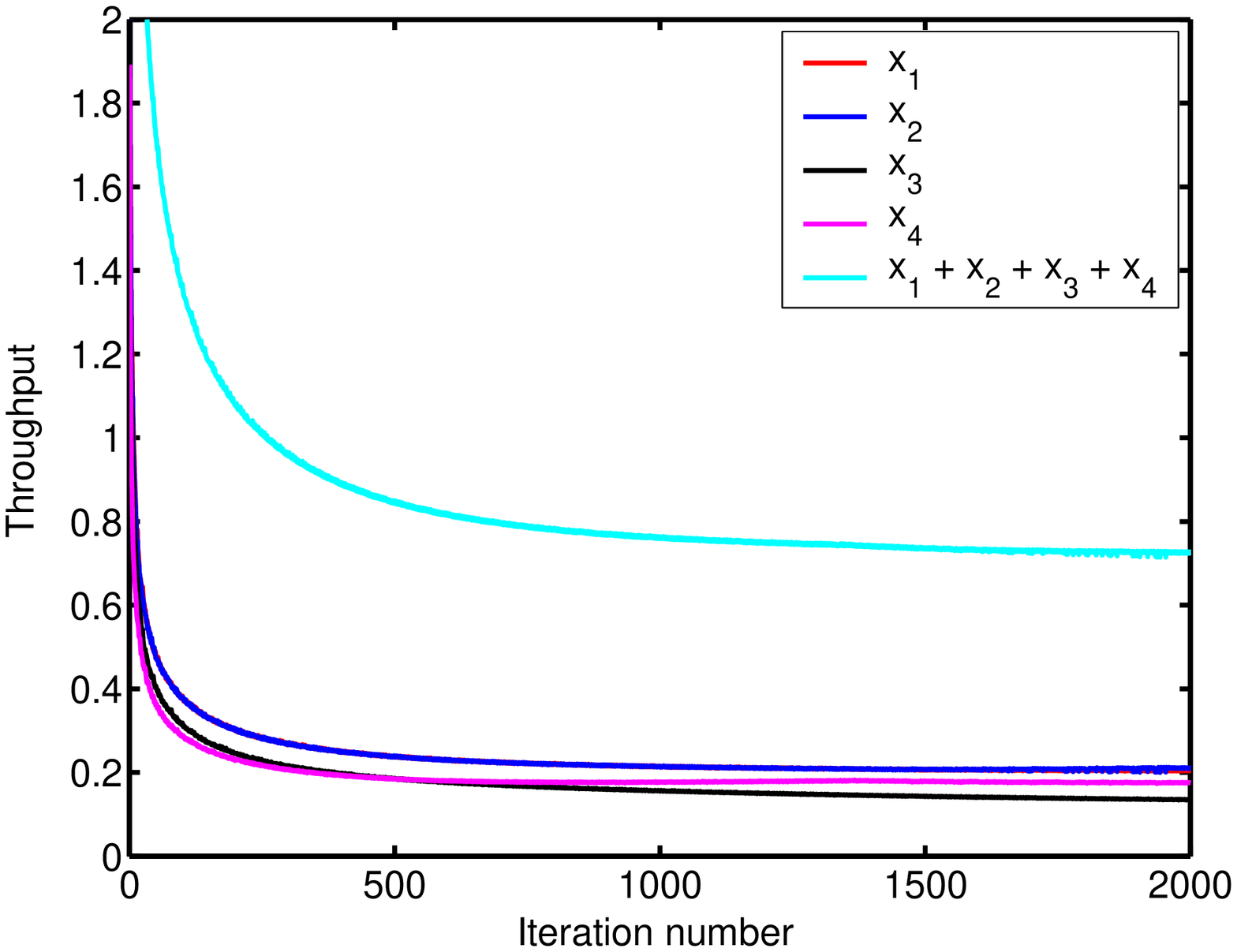}}}
\subfigure[\scriptsize Loss on all links]{{\includegraphics[width=4.3cm]{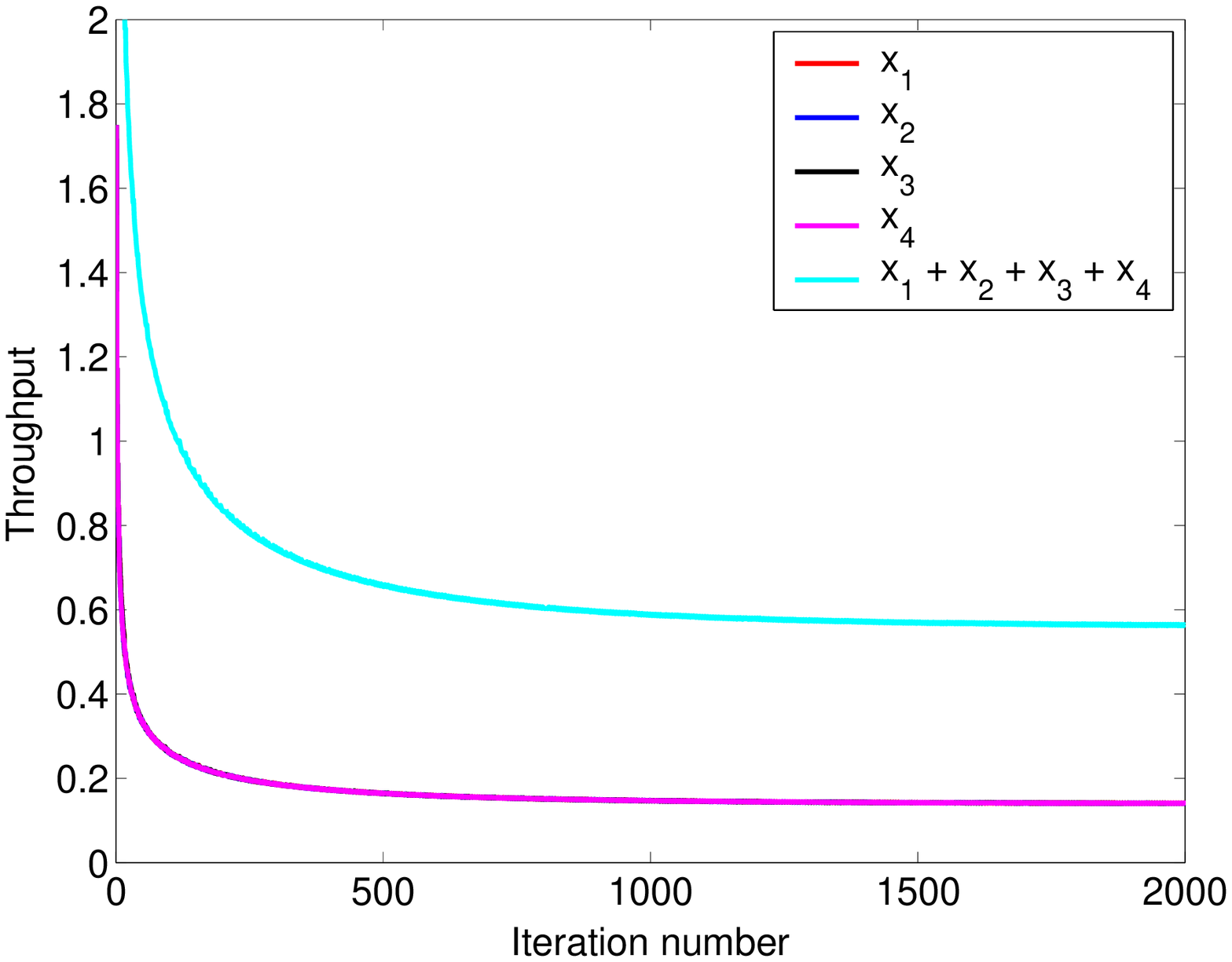}}}
%\vspace{-15pt}
\end{center}
\begin{center}
%\vspace{-10pt}
\caption{\label{fig:cross_nc_stateless_convergence} \scriptsize Cross topology. Convergence of $x_1$, $x_2$, $x_3$, $x_4$, and $x_1+x_2+x_3+x_4$ for {I$^2$NC-stateless}. Loss rate is 30\%.}
\vspace{-10pt}
\end{center}
\end{figure}

\fi

\end{document}